\documentclass{iopjournal}

\usepackage{upgreek}
\usepackage{amsmath,amssymb,amsfonts}
\usepackage{mathrsfs}
\usepackage{bbm}
\usepackage{stmaryrd}
\usepackage[OMLmathsfit]{isomath}
\usepackage[capitalise]{cleveref}
\usepackage[all]{hypcap}
\usepackage{tikz}
\newcommand{\invcircle}[1]{%
       \tikz[baseline=(char.base)]\node[
       shape=circle,
       draw=black,
       fill=black,
       inner sep=1pt,
       font=\sffamily\bfseries\small,
       text=white
       ] (char) {#1};%
}

\newcommand{\one}{\invcircle{1}}
\newcommand{\two}{\invcircle{2}}
\newcommand{\three}{\invcircle{3}}
\newcommand{\four}{\invcircle{4}}
\newcommand{\five}{\invcircle{5}}

\usepackage{secbacklinktoc}
\usepackage{tocformat}

\begin{document}

\articletype{Article} 

\title{Analytic 3D vector non-uniform Fourier crystal optics in arbitrary $\bar{\bar{\varepsilon}}$ dielectric}

\author{Chenzhu Xie~\raisebox{-0.08\baselineskip}[0pt][0pt]{\includegraphics[height=\baselineskip]{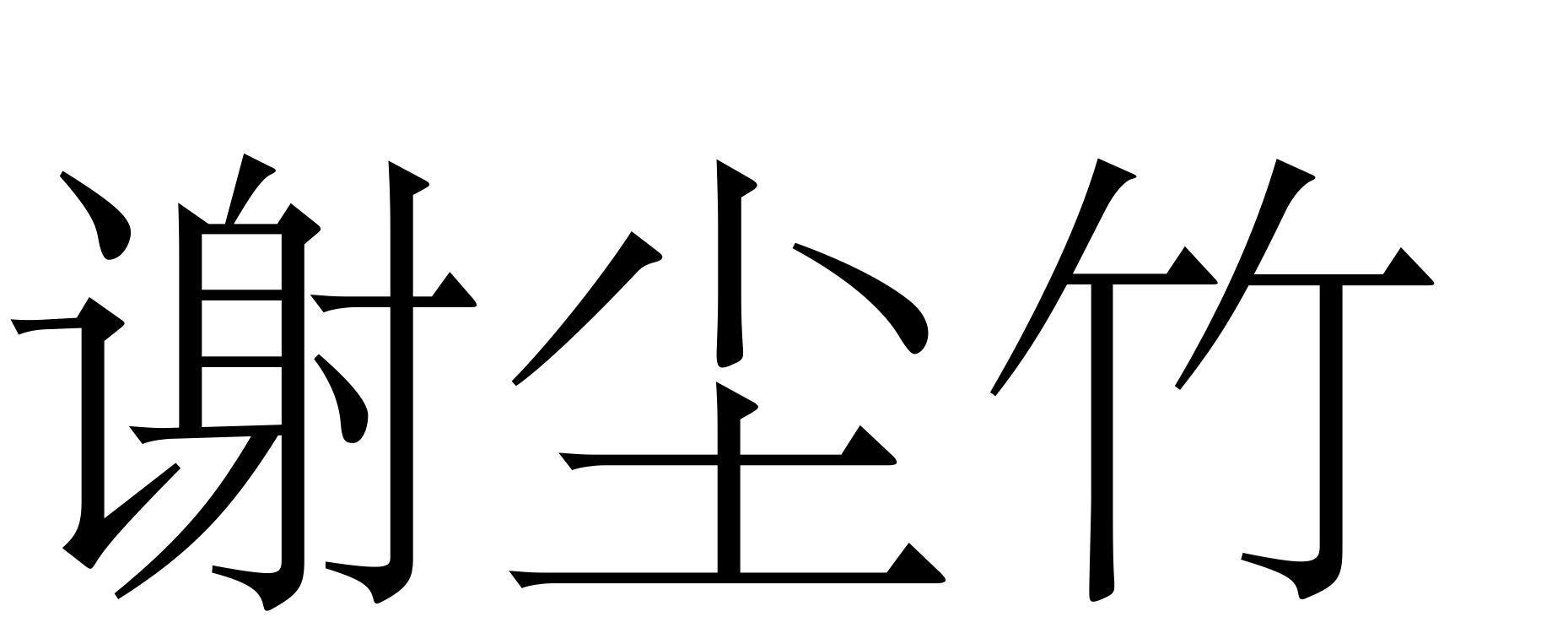}}$^{1,2,3}$\orcid{0009-0003-8987-3640} and Yong Zhang~\raisebox{-0.08\baselineskip}[0pt][0pt]{\includegraphics[height=\baselineskip]{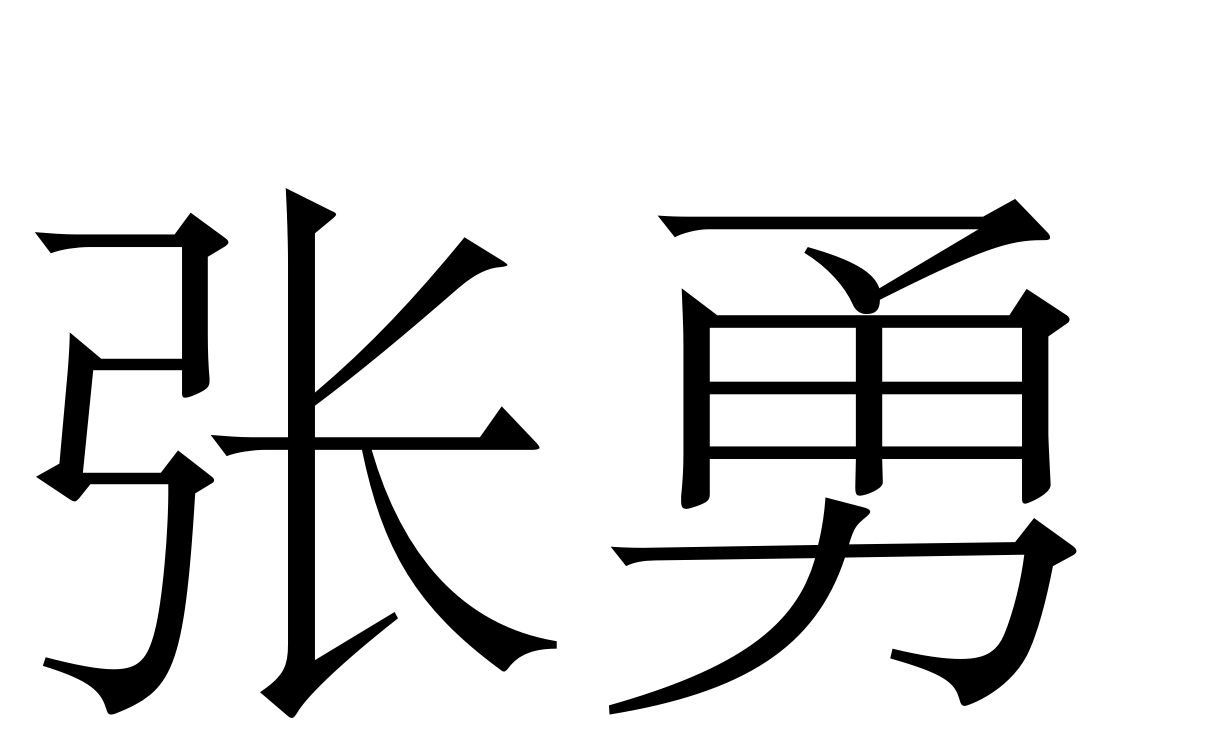}}$^{1,2,3*}$\orcid{0000-0003-1158-2248}}

\affil{$^1$National Laboratory of Solid State Microstructures, Nanjing University, Nanjing, China}

\affil{$^2$Department of Quantum Electronics and Optical Engineering, Nanjing University, Nanjing, China}

\affil{$^3$Collaborative Innovation Center of Advanced Microstructures, Nanjing University, Nanjing, China}

\affil{$^*$Author to whom any correspondence should be addressed.}

\email{zhangyong@nju.edu.cn}


\begin{abstract}
To find a suitable framework for nonlinear crystal optics(NCO), we have revisited linear crystal optics(LCO). At the methodological level, three widely used plane wave bases are compared in terms of eigenanalysis in reciprocal space and light field propagation in real space. Inspired by complex ray tracing, we expand M.V. Berry and M.R. Dennis's 2003 uniform plane wave model to non-uniform Fourier crystal optics and ultimately derive the explicit form of its 3$\times$2 transition matrix, bridging the two major branches of crystal optics in reciprocal space, where either ray direction $\hat{k}$ or spatial frequency $\bar{k}_{\uprho}$ serves as the input variable. Using this model, we create the material-matrix tetrahedral compass to conduct a detailed analysis of how the four fundamental characteristics of materials (linear/circular birefringence/dichroism) influence the eigensystems of the vector electric field in two-dimensional spatial frequency $\bar{k}_\uprho$ domain and its distribution in three-dimensional $\bar{r}$ space with a crystal-2f configuration. Along this journey, we have uncovered new territories in LCO in both real and reciprocal space, such as infinite singularities arranged in disk-, ring-, and crescent-like shapes, ``L shorelines'' resembling hearts, generalized haunting theorem, double conical refraction, and optical knots it induces. We also present our model's early applications in focal engineering and NCO. As the opening chapter in a trilogy, this work interweaves crystal optics, Fourier optics, and nonlinear optics, while integrating theoretical, computational, and experimental physics, advancing all six domains.
\end{abstract}

\tableofcontents

\section{Introduction}\label{sec:i}

\textbf{Threefold purpose guides us}: {\one} to establish crystal optics(CO) and Fourier optics(FO) as the physical and computational bases of nonlinear crystal optics(NCO) in the first part of a trilogy; {\two} to propose that three plane-wave bases form the mathematical core of CO spectral analysis, each offering trade-offs; {\three} to seek and knock the computational boundaries of linear crystal optics(LCO) and NCO via large-scale FO simulations in both three-dimensional(3D) reciprocal and two-dimensional(2D) spatial domains.

Out of the 32 macroscopic crystal symmetries, 21 display non-centrosymmetry, enabling second harmonic generation(SHG) within these point groups. However, among these 21 crystal classes, 18 are birefringent or anisotropic, and more than half($11$) exhibit natural optical rotation, demonstrating intrinsic optical activity(OA) or chirality\cite{ossikovskiConstitutiveRelationsOptically2021}. As a result, NCO often encountered tremendous computational challenges\cite{robertsSimplifiedCharacterizationUniaxial1992,dmitrievEffectiveNonlinearityCoefficients1993,diesperovEffectiveNonlinearCoefficient1997}, due to the complexity of linear crystal optical(LCO) processes in low-symmetry materials with complex dielectric tensor $\bar{\bar{\varepsilon}}$\cite{berryOpticalSingularitiesBirefringent2003}.

Think about the ordinary(o) and exordinary(e) waves in nonlinear crystal optical(NCO) phase-matching types\cite{wuVectorialNonlinearOptics2019,zondyEffectsFocusingTypeI1998,huoEffectiveMethodCalculating2015,zondyTwincrystalWalkoffcompensatedTypeII1994,vaicaitisCherenkovtypePhaseMatching2002,turpinTypeTypeII2013,magnitskiyCharacterizationPolarizationangularSpectrum2015}(e.g. type-I o$+$o$\to$e in \cref{fig:semi-vector}a): what are they? Eigenmodes governed by CO\cite{yaoAccurateCalculationOptimum1992,diesperovEffectiveNonlinearCoefficient1997,wangComplexRayTracing2008a,dineiroComplexUnitaryVectors2007,rafayelyanPlaneElectromagneticWaves2010,yeTheoreticalAnalysisCorrection2024,wangGeometricalRayTracing2018,carnioAnalysisElectricField2018}!! That is to say, the entire \textbf{NCO process is constantly subject to the constrains imposed by CO\cite{yaoAccurateCalculationOptimum1992,berryOpticalSingularitiesBirefringent2003,dmitrievEffectiveNonlinearityCoefficients1993,diesperovEffectiveNonlinearCoefficient1997}, which is itself inherently complex\cite{berryOpticalSingularitiesBianisotropic2005}}. More phenomenologically, for all frequency-mixing phenomena, both the pumps/fundamental waves(FWs) and the newly generated frequencies(NFs) $\{\omega_i\}$ within the crystal, must be first decomposed into the material's intrinsic eigenmodes, then diffract independently (and anisotropically) during subsequently linearly superposition and interference, while potentially enduring absorption or gain\cite{shengNonlinearopticalCalculationsUsing1980,grundmannOpticallyAnisotropicMedia2017}.

Beyond the anisotropy of the uniformly distributed $\bar{\bar{\varepsilon}}$ itself, the micro-nano structures modulated by fabrications on the material's $\bar{\bar{\varepsilon}}(\bar{r})$ and $\bar{\bar{\bar{\chi}}}^{(2)}(\bar{r})$\cite{xuFemtosecondLaserWriting2022,weiExperimentalDemonstrationThreedimensional2018,xuThreedimensionalNonlinearPhotonic2018,keren-zurNewDimensionNonlinear2018,chenOpticallyInducedNonlinear2022,zhangNonlinearPhotonicCrystals2021,liuNonlinearVolumeHolography2020,liuHighlyEfficient3D2023} tensors in recent engineering developments, likewise call for an immediate and precise consideration of the \textbf{linear optical(LO) scattering\cite{neviereElectromagneticResonancesLinear1995,linObservationDisorderedWave2013,goulkovNewParametricScattering2003,liConicalSecondHarmonic2015,rokeNonlinearOpticalScattering2004,katzFocusingCompressionUltrashort2011,liuScatteringassistedSecondHarmonic2016,baudrier-raybautRandomQuasiphasematchingBulk2004,xuConicalSecondHarmonic2004,anConicalSecondHarmonic2013,chenOpticallyInducedNonlinear2022} and diffraction\cite{ebersLightDiffractionSlab2020,popovMaxwellEquationsFourier2001,saltielMultiorderNonlinearDiffraction2009,zhongKdomainMethodFast2020,saltielCerenkovTypeSecondHarmonicGeneration2009,liuNonlinearVolumeHolography2020,liuHighlyEfficient3D2023} effects} caused by the material's (sub)wavelength-scale inhomogeneity, \textbf{both for FWs\cite{liReformulationFourierModal1998,shiPhysicalopticsPropagationCurved2019,gerkeAperiodicVolumeOptics2010,goulkovNewParametricScattering2003,xuConicalSecondHarmonic2004} and NFs $\{\omega_i\}$\cite{xuLargeFieldofviewNonlinear2024,grundmannOpticallyAnisotropicMedia2017,chenQuasiphasematchingdivisionMultiplexingHolography2021b,chenLaserNanoprinting3D2023,linObservationDisorderedWave2013,anConicalSecondHarmonic2013,chenOpticallyInducedNonlinear2022,zhangNonlinearPhotonicCrystals2021,liuNonlinearVolumeHolography2020,liuHighlyEfficient3D2023}}.

Hence, from first principles, NCO is fundamentally built upon LCO\cite{yaoAccurateCalculationOptimum1992,dmitrievEffectiveNonlinearityCoefficients1993,diesperovEffectiveNonlinearCoefficient1997}, which in turn is founded on linear optics(LO) and CO. This implies that \textbf{NCO simultaneously inherits all the challenges arising from both CO and LO}, each of which has undergone an extraordinarily convoluted path of mathematical development spanning nearly two centuries --- for LO, from ray optics\cite{gutierrez-cuevasRayCausticStructure2024,fogretLightrayApproachFractional2023,wangGeometricalRayTracing2018,wyrowskiApproximateSolutionMaxwell2015,avendano-alejoCausticsCausedRefraction2008} to diffraction integrals\cite{shenFastFouriertransformBasedNumerical2006,ochoaAlternativeApproachEvaluate2017,mansuripurDistributionLightFocus1986,khoninaAnalogRayleighSommerfeld2013,iqbalNoncircularlyShapedConical2022,shimobabaEfficientDiffractionCalculations2018} and eventually to FO\cite{zhangPropagationElectromagneticFields2016,pellat-finetSphericalAngularSpectrum2006,odateAngularSpectrumCalculations2011,mansuripurCertainComputationalAspects1989,leuteneggerFastFocusField2006,huEfficientFullpathOptical2020a,heintzmannScalableAngularSpectrum2023,wangApplicationSemianalyticalFourier2019,liuFastGenerationArbitrary2023,matsushimaFastCalculationMethod2003,weiModelingOffaxisDiffraction2023,huDiffractionModelingArbitrary2025}; for CO, from uniform plane-wave $\mathbbm{e}^{\mathbbm{i} k_0^{\;\!\omega} n^{\omega} \hat{k} \cdot \bar{r}}$ models\cite{yaoAccurateCalculationOptimum1992,diesperovEffectiveNonlinearCoefficient1997,chenCoordinateFreeApproach1982,gerardinConditionsVoigtWave2001,ossikovskiExtendedYehsMethod2017,changSimpleFormulasCalculating2001,zuAnalyticalNumericalModeling2022,kirillovEigenvalueSurfacesDiabolic2005,grundmannSingularOpticalAxes2016,brenierLasingConicalDiffraction2016,ossikovskiConstitutiveRelationsOptically2021,berryOpticalSingularitiesBirefringent2003,berryOpticalSingularitiesBianisotropic2005} to non-uniform plane waves $\mathbbm{e}^{\mathbbm{i} \left( \bar{k}^{\;\!\omega}_{\text{R}} + \mathbbm{i} \bar{k}^{\;\!\omega}_{\text{I}} \right) \cdot \bar{r}}$\cite{mackayExorcizingGhostWaves2019,muralidharAlgebraComplexVectors2015,dupertuisGeneralizationComplexSnellDescartes1994,dineiroComplexUnitaryVectors2007,alfonsoComplexUnitaryVectors2004,wangComplexRayTracing2008a,changRayTracingAbsorbing2005,wangComplexRayTracing2008}, $\mathbbm{e}^{\mathbbm{i} \left( \bar{k}_{\uprho} \cdot \bar{\rho} + k^{\;\!\omega}_{\mathrm{z}} z \right)}$\cite{waseerNonuniformPlaneWaves2019,zhangRigorousModelingLaser2015,asoubarSimulationBirefringenceEffects2015,mcleodVectorFourierOptics2014,sturmElectromagneticWavesCrystals2024,zhangFullyVectorialSimulation2016} and finally to matrix exponentials $\mathbbm{e}^{\mathbbm{i} \bar{\bar{k}}^{\;\!\omega}_\mathrm{z} z}$\cite{borzdovWavesLinearQuadratic1996,berremanOpticsStratifiedAnisotropic1972,stallingaBerreman4x4Matrix1999,molerNineteenDubiousWays2003,zarifiPlaneWaveReflection2014,sturmElectromagneticWavesCrystals2024}.

Given that \textbf{the impact of LCO(= LO $\boldsymbol{\times}$ CO) on NCO has been historically either ignored or modeled incorrectly}\cite{grundmannOpticallyAnisotropicMedia2017}, we undertake a thorough reexamination of LCO in Supplementary Note 1. Through this inquiry, we summarize that:

{\one} The computational boundary of LCO, and thus of NCO, terminates at optical singularities, i.e. exceptional points(EPs) in 2D reciprocal space. The optical field along such singular directions is, in principle, uncomputable due to the degeneracy/parallelism of paired eigen-polarization states $\bar{g}^{\pm}_{\;\!\omega}$ when using spectral methods like plane wave $\mathbbm{e}^{\mathbbm{i} \bar{k}^{\;\!\omega} \cdot \bar{r}} = \mathbbm{e}^{\mathbbm{i} k_0^{\;\!\omega} n^{\omega} \hat{k} \cdot \bar{r}} = \mathbbm{e}^{\mathbbm{i} \left( \bar{k}_{\uprho} \cdot \bar{\rho} + k^{\;\!\omega}_{\mathrm{z}} z \right)} = \mathbbm{e}^{\mathbbm{i} \left( \bar{k}^{\;\!\omega}_{\text{R}} + \mathbbm{i} \bar{k}^{\;\!\omega}_{\text{I}} \right) \cdot \bar{r}}$ based FO.

{\two} While matrix exponential\cite{berremanOpticsStratifiedAnisotropic1972,stallingaBerreman4x4Matrix1999,molerNineteenDubiousWays2003,zarifiPlaneWaveReflection2014} $\mathbbm{e}^{\mathbbm{i} \bar{\bar{k}}^{\;\!\omega}_\mathrm{z} z}$ do provide a workaround and Jordan decomposition\cite{hernandezScalableComputationJordan2017,seyranianCouplingEigenvaluesComplex2005,hernandezScalableComputationJordan2017,kirillovUnfoldingEigenvalueSurfaces2005,mailybaevStrongWeakCoupling2005,molerNineteenDubiousWays2003,kirillovGeometricalOpticsStability2025,kirillovDissipationinducedInstabilitiesMagnetized2018,kirillovFindingStrongestStable2021,kirillovLocatingSetsExceptional2018} further allows for an internal probe of these singularities\cite{borzdovWavesLinearQuadratic1996,wiersigDistanceExceptionalPoints2022,wiersigRevisitingHierarchicalConstruction2022,wiersigReviewExceptionalPointbased2020}, the latter becomes invalid when applying large-scale Fourier optical(FO) sampling\cite{molerNineteenDubiousWays2003,hernandezScalableComputationJordan2017}, and both suffer from other inherent computational limitations\cite{stallingaBerreman4x4Matrix1999,sturmElectromagneticWavesCrystals2024,molerNineteenDubiousWays2003,pessoaAvoidingMatrixExponentials2024} (see Supplementary Note 1). As a result, we avoid employing matrix exponentials $\mathbbm{e}^{\mathbbm{i} \bar{\bar{k}}^{\;\!\omega}_\mathrm{z} z}$ as the crystal optical(CO) Fourier basis in this LCO model for massive numerical experiments.

Only two options remain (see ``Methods''): uniform CO basis $\mathbbm{e}^{\mathbbm{i} k_0^{\;\!\omega} n^{\omega} \hat{k} \cdot \bar{r}}$ that contradicts both FO and boundary conditions(BCs), but only requires solving bi-quadratic equations thanks to spherical coordinates($\varominus$)\cite{berryOpticalSingularitiesBirefringent2003}; rectangular($\Yup$) non-uniform CO basis $\mathbbm{e}^{\mathbbm{i} \left( \bar{k}_{\uprho} \cdot \bar{\rho} + k^{\;\!\omega}_{\mathrm{z}} z \right)}$ that fits both physics(FO) and experiments($\approx$BCs), but demands batchly solving $\sim 500 \times 500$ quartics with intractably long formulas\cite{stallingaBerreman4x4Matrix1999}.

\textbf{In this Article}, we propose a non-uniform linear Fourier crystal optical(LFCO) model, whose CO part integrates the advantages of both uniform\cite{gerardinConditionsVoigtWave2001,ossikovskiExtendedYehsMethod2017,changSimpleFormulasCalculating2001,zuAnalyticalNumericalModeling2022,kirillovEigenvalueSurfacesDiabolic2005,grundmannSingularOpticalAxes2016,brenierLasingConicalDiffraction2016,ossikovskiConstitutiveRelationsOptically2021,berryOpticalSingularitiesBirefringent2003,berryOpticalSingularitiesBianisotropic2005} and non-uniform\cite{muralidharAlgebraComplexVectors2015,dineiroComplexUnitaryVectors2007,alfonsoComplexUnitaryVectors2004,wangComplexRayTracing2008a,dupertuisGeneralizationComplexSnellDescartes1994,changRayTracingAbsorbing2005,wangComplexRayTracing2008,barnettParallelNonuniformFast2019,rodriguezvarelaApplicationNonuniformFFT2020} plane wave models $\mathbbm{e}^{\mathbbm{i} k_0^{\;\!\omega} n^{\omega} \hat{k} \cdot \bar{r}} = \mathbbm{e}^{\mathbbm{i} \left( \bar{k}_{\uprho} \cdot \bar{\rho} + k^{\;\!\omega}_{\mathrm{z}} z \right)}$, solving a pure bi-quadratic under FO and boundary constraints. This LFCO model extends the standard non-diagonal/non-positive-definite\cite{ballantineConicalDiffractionDispersion2014,wuBandGapEngineering2020,qianCouplingInteractionsAnisotropic2023,passlerGeneralized442017,passlerLayerresolvedAbsorptionLight2020,harutyunyanAntidiffractionLight2015,highVisiblefrequencyHyperbolicMetasurface2015,ballantineConicalDiffractionDispersion2014,wuStrongExtrinsicChirality2021,wuNarrowbandDirectionalChiral2025,salamaFreeSpaceSuper2020}/non-Hermitian\cite{berryOpticalPolarizationEvolution2011,hernandezExceptionalPointsNonHermitian2011,songBreakupRecoveryTopological2019,shuChiralTransmissionOpen2024,burkeNonhermitianScatteringTightbinding2020,suDirectMeasurementNonhermitian2021,xiaNonlinearTuningPT2021a,tuRenyiEntropiesNegative2022,hanExceptionalEntanglementPhenomena2023,tuGeneralPropertiesFidelity2023}/non-unitary\cite{brenierPhaseDistributionsAccompanying2022,berryConicalDiffractionComplexified2006,ikramovTakagisDecompositionSymmetric2012,houdeMatrixDecompositionsQuantum2024}/non-normal\cite{wiersigDistanceExceptionalPoints2022}/non-(mirror-\cite{berryProximityDegeneraciesChiral2006})symmetric\cite{berryOpticalSingularitiesBirefringent2003,brenierPolarizationPropertiesLasing2014a,brenierVoigtWaveInvestigation2015}/non-reciprocal\cite{chatterjeeRevisitingFresnelCoefficients2003,wangAsymmetricWavefrontShaping2023,berryProximityDegeneraciesChiral2006} 2$\times$2\cite{yehGeneralizedModelWire1982} transfer\cite{zarifiPlaneWaveReflection2014,pereyraTransferMatrixMethod2022}/transition\cite{zarifiPlaneWaveReflection2014,pessoaAvoidingMatrixExponentials2024} matrix($=$ evolution operator\cite{berryChiralConicalDiffraction2006}) in CO to a 3$\times$2 form (see \cref{eq:r-5,fig:3transition2matrix}), enabling explicit fully vectorial\cite{zhangFullyVectorialSimulation2016,mcleodVectorFourierOptics2014,linFastVectorialCalculation2012,liAnalyticalVectorialStructure2013,martinez-herreroEvanescentFieldVectorial2008,chaumetFullyVectorialHighly2006,chenVectorialOpticalFields2018,napeRevealingInvarianceVectorial2022,wuConformalFrequencyConversion2022} $E_\mathrm{x},E_\mathrm{y},E_\mathrm{z}$ computation within arbitrary $\bar{\bar{\varepsilon}}$ materials, and at the same time, allows the LO part to directly operate in the non-uniform\cite{muralidharAlgebraComplexVectors2015,dineiroComplexUnitaryVectors2007,alfonsoComplexUnitaryVectors2004,wangComplexRayTracing2008a,dupertuisGeneralizationComplexSnellDescartes1994,changRayTracingAbsorbing2005,wangComplexRayTracing2008,sturmElectromagneticWavesCrystals2024,dupertuisGeneralizationComplexSnellDescartes1994} FO framework\cite{mcleodVectorFourierOptics2014,zhangPropagationElectromagneticFields2016,asoubarSimulationBirefringenceEffects2015,zhangFullyVectorialSimulation2016,zhangRigorousModelingLaser2015}.

To indirectly validate this LFCO model, and highlight its powerful role in advancing the analytic development of NCO, we offer a roadmap for expanding the LFCO model in \cref{fig:bloem_2-beautiful}c. Following its conceptual architectures --- scalar(, semi-vector,) and (full-)vector NFCO --- we first demonstrate this LFCO model's applications: {\one} in full-vector NFCO, via chiral second-harmonic conical refraction(SHCR) in \cref{fig:bloem_1-beautiful,fig:bloem_2-beautiful}, and harmonic spin-orbit(S-O) angular momentum(AM) cascade in \cref{fig:NaturePhotonics-killed} (see also Figs. SC3-SC5); {\two} in scalar NFCO, through phase-matching-controlled orbital angular momentum(OAM) conversion in \cref{fig:PRL-killed} (see also Figs. SB2-SB6); and {\three} in semi-vector NFCO, via full conical phase matching(FCPM) in \cref{fig:semi-vector} (see also Fig. SC2).

To directly validate this LFCO model, as stated at the beginning of \cref{ssec:r-2}, the procedure must begin with verifying the correctness of the eigenvalue-eigenvector $k^{\;\!\omega\pm}_{\mathrm{z}} \left( \bar{k}_{\uprho} \right), \bar{g}^{\;\!\omega\pm} \left( \bar{k}_{\uprho} \right)$ computations (Fig. S8), followed by confirming the accuracy of the field distributions $\bar{\mathsfit{E}}^{\;\!\omega}_{z} ( \bar{\rho} ) = \bar{\mathsfit{E}}^{\;\!\omega} ( \bar{r} )$ in real $\bar{r}$ space (\cref{fig:MMTC-r}, \cref{fig:high-NA}, Figs. S9-S11, and Figs. SA2-SA10).

Accordingly, we introduce the material-matrix(M-M) tetrahedron compass(TC) in Fig. S8, where scanning parameters along three of its edges depict a theoretical panorama of the adiabatic evolution(AE) of electric field eigenmodes $\bar{g}^{\;\!\omega\pm} \left( \bar{k}_{\uprho} \right) \mathbbm{e}^{\mathbbm{i} k^{\;\!\omega\pm}_{\mathrm{z}} \left( \bar{k}_{\uprho} \right) z}$ in the $\bar{k}_{\uprho}$ domain\cite{gaoTopologicalPhotonicPhase2015}, competing among the three primary material properties, i.e. linear dichroism(LD), circular dichroism(CD), and optical activity(OA). During this process, we observe that CD, adhering to the haunting (C points) theorem\cite{berryOpticalSingularitiesBirefringent2003} as well, can lead to heart-shaped L shorelines (as an upgrade to the L lines in CO\cite{berryOpticalSingularitiesBirefringent2003}) and an infinite array of singularities in 2D $\bar{k}_\uprho$ domain, often arranged in patterns resembling disks, rings\cite{kirillovUnfoldingEigenvalueSurfaces2005,merkulovPossibleConicalSingularities2011}, or crescents (instead of 8 finite EPs\cite{berryOpticalSingularitiesBirefringent2003,sturmSingularOpticalAxes2016,sturmSingularOpticalAxes2016,grundmannAngularPositionSingular2021,grundmannOpticallyAnisotropicMedia2017,sturmElectromagneticWavesCrystals2024,sturmPropagationElectromagneticWaves}). In Fig. S8, certain known phenomena reappeared\cite{kirillovUnfoldingEigenvalueSurfaces2005,merkulovPossibleConicalSingularities2011,brenierPolarizationPropertiesLasing2014}, while the others, to the best of our knowledge, are originally predicted.

Aided by crystal-2f system proposed from Fig. S9, we map out a second, more experimentally relevant\cite{pancharatnamPropagationLightAbsorbing1955,schellLaserStudiesInternal1978b,brenierVoigtWaveInvestigation2015,brenierLasingConicalDiffraction2016,brenierChiralityDichroismCompetition2017} panorama in the 3D $\bar{r}$ domain in \cref{fig:MMTC-r}, freezing the AE of the light field distribution between the three vertices of the M-M TC, corresponding to the material's birefringence(Bi), LD, and OA.

Still in real $\bar{r}$ space, but with higher numerical aperture(N.A.), we offer a unified solution for the forward propagation and inverse design of focal fields in \cref{fig:high-NA}, where a dual-eigenmode decomposition is proposed to explain Raman spikes\cite{berryConicalDiffractionAsymptotics2004} in conical refraction(CR) arising solely from slow modes, multifoci induced by laser processing inside materials, and aberration correction without Zernike polynomials. Along the way, we have observed two surprising new phenomena: double conical refraction(DCR) and the resulting optical field knots.

Moreover, our trilogy forecasts the future of quantum crystal optics(QCO) $\in$ quantum optics(QO) by demonstrating spontaneous parametric down-conversion(SPDC) in 3D nonlinear photonic crystals(NPCs), beyond the extensive faithful reproduction of past sophisticated experimental results in both LCO and NCO.


This work unites mathematical elegance(see Supplementary Notes 2-5), physical intuition(see Supplementary Notes 7), and experimental accuracy(see Supplementary Note 8-9). Even so, it too currently suffers computational limits(see Supplementary Notes 5-6).

\section{Results}\label{sec:r}

\subsection{By 2025, it still remains a great challenge to reproduce chiral SHCR}\label{ssec:canyou?}

To establish the technical soundness of this LFCO approach, we first present \textbf{results from the third-stage development} --- beyond the scope of, yet based on the LFCO model itself --- namely, full-vector nonlinear Fourier crystal optics(NFCO) \textbf{simulations versus experiments}.

\begin{figure}[htbp!]
       \centering
       \includegraphics[width=1\textwidth]{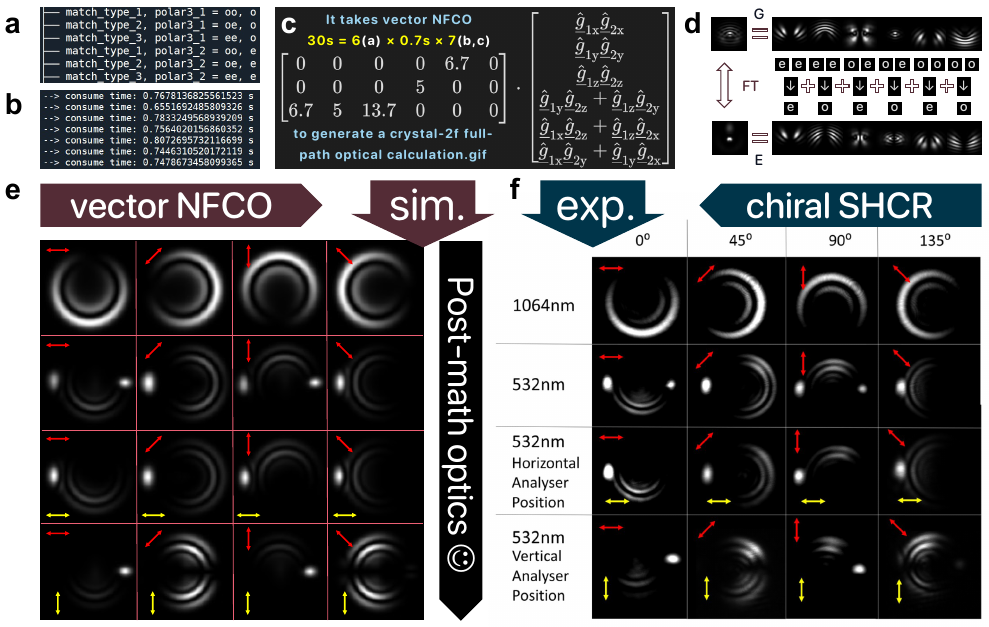}
       \caption{\textbf{Our full-vector nonlinear Fourier crystal optics(NFCO) simulations (\textbf{e}) versus Grant et al.'s experiment Fig. 3 (\textbf{f}) for \textbf{chiral second-harmonic conical refraction(SHCR)}\cite{grantFrequencydoubledConicallyrefractedGaussian2014}.} \textbf{a} All 6 NCO phase-matching types($\bar{E}_{\;\!\omega}^{\text{o,e}} \cdot \bar{E}_{\;\!\omega}^{\text{o,e}} \to \bar{E}_{\;\!2\omega}^{\text{o,e}}$ in \textbf{d}) are, for chiral\cite{zolotovskayaSecondharmonicConicalRefraction2011} SHCR, phase-mismatched and almost non-degenerate. \textbf{b} All $7 = 2 + 2 + 3$ types of nonlinear wave sources from (\textbf{c}) requiring computation. \textbf{c} All 5 non-zero components of $\mathcal{C}$-frame tensor $\bar{\bar{\underline{d}}}^{\;\!2\omega}_{\left[3 \times 6\right]}$\cite{kroupaSecondharmonicConicalRefraction2010} and 3 components of normalized $\mathcal{C}$-frame unit eigenvector(s) (fields) $\hat{\underline{g}}^{\;\!\omega}_{\pm} \left( \bar{k}_{\uprho} \right) = \hat{\underline{g}}^{\;\!\omega;\text{o,e}}_{\left[3 \times 1\right]} \left( \bar{k}_{\uprho} \right)$ of the fundamental wave(FW)(s) are involved in SHCR. \textbf{d} Intentionally picked, distorted, defocused $\mathcal{Z}$-frame second harmonic(SH) fields' intensity patterns $|\bar{\mathsfit{E}}^{\;\!2\omega}|^2 (\bar{\rho}), |\bar{\mathsfit{G}}^{\;\!2\omega}|^2 (\bar{k}_{\uprho})$ in 2D real $\bar{\rho}$ space and reciprocal $\bar{k}_{\uprho}$ space, with six phase-matching components $|\bar{E}^{\;\!2\omega}_{\pm\cdot\pm\to\pm}|^2 (\bar{\rho}), |\bar{G}^{\;\!2\omega}_{\pm\cdot\pm\to\pm}|^2 (\bar{k}_{\uprho})$, to show low field symmetry, tracing back to the material. \textbf{e},\textbf{f} A clean linearly polarized(LP) Gaussian goes in, a kaleidoscopic second harmonic wave(SHW) exits after analyzer --- the most elaborate second harmonic generation(SHG) so far: material-wise --- 1 cm($>$ $10^4$ $\lambda$) long crystal, all nonzero tensor elements($\underline{d}_{31}=\underline{d}_{15},\underline{d}_{32}=\underline{d}_{24},\underline{d}_{33}$); field-wise --- all frequencies($\omega,2\omega$), eigen-polarizations(o,e), and vector components(x,y,z), all undergoing conical diffraction(-accompanied birefringence), walk-off, and chirality-driven polarization rotation. } \label{fig:bloem_1-beautiful}
\end{figure}

The optically active($=$ chiral\cite{brenierChiralityDichroismCompetition2017}) SHCR\cite{schellLaserStudiesInternal1978b,schellLaserStudiesInternal1978a,schellLaserStudiesInternal1978,grantFrequencydoubledConicallyrefractedGaussian2014,zolotovskayaSecondharmonicConicalRefraction2011} experiment in \cref{fig:bloem_1-beautiful}f\cite{grantFrequencydoubledConicallyrefractedGaussian2014} represents an NCO experiment that, in principle, cannot be well-reproduced without a mature LCO framework\cite{zolotovskayaSecondharmonicConicalRefraction2011,kroupaSecondharmonicConicalRefraction2010,alekseevaShadowConicalRefraction1999,shihConicalRefractionSecondHarmonic1969,schellLaserStudiesInternal1978,velichkinaDemonstrationPhenomenaConical1980,stroganovConicalRefractionSecond1980,illarionovExperimentalObservationConical1979,grantFrequencydoubledConicallyrefractedGaussian2014,maSumfrequencyGenerationFemtosecond2018,peetFrequencyDoublingLaser2011}. Almost all previous NCO modeling attempts\cite{stroganovConicalRefractionSecond1980,shihConicalRefractionSecondHarmonic1969,shihConicalRefractionSecondHarmonic1969,schellLaserStudiesInternal1978,alekseevaShadowConicalRefraction1999} have invariably failed to capture the experimental phenomena at a pixel-level resolution far beyond phenomenology. Other works refrain from modeling altogether\cite{grantFrequencydoubledConicallyrefractedGaussian2014,zolotovskayaSecondharmonicConicalRefraction2011,illarionovExperimentalObservationConical1979}.

As the concluding chapter\cite{schellLaserStudiesInternal1978} of Bloembergen's trilogy\cite{schellLaserStudiesInternal1978b,schellLaserStudiesInternal1978a,schellLaserStudiesInternal1978}, \textbf{SHCR is nothing less than a Holy-Grail level\cite{olyslagerElectromagneticsExoticMedia2002} modeling challenge as an open benchmark in NCO computation, demanding equally deep command of LCO and NCO alike}. The absence of either renders the process inherently intractable.

\subsection{The union of CO, NO, and FO: a theoretical, experimental, and computational Holy Grail}\label{ssec:NCO-validate}

\textbf{The built-in difficulty of (chiral-dichroic) SHCR simulation is discussed through four lenses --- LCO, NCO, phenomenology, and computation.} As to its mathematical/theoretical/modeling challenges, for now, must be deferred, to the latter two parts, of the trilogy.

\begin{figure}[htbp!]
       \centering
       \includegraphics[width=1\textwidth]{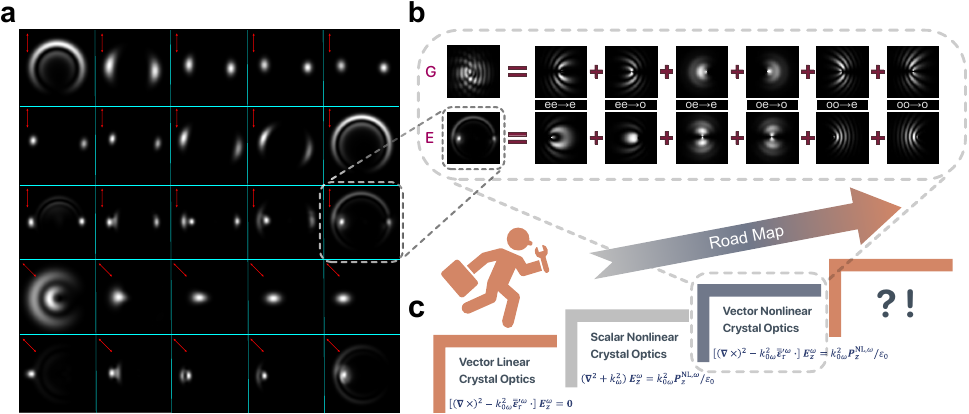}
       \caption{\textbf{Towards stage III of the trilogy: full-vector nonlinear Fourier crystal optics(NFCO).} \textbf{a} Reconstructed experiment results from Grant et al.\cite{grantFrequencydoubledConicallyrefractedGaussian2014} in their Fig. 4 on chiral second-harmonic conical refraction(SHCR) by utilizing the third stage (in \textbf{c}) of this LCO model: vector NFCO. \textbf{b} The real- and reciprocal-space distributions $|\bar{\mathsfit{E}}^{\;\!2\omega}|^2 (\bar{\rho}), |\bar{\mathsfit{G}}^{\;\!2\omega}|^2 (\bar{k}_{\uprho})$ of the 532 nm second harmonic wave(SHW) at the focal plane, generated by pumping a vertically polarized 1064 nm Gauss fundamental wave(FW) along KTP's optic axis at 532 nm, with its decomposition into six phase-matching types o,e$+$o,e$\to$o,e. } \label{fig:bloem_2-beautiful}
\end{figure}

{\one} \textbf{On a phenomenological level}, this NCO process(i.e. SHCR) involves synchronous but distinct LCO anisotropic diffraction for both the fundamental and harmonic waves(HWs). In \cref{fig:bloem_1-beautiful}f, the FW($\omega$) conical refracts(CR) along its optic axis, while the HW($2\omega$) tends to double refract(DR) (yet mixed with CR, see Fig. S9 in Supplementary Note 8), forming a CR$^\omega$$+($CR-DR\cite{berryConicalDiffractionComplexified2006})$^{2\omega}$ combination. In \cref{fig:bloem_2-beautiful}, pumped along the harmonic's axis, the configuration shifts to DR$^{\omega}$$+($DR-CR\cite{berryConicalDiffractionObservations2006})$^{2\omega}$. Each Fourier component of each $\pm$ eigenmode, in both 2D spatial and one-dimensional(1D) angular frequency domains $\bar{k}_{\uprho}; \omega(\text{or}~ 2\omega)$,  exhibits wave vector $\bar{k}^{\pm}_\omega(\text{or}~ \bar{k}^{\pm}_{2\omega})$ double(DR)/conical(CR)/double-conical(DR-CR) or even double conical refraction(DCR) (see \cref{fig:high-NA}d) at the material interfaces, superimposed on walk-off between Poynting vector $\bar{S}^{\pm}_\omega(\text{or}~ \bar{S}^{\pm}_{2\omega})$ and wave vector $\bar{k}^{\pm}_\omega(\text{or}~ \bar{k}^{\pm}_{2\omega})$ inside material.

Challenge deepens: what if the material is also chiral\cite{berryChiralConicalDiffraction2006,zolotovskayaSecondharmonicConicalRefraction2011,grantFrequencydoubledConicallyrefractedGaussian2014,saadGeneralStudyInternal2016,brenierChiralityDichroismCompetition2017,brenierPhaseDistributionsAccompanying2022,brenierLightPropagationProperties2019} and dichroic\cite{berryConicalDiffractionComplexified2006,shengNonlinearopticalCalculationsUsing1980,grundmannOpticallyAnisotropicMedia2017,brenierChiralityDichroismCompetition2017,brenierVoigtWaveInvestigation2015,brenierRevealingModesControlling2017,brenierPolarizationPropertiesLasing2014,brenierLasingConicalDiffraction2016}, beyond being birefringent\cite{grundmannOpticallyAnisotropicMedia2017,favaroNonbirefringentLimitAll2011,berryConicalDiffractionComplexified2006}? The corresponding LCO-based NCO process lies at \textbf{the pinnacle of linear crystal optics(LCO)}, a domain shaped over nearly 200 years by the world's leading LCO theorists\cite{borzdovWavesLinearQuadratic1996,eimerlQuantumElectrodynamicsOptical1988,sturmElectromagneticWavesCrystals2024,berryChapterConicalDiffraction2007,mackayElectromagneticAnisotropyBianisotropy2010,lakhtakiaCovariancesInvariancesMaxwell1995,berryChapterConicalDiffraction2007,favaroRecentAdvancesClassical2012,gerardinConditionsVoigtWave2001,lakhtakiaWhenDoesChoice2007,mackayModernAnalyticalElectromagnetic2020,baeklerKummerTensorDensity2014,favaroLightPropagationLocal2016,favaroElectromagneticWavePropagation2016,laxLinearNonlinearElectrodynamics1971,chenWavevectorspaceMethodWave1993,matosConicalRefractionGeneralized2011,berryOpticalSingularitiesBirefringent2003,berryOpticalSingularitiesBianisotropic2005,kirillovUnfoldingEigenvalueSurfaces2005,zeunerOpticalAnaloguesMassless2012,berryConicalDiffractionNcrystal2010,raabMultipoleTheoryElectromagnetism2004}, experimentalists\cite{pancharatnamPropagationLightAbsorbing1955,schellLaserStudiesInternal1978a,schellLaserStudiesInternal1978b,mikhailichenkoConicalRefractionExperiments2007,brenierPhaseDistributionsAccompanying2022,peetFarfieldStructureGaussian2013,peetExperimentalStudyInternal2014,mylnikovPartiallyCoherentConical2022,mylnikovCloseRelationshipBessel2020,sokolovskiiConicalRefractionNew2013,berryChapterConicalDiffraction2007,abdolvandConicalRefractionNd2010,brenierVoigtWaveInvestigation2015,brenierRevealingModesControlling2017,brenierPolarizationPropertiesLasing2014,brenierLasingConicalDiffraction2016,phelanConicalDiffractionBessel2009,peinadoInterferometricCharacterizationStructured2015}, applied\cite{odwyerCreationAnnihilationOptical2011,odwyerGenerationContinuouslyTunable2010,odwyerOpticalTrappingUsing2012,qianQuasiAiryBeamsTunable2016,peinadoConicalRefractionTool2013,turpinPolarizationTailoredNovel2015,ribes-pleguezueloMethodSimulateAnalyse2017,peetImprovingDirectivityLaser2010,peetBiaxialCrystalVersatile2010,turpinConicalRefractionFundamentals2016,ciattoniCircularlyPolarizedBeams2003,ciattoniAngularMomentumDynamics2003,berryOrbitalSpinAngular2005,odwyerConicalDiffractionLinearly2010,luSpinorbitInteractionsGaussian2012,qianGenerationAcceleratingBeams2019,belyiPropagationHighorderCircularly2011,khoninaComparativeInvestigationNonparaxial2015,belyiSpintoorbitalAngularMomentum2013,brenierAspectsScalingOrbital2021,brenierInvestigationSumOrbital2020,brenierEvolutionVorticesCreated2020,ahnConicalRefractionElastic2017,mohammadiDesignOpticalDiode2019,iqbalSituHologramsTwowave2024}, and computational\cite{jonesNewCalculusTreatment1941,berremanOpticsStratifiedAnisotropic1972,landryCompleteMethodDetermine1995,liReformulationFourierModal1998,borzdovWavesLinearQuadratic1996,stallingaBerreman4x4Matrix1999,zarifiPlaneWaveReflection2014,asoubarEfficientSemianalyticalPropagation2014,asoubarSimulationBirefringenceEffects2015,ribes-pleguezueloMethodSimulateAnalyse2017,zhongKdomainMethodFast2020,yangLightShapingFreeform2020a,pflaumFarFieldCalculation2022a,yehElectromagneticPropagationBirefringent1979,ariyawansaObliquePropagationLight2018,zhangAnalysisPulseFront2014,zhangAlgorithmPropagationElectromagnetic2017,wyrowskiApproximateSolutionMaxwell2015,zhangNonparaxialIdealizedPolarizer2018,abdulhalimExactMatrixMethod1999,grechinFourierSpaceMethod2014,zhangPropagationElectromagneticFields2016,zhangRigorousModelingLaser2015,mcleodVectorFourierOptics2014,zhangEfficientRigorousPropagation2012,zhangAlgorithmAccurateEfficient2024,anelliOriginDependenceMaterial2015} physicists.

Challenge deepens twice again: above LCO processes repeat itself, acting concurrently and independently across all wavelength components --- discrete $\{ \lambda^{\;\!\omega}_i \}$ for continuous-wave pumping, continuous $\{ \lambda^{\;\!\omega} \}$ for ultrafast excitation. Consider SHG, the hallmark of NCO: fundamental diffraction($\in$ LCO), harmonic diffraction($\in$ LCO), up-conversion($\in$ NCO), and down-conversion(energy backflow $\in$ NCO) --- each persists in isolation, requiring none of the other three. These 4 dynamics are not stepwise\cite{ebersLightDiffractionSlab2020}, not sequential\cite{wangSequentialThreeDimensionalNonlinear2023,jinCompactEngineeringPathEntangled2013}, not cascading\cite{tangHarmonicSpinOrbit2020,berryConicalDiffractionNcrystal2010,kalkandjievConicalRefractionExperimental2008,abdolvandConicalDiffractionMulticrystal2011,odwyerOpticalTrappingUsing2012,mateosSimultaneousGenerationSecond2012,phelanConicalDiffractionGaussian2012,fangConicalThirdharmonicGeneration2017,jangMulticycleTerahertzPulse2020}, but fully simultaneous\cite{chenQuasiphasematchingdivisionMultiplexingHolography2021,pinheirodasilvaSpinOrbitalAngular2022,mateosSimultaneousGenerationSecond2012,tangHarmonicSpinOrbit2020,liNonlinearMetasurfaceSimultaneous2017,yuSimultaneousForwardBackward2008,zhangSimultaneousNegativeRefraction2015,zhangAnalysisPulseFront2014}. From the very first moment the pump touches the crystal, all four unfold everywhere, at every instant, until the harmonics exit through both end faces, or the pumping stops.

So, we ask, how, exactly, to calculate the Chiral SHCR process in \cref{fig:bloem_1-beautiful}? Most previous efforts have ignored the foundation of NCO --- LCO, and unceasingly phenomenologically to decouple LCO(anisotropic diffraction) from NCO(frequency conversion processes), both of which occur in parallel across all spatiotemporal spectral components, all eigenmodes, and all tensor \& vector components. Some ask, parallel computation? Yes, but still either mathematically incorrect or numerically inefficient.


{\two} \textbf{On a computational level}, each of the six NCO phase-matching types($\bar{E}_{\;\!\omega}^{\pm} \cdot \bar{E}_{\;\!\omega}^{\pm} \to \bar{E}_{\;\!2\omega}^{\pm}$) in \cref{fig:bloem_1-beautiful}a,d (and \cref{fig:bloem_2-beautiful}b) is, for chiral SHCR, \textbf{phase-mismatched} and ``very likely'' \textbf{non-degenerate}. Wherein, \textbf{phase mismatch} implies that none of these processes dominates the NCO in terms of energy conversion efficiency, necessitating the simultaneous consideration of all six. \textbf{Non-degeneracy} means that outcomes of the six (eigen)mode combinations($\bar{E}_{\;\!\omega}^{\pm} \cdot \bar{E}_{\;\!\omega}^{\pm} \to \bar{E}_{\;\!2\omega}^{\pm}$) are usually distinct(see \cref{fig:bloem_1-beautiful}d), requiring separate calculations for each.

\begin{figure}[htbp!]
       \centering
       \includegraphics[width=1\textwidth]{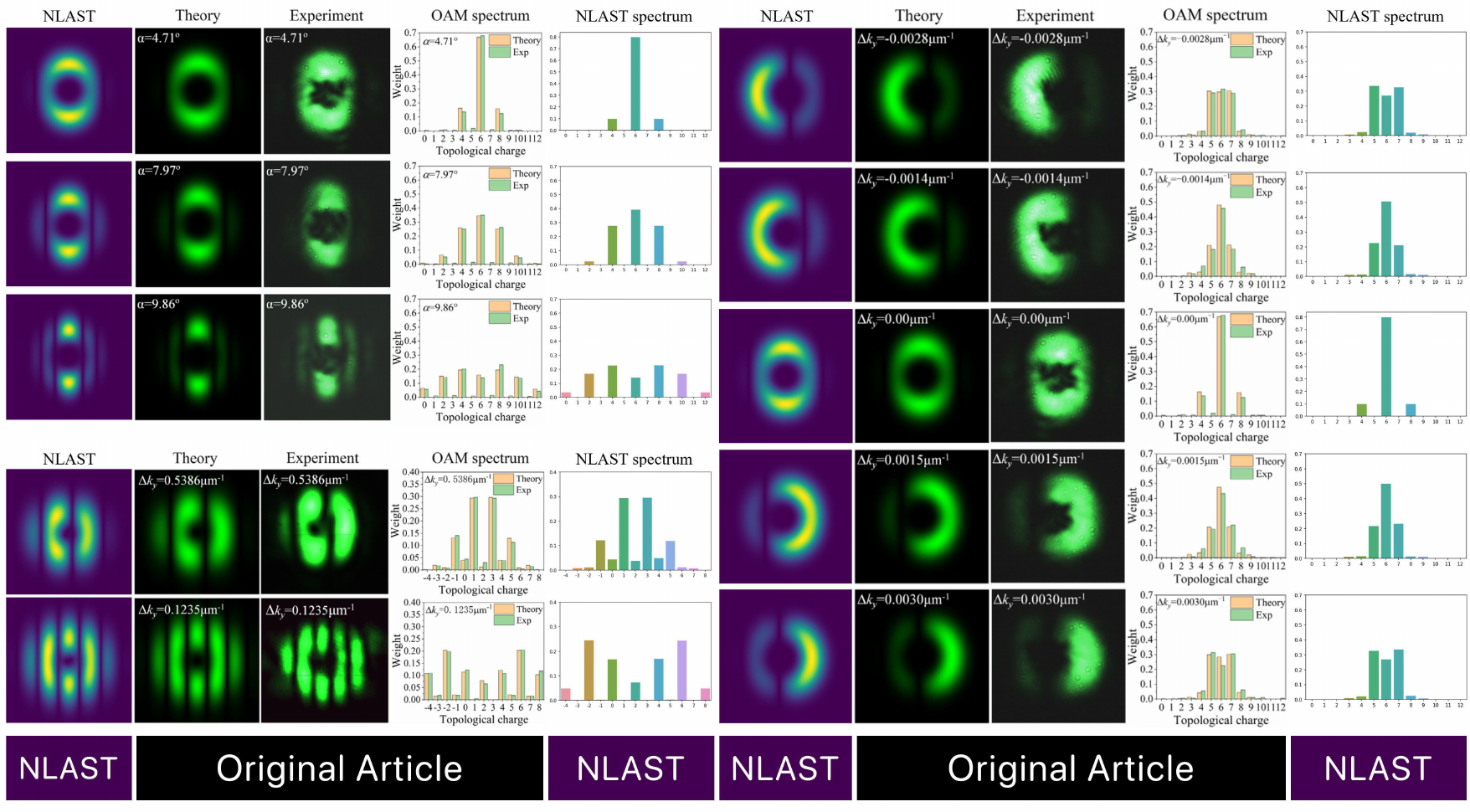}
       \caption{\textbf{Stage II of our trilogy: a scalar nonlinear Fourier crystal optics(NFCO) model, namely, the Nonlinear Angular Spectrum Theory(NLAST) reproduces Chen et al.'s all experimental figures\cite{chenPhaseMatchingControlledOrbital2020a}.}} \label{fig:PRL-killed}
\end{figure}

Except for computing all phase-matching types $\pm\cdot\pm\to\pm$ (6 for KTP), a (full-)vector, (phase-)mismatched NCO process also involves all non-zero second-order (nonlinear coefficient) tensor elements $\underline{d}^{\;\!2\omega}_{ij}$ (5 for KTP), and all nonlinear source terms (7 for KTP) from pairwise products of the fundamental wave(FW)s' eigen-polarizations' x, y, z components. --- The $\pm$($=$ o, e for KTP) eigen-polarizations and x, y, z Cartesian components are independent, leading to multiplicative complexity.

\begin{figure}[htbp!]
       \centering
       \includegraphics[width=1\textwidth]{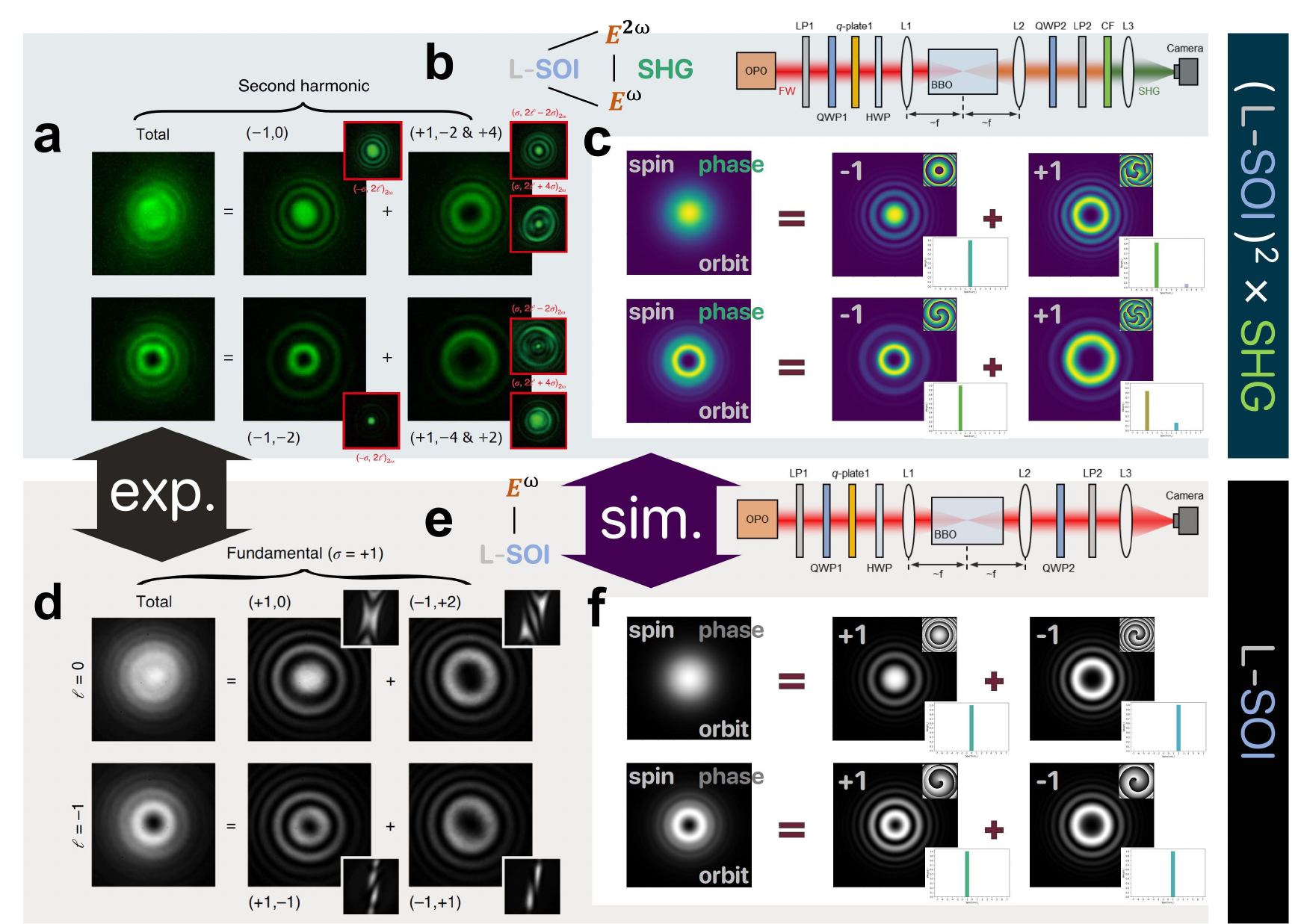}
       \caption{\textbf{Full-vector nonlinear Fourier crystal optics(NFCO) Case 2: linear and nonlinear spin-orbit interaction(SOI) with femtosecond fundamental-wave pumping along the optic axis of BBO.} \textbf{a,e} Main experimental results from Tang et al.\cite{tangHarmonicSpinOrbit2020} \textbf{c,f} The corresponding simulation using our vector NFCO model. \textbf{a,c} Emitted second harmonic wave(SHW) from BBO, and its spin-orbit (spectral) decomposition, under the simulation/laboratory setup (\textbf{b}). \textbf{d,f} Emitted fundamental wave(FW), and its spin-orbit decomposition, under the setup (\textbf{e}). } \label{fig:NaturePhotonics-killed}
\end{figure}

For KTP, one vector NFCO process takes $6 \times 7 = 42$ scalar NFCO runs (\cref{fig:bloem_1-beautiful}c). If one further scans wavelength $\lambda$ (pulse injection, see \cref{fig:NaturePhotonics-killed,fig:bloem_1-beautiful,fig:bloem_2-beautiful,fig:PRL-killed} and SC2), propagation distance $z$ (dynamical evolution, see Figs. S9-S11, \cref{fig:high-NA} and \cref{fig:bloem_1-beautiful}c), pump power, beam waist (see Fig. S11c), incident angle $\theta$ (see \cref{fig:bloem_2-beautiful}, \cref{fig:PRL-killed}, and Fig. S9), N.A. (see \cref{fig:high-NA} and Fig. S11c), temperature $T$, external magnetic field $\bar{H}_{\text{ex}}$, $\mathcal{C}$-frame orientation $\bar{O}_{\mathcal{C}}$ (see Figs. SA4-SA5), or material coefficients $\underline{\varepsilon}_{ij},\underline{d}_{ij}$ (for adiabatic tuning, see \cref{fig:MMTC-r} and Fig. S8) --- each adds a for-loop layer, and each layer may grow due to low crystallographic symmetry, or dense parameter sampling.

Therefore, to comprehensively investigate the high-dimensional parameter space of light-matter interactions, as done by Berry et al.\cite{berryOpticalSingularitiesBirefringent2003,berryOpticalSingularitiesBianisotropic2005}, Mcleod et al.\cite{mcleodVectorFourierOptics2014}, A.Favaro et al.\cite{baeklerKummerTensorDensity2014}, Hehl et al.\cite{hehlSpacetimeMetricLocal2006}, this study, and many others\cite{favaroRecentAdvancesClassical2012,mackayElectromagneticAnisotropyBianisotropy2019,berryJohnFrederickNye2020,nyePhysicalPropertiesCrystals2012,raabMultipoleTheoryElectromagnetism2004,olyslagerElectromagneticsExoticMedia2002,gaoTopologicalPhotonicPhase2015}, both linear and nonlinear crystal optics(L/NCO) call for theories/models/algorithms with a high speed-accuracy product.


The incoming second part of this trilogy (our scalar NFCO model) points out: with a $1$:$10^4$ scale mismatch --- crystal macro(10 mm\cite{grantFrequencydoubledConicallyrefractedGaussian2014}), lightwave meso(1 $\mu$m) --- Green-function formalism\cite{yaoWavefrontPhasemodulationControl2013,chenPhaseMatchingControlledOrbital2020} struggles transversely/in-plane(x-y), while split-step Fourier\cite{zoldiParallelImplementationsSplitStep1997,trajtenberg-millsSimulatingCorrelationsStructured2020,barsiImagingNonlinearMedia2009,ellenbogenNonlinearGenerationManipulation2009,rozenbergInverseDesignSpontaneous2022,yesharimObservationAllopticalStern2022} and pseudo-spectral method\cite{treebyNonlinearUltrasoundSimulation2020,treebyModelingNonlinearUltrasound2012,firouziSpacePseudospectralMethod2017} (with Runge-Kutta scheme\cite{zhongFastPropagationElectromagnetic2018,zhongKdomainMethodFast2020,zhaoNontrivialPhaseMatching2022,jangMulticycleTerahertzPulse2020}) strain longitudinally/out-of-plane(z), making forward computation of even one scalar NFCO hard, let alone all 42 in a full-vector NFCO. To thoroughly resolve the scalar NFCO process, we formulate the imminent Nonlinear Angular Spectrum Theory(NLAST), with its validity demonstrated in \cref{fig:PRL-killed} (and Figs. SB2-SB6).

\begin{figure}[htbp!]
       \centering
       \includegraphics[width=1\textwidth]{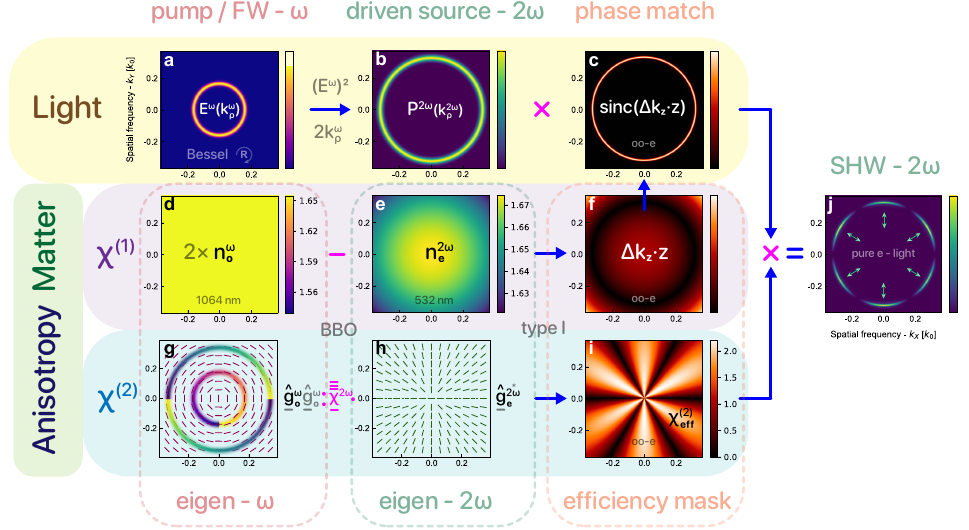}
       \caption{\textbf{Reinterprete type-I o$+$o$\to$e full conical phase matching(FCPM)\cite{belyiPropagationHighorderCircularly2011} along the optic axis of BBO crystal for second harmonic generation(SHG) within the framework of semi-vector nonlinear Fourier crystal optics(NFCO).} On the first `Light' row, by squaring the right-handed circularly polarized(RHCP) Bessel fundamental wave (\textbf{a}) and expanding its field of view with a twofold interpolation in the $\bar{k}_{\uprho}$ domain, the intracrystal nonlinear driven source $\bar{P}^{(2)}_{2\omega}$ (\textbf{b}) is obtained. This traveling field, entirely determined by the pump, is multiplied by the ``eigenvalue efficiency mask'' sinc$(\Delta k_\mathrm{z} z)$ = longitudinal phase matching coherence level (\textbf{c}) derived from the path (\textbf{d},\textbf{e}$\to$\textbf{f}$\to$\textbf{c}), followed by the ``eigenvector efficiency mask'' $\chi^{(2)\text{ooe}}_{2\omega;\text{eff}}$ = effective nonlinear coefficient distribution\cite{midwinterEffectsPhaseMatching1965,yaoAccurateCalculationOptimum1992,dmitrievEffectiveNonlinearityCoefficients1993,diesperovEffectiveNonlinearCoefficient1997} (\textbf{i}) from the path (\textbf{g},\textbf{h}$\to$\textbf{i}). The resulting output in (\textbf{j}) is a hexagonal conical radial vector light field purely composed of extraordinary light of BBO at $2\omega$.}\label{fig:semi-vector}
\end{figure}

All forms of N(F)CO modeling --- scalar (\cref{fig:PRL-killed} and Figs. SB2-SB6), semi-vector (\cref{fig:semi-vector} and Figs. SC2), or full-vector (\cref{fig:bloem_1-beautiful,fig:bloem_2-beautiful,fig:NaturePhotonics-killed} and Figs. SC3-SC5) --- depend on a vector L(F)CO model (\cref{fig:3transition2matrix,fig:MMTC-r,fig:high-NA}, Figs. S8-S11, and Figs. SA2-SA10) to yield the requisite eigenvectors (eigen-polarizations) $\hat{\underline{g}}^{\pm}_{\omega}$ and eigenvalues (eigen-wavevector-z-components for FO) $k^{\omega\pm}_{\mathrm{z}}$, necessary for evaluating the initial complex amplitudes (i.e. modal coefficients) $\mathtt{g}^{\;\!\omega}_{\pm}$ in Eq. (S56c) to calculate frequency-mixing dynamics, and for determining the nonlinear conversion efficiency $=$ ``eigenvalue mask'' sinc$(\Delta k_\mathrm{z} z)$ $\cdot$ ``eigenvector mask'' $\chi^{(2)\text{ooe}}_{2\omega;\text{eff}}$ in \cref{fig:semi-vector}.

To showcase the basics of semi-vector NFCO, \cref{fig:semi-vector}j reproduces the full conical phase matching(FCPM) second harmonic generation(SHG) proposed by Belyi et al.\cite{belyiPropagationHighorderCircularly2011}, illustrating the calculation of $\omega \to 2\omega$ conversion in a uniaxial BBO crystal through \cref{fig:semi-vector}a-i, where both the first- and second-order susceptibilities $\bar{\bar{\chi}}^{(1)}_{\omega},\bar{\bar{\chi}}^{(1)}_{2\omega}, \bar{\bar{\bar{\chi}}}^{(2)}_{2\omega}$ are anisotropic.

In subsequent (both scalar and vector) NFCO, transverse wave vector conservation (\cref{fig:semi-vector}a,b) is elevated to the same fundamental level as $\omega$ conservation and phase continuity at the boundaries, making it a prerequisite that must take priority. The conversion efficiency, determined solely by the longitudinal phase mismatch $\Delta k_\mathrm{z} z$ (\cref{fig:semi-vector}f) and effective nonlinear coefficient (\cref{fig:semi-vector}i)\cite{midwinterEffectsPhaseMatching1965,yaoAccurateCalculationOptimum1992,dmitrievEffectiveNonlinearityCoefficients1993,diesperovEffectiveNonlinearCoefficient1997}
\begin{align} \label{eq:chi_eff}
       \chi^{(2)\mathrm{p}_{312}}_{3\text{eff}} := \hat{\underline{g}}^{\mathrm{p}_{3}*}_{3{\upmu}_{3}} \underline{\chi}^{(2)}_{3{\upmu}_{312}} \hat{\underline{g}}^{\mathrm{p}_{1}}_{1{\upmu}_1} \hat{\underline{g}}^{\mathrm{p}_{2}}_{2{\upmu}_2} =: \hat{\underline{g}}^{\mathrm{p}_{3}*}_{{\omega}_3} \cdot \bar{\bar{\bar{\underline{\chi}}}}^{(2)}_{{\omega}_3} \colon \hat{\underline{g}}^{\mathrm{p}_{1}}_{{\omega}_1} \hat{\underline{g}}^{\mathrm{p}_{2}}_{{\omega}_2},
\end{align}
is subordinated to the fulfillment of transverse momentum conservation. Notably, $\mathrm{p}_1$, $\mathrm{p}_2$, and $\mathrm{p}_3$ in \cref{eq:chi_eff} represents the eigen-polarization states $\pm$ of $\omega_1$, $\omega_2$, and $\omega_3$ in the $\mathcal{C}$ frame, whose combination are fixed under certain phase-matching type. For example, phase-matching type-I o$+$o$\to$e from \cref{fig:semi-vector}g,h needs to compute $\hat{\underline{g}}^{\mathrm{p}_1}_{\omega_1}, \hat{\underline{g}}^{\mathrm{p}_2}_{\omega_1}, \hat{\underline{g}}^{\mathrm{p}_3}_{\omega_3}$ = $\hat{\underline{g}}^{\text{o}}_{\omega}, \hat{\underline{g}}^{\text{o}}_{\omega}, \hat{\underline{g}}^{\text{e}}_{2\omega}$ from \cref{eq:chi_eff}. Above $\mathcal{C}$-frame eigenvectors are almost directly provided by Berry \& Dennis's 2003 model\cite{berryOpticalSingularitiesBirefringent2003} in \cref{fig:m-1}b, though a few minor adjustments in \cref{fig:m-1}c are still required.

As two masks determining the angular distribution of frequency conversion efficiency, the degree of longitudinal phase matching sinc$(\Delta k_\mathrm{z} z)$ from \cref{fig:semi-vector}c and the effective nonlinear coefficient $\chi^{(2)\text{ooe}}_{2\omega;\text{eff}}$ from \cref{fig:semi-vector}i, respectively, depend on the eigenvalues and eigenvectors in linear Fourier crystal optics(LFCO), reaffirming that NCO is fundamentally grounded in LCO, and both linear(LO) and nonlinear optics(NO) should ultimately be incorporated into the framework of Fourier optics(FO).

\subsection{NO's two pillars: CO (3$\times$2 transition matrix) and FO (FT pairs $+$ OTFs)}\label{ssec:FCO}


Our LFCO story began with a fleeting glimpse of CO: plugging a plane-wave trial for the electric vector field $\bar{E}^{\;\!\omega} (\bar{r}) = \bar{g}^{\;\!\omega} \cdot \mathbbm{e}^{\mathbbm{i} \bar{k}^{\;\!\omega} \cdot \bar{r}}$ into the monochromatic wave equation of purely electro-anisotropic medium yields a characteristic equation\cite{mcleodVectorFourierOptics2014} (see Supplementary Note 2)
\begin{align} \label{eq:r-2}
       \left( \bar{k}^{\intercal}_{\omega}\bar{k}^{\;\!\omega} - \bar{k}^{\;\!\omega}\bar{k}^{\intercal}_{\omega} - k^{2}_{0\omega} \bar{\bar{\varepsilon}}^{\;\!\prime\omega}_{\mathrm{r} z} \right) \cdot \bar{g}^{\;\!\omega} = \bar{0}.
\end{align}
for both coordinate-free\cite{chenCoordinateFreeApproach1982,chenCoordinatefreeApproachWave1981} wave eigenvector $\bar{k}^{\;\!\omega}$ field(as FO $=$ plane-wave ensembles) and electric eigenvector field $\bar{g}^{\;\!\omega}$, with no explicit independent/free/input variable $\hat{k}$(ray direction) or $\bar{k}_{\uprho} := (k_{\mathrm{x}},k_{\mathrm{y}})^\intercal$(spatial frequency in FO), admitting both uniform spherical solution $\bar{g}^{\;\!\omega} (\hat{k}) \cdot \mathbbm{e}^{\mathbbm{i} \bar{k}^{\;\!\omega} (\hat{k}) \cdot \bar{r}}$ and non-uniform Cartesian solution $\bar{g}^{\;\!\omega} (\bar{k}_{\uprho}) \cdot \mathbbm{e}^{\mathbbm{i} \bar{k}^{\;\!\omega} (\bar{k}_{\uprho}) \cdot \bar{r}}$. Should $\bar{g}^{\;\!\omega}, \bar{k}^{\;\!\omega}, \bar{r}$ be expressed in spherical($\varominus$) coordinates, i.e. $\bar{g}^{\;\!\omega} ( \hat{k} ), \bar{k}^{\;\!\omega}( \hat{k} ), r \hat{r}$, or Cartesian($\Yup$) coordinates, i.e. $\bar{g}^{\;\!\omega}(\bar{k}_{\uprho}), \bar{k}^{\;\!\omega}(\bar{k}_{\uprho}), (x,y,z)^\intercal$?

As discussed, these two Fourier eigenbases each have their strengths and limitations (see ``Introduction'', ``Method'', and Supplementary Note 1,2): $\bar{k}_{\uprho}$-based CO is a ``black box'' --- for lacking explicitness; $\hat{k}$-based CO is white-boxed, yet fails FO and BCs.

In order to directly utilize the closed-form eigenvalues, i.e. refractive index $n^{\omega} (\hat{k})$, to the bi-quadratics from existing uniform-plane-wave CO models\cite{chenCoordinateFreeApproach1982,gerardinConditionsVoigtWave2001,ossikovskiExtendedYehsMethod2017,changSimpleFormulasCalculating2001,zuAnalyticalNumericalModeling2022,kirillovEigenvalueSurfacesDiabolic2005,grundmannSingularOpticalAxes2016,brenierLasingConicalDiffraction2016,ossikovskiConstitutiveRelationsOptically2021,berryOpticalSingularitiesBirefringent2003,berryOpticalSingularitiesBianisotropic2005}, we express the real wave vector direction $\hat{k}$ as a function of spatial frequency $\bar{k}_{\uprho}$ as follows
\begin{align} \label{eq:r-3}
       \hat{k} \left( \bar{k}_{\uprho} \right) = \mathcal{N} \left\{ \text{Re} \left[ \begin{pmatrix} \bar{k}_{\uprho}, & \sqrt{k^2_{0\omega} n^{2}_{\;\!\omega} ( \hat{k} ) - k^2_\uprho} \end{pmatrix}^{\intercal} \right] \right\},
\end{align}
which is a transcendental equation of $\hat{k}$ with $n^{2}_{\;\!\omega} ( \hat{k} )$ being provided by various even-spectrum($=$ biquadratic) LCO models\cite{chenCoordinateFreeApproach1982,gerardinConditionsVoigtWave2001,ossikovskiExtendedYehsMethod2017,changSimpleFormulasCalculating2001,zuAnalyticalNumericalModeling2022,kirillovEigenvalueSurfacesDiabolic2005,grundmannSingularOpticalAxes2016,brenierLasingConicalDiffraction2016,ossikovskiConstitutiveRelationsOptically2021,berryOpticalSingularitiesBirefringent2003,berryOpticalSingularitiesBianisotropic2005}. \cref{eq:r-3} can be solved through direct iteration (see Method), Newton's iteration\cite{grechinFourierSpaceMethod2014}, or other methods (see Discussion and Supplementary Note 6.2), with convergence typically occurring within two iterations when anisotropy is weak and the N.A. is small.

\begin{figure}[htbp!]
       \centering
       \includegraphics[width=1\textwidth]{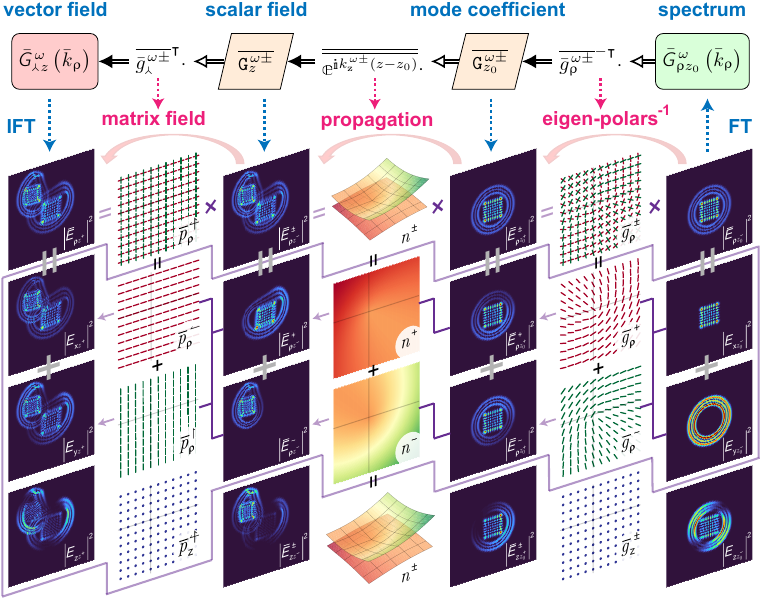}
       \caption{\textbf{The core procedure for computing the optical vector fields between any two sections within an arbitrary $\boldsymbol{\bar{\bar{\varepsilon}}}$ material, i.e., sequentially left-multipling three eigensystem matrix fields.} Below the first row, an example is provided for calculating the x, y, z components of the output vector optical field distribution (in the leftmost column). Initial condition: the known x, y components of the input 1064 nm pump (from the rightmost column), incident normally on the 15-mm-long KTP crystal with a 2$^{\circ}$ deviation off its optic axis. The vector pump is composed of vertically polarized $\text{LG}^{p=2}_{l=50}$ and horizontally polarized $\text{HG}_{6,6}$. }\label{fig:3transition2matrix}
\end{figure}

The eigenvalues $k^{\;\!\omega}_{\mathrm{z}}$ and eigenvectors (i.e., eigen-polarization states) $\bar{g}^{\;\!\omega}$ of non-uniform LFCO can be obtained by respectively substituting $\hat{k}$ obtained through \cref{eq:r-3} into
\begin{align} \label{eq:r-4}
       k^{\;\!\omega}_{\mathrm{z}} (\hat{k}) = \sqrt{k^2_{0\omega} n^{2}_{\;\!\omega} ( \hat{k} ) - k^2_\uprho}
\end{align}
and $\bar{g}^{\;\!\omega} ( \hat{k} )$ in Method. Then, as depicted in \cref{fig:3transition2matrix} and Method, the eigenvalue pairs $k^{\;\!\omega\pm}_{\mathrm{z}}$ construct the propagation matrix $\overline{\overline{\mathbbm{e}^{\mathbbm{i} k^{\;\!\omega\pm}_{\mathrm{z}} \left( z - z_0 \right)}}}_{[2 \times 2]}$, while the eigenvector pairs $\bar{g}^{\;\!\omega\pm}$ form the eigen-polarization state matrices $\overline{\bar{g}^{\;\!\omega\pm}_{\Yup}}^{\intercal}_{[3 \times 2]}$ and $\overline{\bar{g}^{\;\!\omega\pm}_{\uprho}}^{-\intercal}_{[2 \times 2]}$. A sandwich multiplication of these three matrices yields the 3$\times$2 transition matrix (see Supplementary Note 5)
\begin{subequations} \label{eq:r-5}
       \begin{align}
              \bar{\bar{\mathsfit{T}}}^{\;\!\omega}_{\Yup z \uprho} &= \overline{\bar{g}^{\;\!\omega\pm}_{\Yup}}^{\intercal} \cdot \overline{\overline{\mathbbm{e}^{\mathbbm{i} k^{\;\!\omega\pm}_{\mathrm{z}} \left( z - z_0 \right)}}} \cdot \overline{\bar{g}^{\;\!\omega\pm}_{\uprho}}^{-\intercal} \label{eq:r-5a} \\ &= \begin{pmatrix} g^{\;\!\omega+}_{\mathrm{x}} & g^{\;\!\omega-}_{\mathrm{x}} \\ g^{\;\!\omega+}_{\mathrm{y}} & g^{\;\!\omega-}_{\mathrm{y}} \\ g^{\;\!\omega+}_{\mathrm{z}} & g^{\;\!\omega-}_{\mathrm{z}} \end{pmatrix} \cdot \begin{pmatrix} \mathbbm{e}^{\mathbbm{i} k^{\;\!\omega+}_{\mathrm{z}} \left( z - z_0 \right)} & 0 \\ 0 & \mathbbm{e}^{\mathbbm{i} k^{\;\!\omega-}_{\mathrm{z}} \left( z - z_0 \right)} \end{pmatrix} \cdot \begin{pmatrix} g^{\;\!\omega+}_{\mathrm{x}} & g^{\;\!\omega-}_{\mathrm{x}} \\ g^{\;\!\omega+}_{\mathrm{y}} & g^{\;\!\omega-}_{\mathrm{y}} \end{pmatrix}^{-1}. \label{eq:r-5b}
       \end{align}
\end{subequations}
which structurally resembles the matrix exponential $\mathbbm{e}^{\mathbbm{i} \bar{\bar{k}}^{\;\!\omega}_\mathrm{z} z}$ after Jordan decomposition $\mathbbm{e}^{\mathbbm{i} \bar{\bar{k}}^{\;\!\omega}_\mathrm{z} z} = \overline{\bar{v}^{\;\!\omega}_i}^{\intercal} \cdot \left( \overline{\overline{\mathbbm{e}^{\mathbbm{i} {\lambda}^{\;\!\omega}_i z}}} + \mathbbm{e}^{\mathbbm{i} \bar{\bar{N}}^{\;\!\omega} z} \right) \cdot \overline{\bar{v}^{\;\!\omega}_i}^{-{\intercal}}$ (see Supplementary Note 1), realizing the transition $\bar{\mathsfit{G}}^{\;\!\omega}_{\uprho z_0} \to \bar{\mathsfit{G}}^{\;\!\omega}_{\Yup z}$ from the two-component vector field $\bar{\mathsfit{G}}^{\;\!\omega}_{\uprho} = \begin{pmatrix}
       \mathsfit{G}^{\;\!\omega}_{\mathrm{x}}, \mathsfit{G}^{\;\!\omega}_{\mathrm{y}}
\end{pmatrix}^\intercal$ on the input plane at $z_0$ to the three-component vector field $\bar{\mathsfit{G}}^{\;\!\omega}_{\Yup} = \begin{pmatrix}
       \mathsfit{G}^{\;\!\omega}_{\mathrm{x}}, \mathsfit{G}^{\;\!\omega}_{\mathrm{y}},
       \mathsfit{G}^{\;\!\omega}_{\mathrm{z}}
\end{pmatrix}^\intercal$ on the output plane at $z$ through $\bar{\mathsfit{G}}^{\;\!\omega}_{\Yup z} = \bar{\bar{\mathsfit{T}}}^{\;\!\omega}_{\Yup z \uprho} \cdot \bar{\mathsfit{G}}^{\;\!\omega}_{\uprho z_0}$. Coupled with Fourier transform(FT) pairs (defined in Eqs. (S4, S5)) $\bar{\mathsfit{E}}^{\;\!\omega}_{\Yup} ( \bar{\rho} ) = \mathcal F^{-1} \left[ \bar{\mathsfit{G}}^{\;\!\omega}_{\Yup} ( \bar{k}_{\uprho} ) \right]$, \cref{eq:r-5} allows us to study the evolution and 3D distribution $\bar{\mathsfit{E}}^{\;\!\omega}_{\Yup z} ( \bar{\rho} ) = \bar{\mathsfit{E}}^{\;\!\omega}_{\Yup} ( \bar{r} )$ of the vector electric field $\bar{\mathsfit{E}}^{\;\!\omega}_{\Yup} = \begin{pmatrix}
       \mathsfit{E}^{\;\!\omega}_{\mathrm{x}}, \mathsfit{E}^{\;\!\omega}_{\mathrm{y}},
       \mathsfit{E}^{\;\!\omega}_{\mathrm{z}}
\end{pmatrix}^\intercal$ in arbitary anisotropic dielectrics in real $\bar{r}$ space, as demonstrated in \cref{fig:3transition2matrix}.

Apart from CO's \textbf{in-crystal} 3$\times$2 transition matrix in \cref{eq:r-5} and FO's FT pairs in \cref{fig:3transition2matrix}, this LFCO model and the full trilogy, make broad use of self-built \textbf{out-of-crystal} optical transfer functions(OTFs) for lenses (\cref{fig:bloem_1-beautiful,fig:bloem_2-beautiful,fig:NaturePhotonics-killed,fig:PRL-killed,fig:MMTC-r,fig:high-NA} and Fig. S9), objectives (\cref{fig:high-NA,fig:NaturePhotonics-killed}), half- and quarter-wave plates (\cref{fig:NaturePhotonics-killed}), q-plates (\cref{fig:NaturePhotonics-killed}), linear polarizers/analyzers (Fig. S11), etc., all within a custom FO framework.

\subsection{Material-$\bar{\bar{\varepsilon}}$ tetrahedron compass $+$ Crystal-2f setup: guidances for batch LFCO numerical experiments in 2D reciprocal and 3D real space}\label{ssec:r-2}

\begin{figure}[htbp!]
       \centering
       \includegraphics[width=1\textwidth]{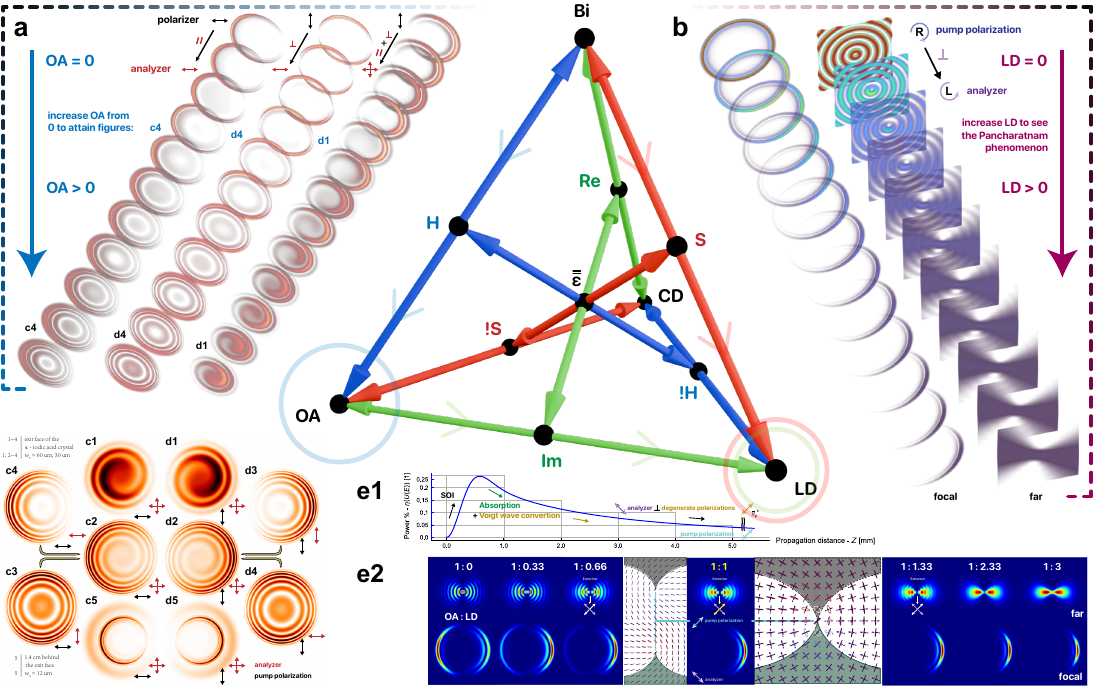}
       \caption{\textbf{Scan Bi, OA, and LD along 3 edges (`H', `S', and `Im') of the material-matrix(M-M) tetrahedron compass(TC) to see the AE of optical fields in the 3D real $\boldsymbol{\bar{r}}$ space.} \textbf{a} Increase OA to see chiral CR evolution at the focal plane, which ultimately results in (\textbf{c1}-\textbf{c5}) and (\textbf{d1}-\textbf{d5}), matching Bloembergen et al.'s experiment\cite{schellLaserStudiesInternal1978b} in their FIGs. 5A-9A and 5B-9B respectively. \textbf{b} Increase LD to see Pancharatnam phenomenon\cite{pancharatnamPropagationLightAbsorbing1955} and Brenier's anisotropic absorbing spectrum\cite{brenierVoigtWaveInvestigation2015,brenierLasingConicalDiffraction2016} (in his Fig. 6b). \textbf{e2} By Increasing LD while keeping OA constant, the competition between LD and OA is examined, providing extinction in the far field as the experimental criterion\cite{brenierChiralityDichroismCompetition2017} for OA $=$ LD under the setup where the pump's polarization $\parallel$ eigenvectors $\perp$ analyzer. \textbf{e1} The corresponding evolution of $45^\circ$ $\to$ $135^\circ$ LP SOI efficiency when OA $=$ LD.}\label{fig:MMTC-r}
\end{figure}

For this LFCO spectral method, the investigative and computational agenda unfolds according to the following sequence: different optical materials $\xrightarrow[\text{\cref{fig:MMTC-r} and Fig. S8}]{\text{Consult M-$\bar{\bar{\varepsilon}}$ TC}}$ different $\bar{\bar{\varepsilon}}$ tensors $\xrightarrow[\text{\cref{fig:m-1} and Fig. S8}]{\text{Solve \cref{eq:r-2}}}$ different eigenmodes/eigensystems/eigen-value-vector-pairs $k^{\;\!\omega\pm}_{\mathrm{z}} \left( \bar{k}_{\uprho} \right), \bar{g}^{\;\!\omega\pm} \left( \bar{k}_{\uprho} \right)$ $\xrightarrow[\text{\cref{fig:m-2,fig:3transition2matrix}}]{\text{Build \cref{eq:r-5}}}$ different 3$\times$2 transition matrices $\bar{\bar{\mathsfit{T}}}^{\;\!\omega}_{z} \left( \bar{k}_{\uprho} \right)$ $\xrightarrow[\text{\cref{fig:MMTC-r,fig:3transition2matrix}}]{\text{FT in Eqs. (S4, S5)}}$ different in-crystal field distributions $\bar{\mathsfit{E}}^{\;\!\omega}_{z} ( \bar{\rho} ) = \bar{\mathsfit{E}}^{\;\!\omega} ( \bar{r} )$ $\xrightarrow[\text{\cref{fig:NaturePhotonics-killed}, \cref{fig:high-NA} and Fig. S9}]{\text{OTFs, e.g. Single-lens 2f system}}$ different out-of-crystal field evolution $\bar{\mathsfit{E}}^{\;\!\omega}_{z} ( \bar{\rho} ) = \bar{\mathsfit{E}}^{\;\!\omega} ( \bar{r} )$.

The field itself varies over space (2D reciprocal space, for eigensystems in Fig. S8; 3D real space, for wave diffraction in \cref{fig:MMTC-r}). This work first explores the field's adiabatic evolution(AE) (Fig. S8 and \cref{fig:MMTC-r}) or dynamic evolution (Fig. S9 and \cref{fig:high-NA}) under parameter shifts, using the proposed M-M TC($=$ M-$\bar{\bar{\varepsilon}}$ TC in this work), implying a ``change of the change''.

Arbitrary complex dielectric tensors $\bar{\bar{\varepsilon}}^{\;\!\prime\omega}_{\mathrm{r}}$ are typically associated with birefringent-chiral-dichroic dielectrics\cite{merkulovPossibleConicalSingularities2011,samlanChiralDynamicsExceptional2018,takenakaUnifiedFormalismPolarization1973}. In order to relate the mathematical properties of the $\bar{\bar{\varepsilon}}$ matrix to the optical/physical properties of the material, we propose the M-M TC in \cref{fig:MMTC-r} and Fig. S8.

The M-M TC consists of 6 edges: Hermitian(H), anti-Hermitian(!H), symmetric(S), anti-symmetric(!S), real part(Re), and imaginary part(Im) --- corresponding to the six mathematical attributes of a complex matrix ($\bar{\bar{\varepsilon}}$).

The M-M TC consists of 4 vertices: birefringence(Bi), optical activity(OA), linear dichroism(LD), and circular dichroism(CD) --- describing the four optical properties of matter.

The potentially complex $\bar{\bar{\varepsilon}}^{\;\!\prime\omega}_{\mathrm{r}}$ in \cref{eq:r-2} is generally non-diagonalizable in 7 out of 11 cases, unless $\bar{\bar{\varepsilon}}^{\;\!\prime\omega}_{\mathrm{r}}$ is a normal matrix\cite{wiersigDistanceExceptionalPoints2022}, corresponding to the 4 possible combinations, that is, OA + Bi, LD + CD, Bi + LD, OA + CD, i.e. two endpoints of the M-M TC's 4 edges `H', `!H', `S', `!S', respectively.

Leveraging the reciprocal-space M-M TC in Fig. S8, we analyze the pairwise competitions among CD, OA, and LD, and how it affects of the eigensystem pairs' distribution in 2D $\bar{k}_{\uprho}$ domain in Supplementary Note 7. {\one} We established a one-to-one extension of the three principal concepts proposed by Berry \& Dennis\cite{berryOpticalSingularitiesBirefringent2003} --- optical singularities(i.e. EPs or singular axes), L lines, and C points --- by (1) generalizing the finite set of eight optical singularities into infinite families with disk-like, annular, and crescent geometries; (2) extending L lines into L lakes and their contours, namely the L shorelines; and (3) broadening the haunting theorem associated with C points from a narrow interpretation (restricted to varing LD) to a general one (also valid for varing CD). {\two} We formulated criteria for the equivalences CD $=$ OA, OA $=$ LD, and CD $=$ LD. {\three} Ultimately, all of these confirm that eigensystems of any $\bar{\bar{\varepsilon}}^{\;\!\prime\omega}_{\mathrm{r}}$ material are computable under our LFCO framework, foreshadowing general field reconstructions in 3D $\bar{r}$ domain.

Having mastered the computation of eigensystems within arbitrary $\bar{\bar{\varepsilon}}^{\;\!\prime\omega}_{\mathrm{r}}$ materials, the next task is to batch-reproduce camera-recorded data in real $\bar{r}$ space. For this purpose, both the in-crystal CO eigenmodes with their 3$\times$2 transition matrix (\cref{eq:r-5,fig:3transition2matrix}) in reciprocal $\bar{k}_{\uprho}$ space and specific out-of-crystal FO OTFs are essential.

We thus design a crystal-2f system in Fig. S9, placing a single lens at 1f(one focal) length from the rear face of the crystal, thereby performing a 2D FT between the two focal planes of the lens. The output field at the crystal's back surface --- calculated via the 3$\times$2 transition matrix --- is projected onto the XY plane at 2f, representing the strict far-field limit of the Fraunhofer diffraction.

To quantitatively validate the proposed crystal-2f system, we scanned the pump's off-axis angle $\theta$ and the propagation distance $z$ in Fig. S9, thereby identifying the precise parameters required to reproduce Peet's experimental results\cite{peetExperimentalStudyInternal2014} on internal conical diffraction with Laguerre-Gauss(LG) light beams.

We now revisit the real-space M-M TC in \cref{fig:MMTC-r}, upon completing the quantitative validation in Fig. S9 around the optic axis, i.e. the diabolic point of a typical biaxial material, where the eigenvalue degeneracy does not extend to the eigenvectors.

Along its three edges, namely `H', `S', and `Im', we broadly reproduce experimental results\cite{ballantineConicalDiffractionDispersion2014,zhangNonparaxialIdealizedPolarizer2018,berryOrbitalSpinAngular2005,tangHarmonicSpinOrbit2020,schellLaserStudiesInternal1978b,pancharatnamPropagationLightAbsorbing1955,brenierVoigtWaveInvestigation2015,brenierLasingConicalDiffraction2016,brenierChiralityDichroismCompetition2017} of Pancharatnam\cite{pancharatnamPropagationLightAbsorbing1955}, Bloembergen et al.\cite{schellLaserStudiesInternal1978b}, Brenier et al.\cite{brenierVoigtWaveInvestigation2015,brenierLasingConicalDiffraction2016,brenierChiralityDichroismCompetition2017} in 3D real space, focusing on three optical properties of matter, i.e. Bi, OA, and LD.

Perform a counterclockwise scan along the three edges of the M-M TC. First, increasing OA from 0 along the `H' edge of the M-M TC yields the AE diagram of the optical rotation field from vertice `Bi' to `OA' in \cref{fig:MMTC-r}e (pointed by the light blue arrows), ultimately converging to Bloembergen et al.'s experiment: chiral CR\cite{schellLaserStudiesInternal1978b} through $\alpha$-HIO$_3$. Next, while maintaining OA, as LD increases along the `Im' edge from 0 (denoted by the light green arrows), Brenier's experimental results on chirality versus dichroism\cite{brenierChiralityDichroismCompetition2017} of acentric chiral Nd$^{3+}$-doped BZBO are obtained, with a clear criterion for OA $=$ LD. That is, when pump's polarization is aligned parallel to the polarization of Voigt wave while the analyzer is set perpendicular, central extinction is observed in the far field. Following this, by canceling OA and keeping only LD, the angular absorption distribution of laser crystal KGd(WO$_4$)$_2$ is obtained\cite{brenierVoigtWaveInvestigation2015,brenierLasingConicalDiffraction2016}. Then, by gradually reducing LD along the `S' edge (antiparallel to the light red arrows), the Pancharatnam phenomenon\cite{pancharatnamPropagationLightAbsorbing1955} emerges.

Finally, as LD is reduced to 0, only Bi remains, including uniaxiality (Fig. S10b and Fig. S11), biaxiality (Fig. S10a and Fig. S11b1), and their hyperbolic counterparts (Fig. S10a,b). As an effect caused by the off-diagonal elements of the 2$\times$2 eigen-polarization matrix $\overline{\bar{g}^{\;\!\omega\pm}_{\uprho}}^{-\intercal}$ and an embodiment of conservation law, the global grasp and detailed calculation of SOI play a significant role in not only LCO\cite{ciattoniCircularlyPolarizedBeams2003,ciattoniAngularMomentumDynamics2003,berryOrbitalSpinAngular2005,luSpinorbitInteractionsGaussian2012,belyiPropagationHighorderCircularly2011,belyiSpintoorbitalAngularMomentum2013,khoninaComparativeInvestigationNonparaxial2015,brenierAspectsScalingOrbital2021,brenierInvestigationSumOrbital2020,brenierEvolutionVorticesCreated2020} (see \cref{fig:3transition2matrix} and Figs. S9, S11 and SA3-SA5) but also NCO\cite{tangHarmonicSpinOrbit2020,liSpintoorbitalAngularMomentum2020,wuVectorialNonlinearOptics2019,liNonlinearMetasurfaceSimultaneous2017} (see \cref{fig:NaturePhotonics-killed}).


\subsection{Tightly focused light in highly anisotropic materials}\label{ssec:r-4}

\begin{figure}[htbp!]
       \centering
       \includegraphics[width=1\textwidth]{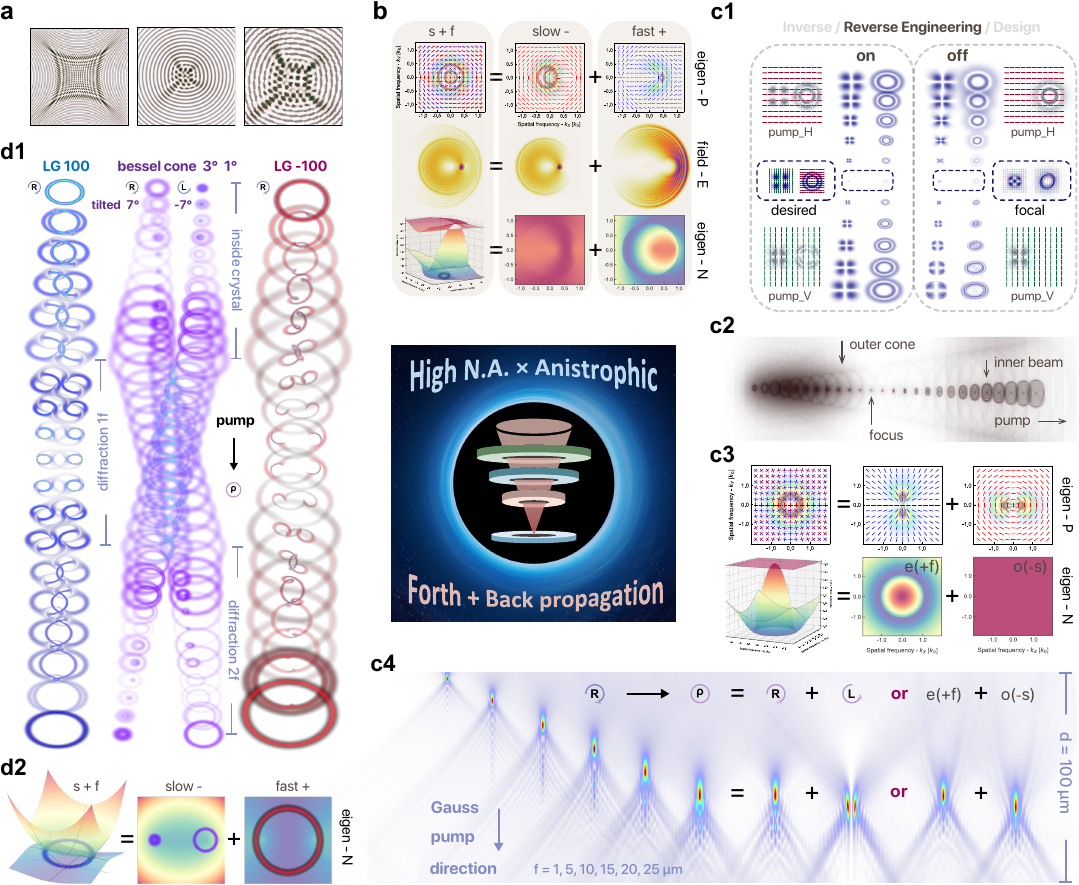}
       \caption{\textbf{Forward and backward propagation in anisotropic materials with high N.A. pump.} \textbf{a} Bessel caustics reconstructed from Zusin et al.\cite{zusinBesselBeamTransformation2010}. \textbf{b} Raman spike\cite{berryConicalDiffractionAsymptotics2004} of CR origins solely from the slow mode. The vertically polarized off-optical-axis LG$_1$ pump is chosen to pump $\alpha$-HIO$_3$. \textbf{c1} Inverse design of tightly focused focal fields: The backward propagation(BP) of the superposed $\pm$ eigenmodes (left) eliminates the aberrations introduced by the direct isotropic BP (right) of the target desired vector field in the focal plane: horizontally polarized LG$_{l=-10}$\cite{linFastVectorialCalculation2012} $+$ vertically polarized HG$_{11}$, without the need for any Zernike polynomial compensation. \textbf{c2} In the case of high N.A., outside the axial field that first focuses and then diverges, a ``Cherenkov cone'' as an aliasing error that continuously diverges exists independently. \textbf{c3} Lithium niobate(LN)'s eigensystems used to simulate (\textbf{c1}, \textbf{c2}, and \textbf{c4}), where vertically polarized Gaussian is used to simulate (\textbf{c2}). \textbf{c4} RHCP Gauss pump travels along the optic axis of LN, with the objective aperture at the surface of the material. The first 6 spots from the left: By switching focal lengths, the intensity section distribution within the material at different depths without wavefront correction. Spots 7-10: Decomposing the sixth spot into L/RHCP or eigenmodes: respectively reveals the transverse SOI and the nature of axial multifocality when no aberration correction is applied --- interference between mismatched o- and e-waves. \textbf{d1} A single light beam (left \& right), or two coherent beams (middle), whose angular spectrum distributions in $\bar{k}_{\uprho}$ domain cover both optical axes of the (hyperbolic) biaxial crystal, undergo two separate conical refractions that superimpose and interfere. --- In the case of a single beam, light field self-twisting behavior is likely to occur, especially when the reciprocal space coverage of the pump's angular spectrum is greater than (right) or equal to (left) the angle of bi-axes. \textbf{d2} The distributions of the three pumps from (\textbf{d1}) in reciprocal space relative to the refractive index.}\label{fig:high-NA}
\end{figure}

In the extreme case of strong linear interaction between highly anisotropic materials and tightly focused light fields --- an area at the forefront of both industry and academia in laser processing, aberration correction, and inverse focal engineering --- our LFCO model provides a unified solution, as shown in \cref{fig:high-NA}. Here, we show the underlying mechanism of Raman spike\cite{berryConicalDiffractionAsymptotics2004} (\cref{fig:high-NA}b) in CR, the reverse engineering technique for the targeted vector complex field at the focal plane free of Zernike polynomial (\cref{fig:high-NA}c1), the ``Cherenkov cone'' outside the inner beam (\cref{fig:high-NA}c2) marking the computational boundary, the R-L and o-e decomposition analysis for high-N.A. pump (\cref{fig:high-NA}c4), and double conical refraction and the optical field self-twisting effects it induces (\cref{fig:high-NA}d1).

The interference of complex fields gives rise to the intricate and distorted shapes of light spots, especially in the case of tightly focusing. This phenomenon, emblematic of wave optics (paralleling quantum mechanics), as its most distinctive feature differentiating ray optics (analogous to classical mechanics), not only represents the mathematical frontier of highly oscillatory partial differential equations(PDEs), where both the inner integrand and the integral outcome are potentially highly oscillatory\cite{agocsAdaptiveSpectralMethod2024}, but also pushes the computational limits in raising the speed-accuracy product under the constraints of the Nyquist(-Shannon) sampling theorem\cite{wangTheoryAlgorithmHomeomorphic2020}, thus stands as one of the most promising candidates for fully challenging the performance of all neural networks claiming that they are physical\cite{wrightDeepPhysicalNeural2022}.

The strong anisotropy of the material further heightens the computational demands for calculating the light field distribution. This arises from the enhanced distortion of the electromagnetic field's eigenvalue surfaces and their associated wavefronts (equiphase surfaces), compounded by the faster variation of eigenvectors controlling polarization directions which ultimately implement interference along x, y, z axes, hastening the premature arrival of the nightmare where phase differences exceeding $\pi$ between adjacent pixels\cite{leuteneggerFastFocusField2006,heintzmannScalableAngularSpectrum2023} on a typical planar interference pattern at $z=z_0$.

Using the crytal-version vector angular spectrum method(ASM), i.e. this LFCO model, without any optimization, we explored various tightly focused field distributions while operating at the edge of the sampling theorem, with each subfigure (except \cref{fig:high-NA}c2) in \cref{fig:high-NA} nearing the computational limits just before aliasing errors occur. Initially, we believed that the origin of the outer cone in \cref{fig:high-NA}c2 was physical because: (1) it began forming even before propagation, and (2) it appeared at the center of the image rather than being reflected from the edges. However, we later discovered that it was still caused by aliasing errors originating from circular convolution. This conclusion was drawn because, in order to shift the focal plane of \cref{fig:high-NA}c2 inside the LN materiral, we set the incident field distribution at the material's entrance to a LP Gaussian beam with a N.A. close to 1 (see the top row of \cref{fig:high-NA}c3), propagating backward by $z_0=-0.2$ mm. However, when $z_0$ was set to zero, the outer cone disappeared. In contrast, \cref{fig:high-NA}c4 uses the transfer function of the objective lens instead of backward propagation to shift the focal plane below the upper surface of LN, completely eliminating the aliasing error.

We find that by setting $\underline{\varepsilon}_{\text{zz}}$ negative, thereby transforming the refractive index surface from an ellipsoid to a dielectric-type two-sheet hyperboloid\cite{wuBandGapEngineering2020,harutyunyanAntidiffractionLight2015,highVisiblefrequencyHyperbolicMetasurface2015,ballantineConicalDiffractionDispersion2014,salamaFreeSpaceSuper2020}, the two optical axes can be brought closer together, as shown in \cref{fig:high-NA}d2. For KTP crystal at 1064 nm, the angle between the two optical axes decreases from 34.6$^\circ \times 2$ in the ellipsoidal case to 7.7$^\circ \times 2$ in the hyperboloidal case, allowing us to compute the full evolution inside and outside the crystal in a non-paraxial manner over relatively longer propagation distances of the entire crystal-2f system. Optical field knotting in the 1st and 3rd column of \cref{fig:high-NA}d1 are beyond our present mathematical understanding and techniques for explanation.

\subsection{Superstructure of this LCO model: NCO $\to$ QNCO}\label{ssec:r-5}

Nonlinear optics(NO), as a profound gateway for exploring and understanding light-matter interactions, introduces all higher-order nonlinear terms $\bar{P}^{(2)}_{\omega} + \bar{P}^{(3)}_{\omega} + \cdots$ in the CRs beyond the first-order linear electric susceptibility $\bar{\bar{\chi}}^{(1)}_{\omega} = \bar{\bar{\varepsilon}}^{\;\!\prime\omega}_{\mathrm{r}} - 1$. These higher-order terms represent the nonlinear response of bound electric dipoles to external optical fields. Ultimately, they act as cross-band light sources that coherently generate NFs within crystals through parametric radiation (there are also non-parametric/inelastic cases where phonons or molecules are involved\cite{grundmannOpticallyAnisotropicMedia2017,boydNonlinearOptics2019}).

Both the NFs $\{\omega_i\}$ and the pumps involved in the interaction, are constrained by their own monochromatic passive LCO wave equation, necessitating independent diffraction as the crystal's eigenmodes (\cref{fig:semi-vector}d,e,g,h).

The NCO parametric frequency conversion process must first satisfy energy conservation, followed by momentum conservation (often described as wave vector or phase matching), which further tests the precision of LCO eigenvalue calculations (\cref{fig:semi-vector}d,e). If the anisotropy of the second-order nonlinear coefficient tensor $\bar{\bar{\bar{\chi}}}^{(2)}_{\omega}$ is additionally involved, the accurate computation of eigenvectors (\cref{fig:semi-vector}g,h) in the $\mathcal{C}$ frame\cite{midwinterEffectsPhaseMatching1965,yaoAccurateCalculationOptimum1992,dmitrievEffectiveNonlinearityCoefficients1993,diesperovEffectiveNonlinearCoefficient1997} within LCO must also be ensured as a prerequisite.

These two core principles, i.e, passive independent diffraction and active coupled conversion of all $\{\omega_i\}$ that participate in NCO processes, anchor all NCO phenomena within the framework of LCO. The precise theoretical modeling of NCO processes in anisotropic materials becomes naturally, a more rigorous test for all established LCO models.

\begin{figure}[htbp!]
       \centering
       \includegraphics[width=1\textwidth]{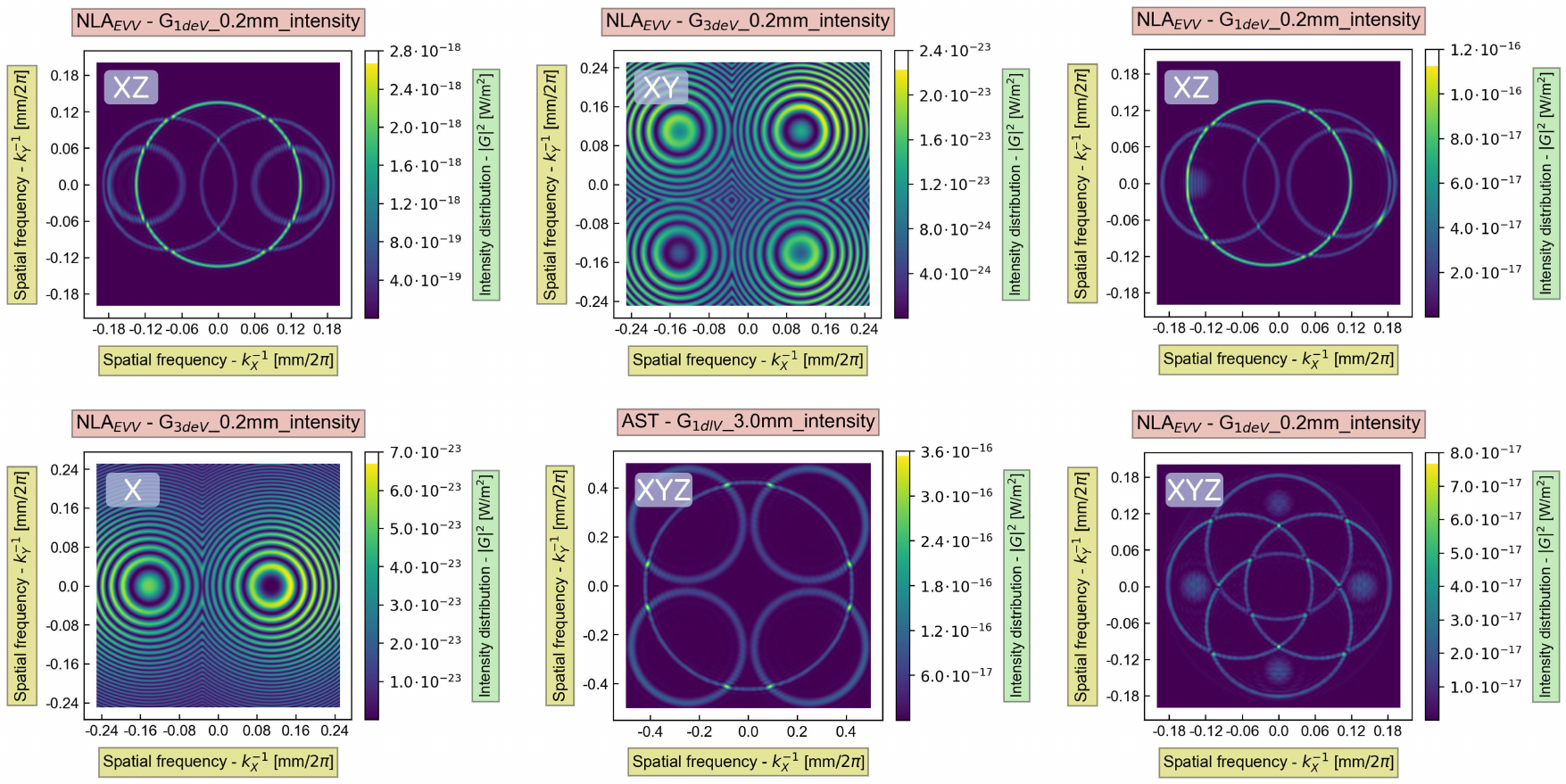}
       \caption{\textbf{The trilogy done right\cite{matosAnisotropyDoneRight2010}: quantum nonlinear Fourier crystal optics(QNFCO) in 1D, 2D, and 3D nonlinear photonic crystals(NPCs).} The top-left label of each subplot indicates the directions provided by the reciprocal lattice vectors $\bar{A}$; for example, `XZ' denotes that both $A_{\mathrm{x}}$ and $A_{\mathrm{z}}$ are non-zero.} \label{fig:quantum-killed}
\end{figure}

Having conducted large-scale numerical validation of known NCO phenomena in \cref{ssec:NCO-validate}, thereby indirectly substantiating the LCO model advanced in this paper, it is imperative to further establish the model's forward-looking predictive capabilities, for the potentially vast range of yet-unknown NCO phenomena.

To this purpose, we propose simulations of spontaneous parametric down-conversion(SPDC) in 1D, 2D, and 3D NPCs in \cref{fig:quantum-killed} (and Fig. SB6), as a preliminary step toward developing a comprehensive theoretical and mathematical FO framework for quantum NCO(QNCO) $\in$ nonlinear quantum optics(NQO) $\in$ QO.

\section{Discussion}\label{sec:d}

In arbitrary $\bar{\bar{\varepsilon}}^{\;\!\prime\omega}_{\mathrm{r}}$, namely birefringent-chiral-dichroic dielectrics, complex eigenmodes and their anisotropic diffraction behaviors form the fundamental basis of all advanced studies involving the interaction between light and matter. Of all the CO models created to address this issue, the ultimate plane wave solution in the form of matrix exponential $\mathbbm{e}^{\mathbbm{i} \bar{\bar{k}}^{\;\!\omega}_\mathrm{z} z}$ after Jordan decomposition is regarded as the most likely key to deciphering the internal structure of singularities(EPs) with second-order or higher degeneracy (see Supplementary Note 1). Yet, the batch numerical implementation of this approach is unstable and incapable of managing thicker non-Hermitian materials.

Conversely, numerically stable models that allow for the treatment of thick non-Hermitian slabs fail entirely to resolve EPs. These models share the same structure of classical plane wave $\mathbbm{e}^{\mathbbm{i} \bar{k}^{\;\!\omega} \cdot \bar{r}}$, whose two distinct forms $\mathbbm{e}^{\mathbbm{i} k^{\;\!\omega} (\hat{k}) \hat{k} \cdot \bar{r}}, \mathbbm{e}^{\mathbbm{i} \bar{k}^{\;\!\omega} (\bar{k}_{\uprho}) \cdot \bar{r}}$ based on real spherical/rectangular coordinates, with different independent variables $\hat{k}, \bar{k}_{\uprho}$, both satisfy the same wave/characteristic \cref{eq:r-2}. All models choosing $\bar{k}_{\uprho}$ as input variable typically requires numerical solutions for quartics, while the rest models depends on $\hat{k}$ possess simple closed-form solutions to biquadratics. However, the former satisfies FT and boundary conditions, while the latter does not.

In light of complex ray tracing, we extend Berry \& Dennis's 2003 uniform plane wave LCO model to non-uniform LFCO and ultimately derive the explicit form of its 3$\times$2 transition matrix field between any two sections within a planar slab dielectric, by transforming Berry \& Dennis's eigensystem $\hat{\underline{k}}, \bar{\underline{d}}^{\;\!\omega} (\hat{\underline{k}})$ into $\bar{k}_{\uprho}, \bar{g}^{\;\!\omega} (\bar{k}_{\uprho})$, thus bridging the two major branches of LCO in reciprocal space, where either ray direction $\hat{k}$ or spatial frequency $\bar{k}_{\uprho}$ serves as the input variable.

Using this LFCO model, we have comprehensively explored and revealed a new facet of the LCO from two perspectives: the AE of eigensystems in 2D reciprocal space under the dual competition among the four material properties: linear/circular birefringence/dichroism (via the M-M TC), together with the corresponding evolution of the light field in 3D real space (in collaboration with OTFs exemplified by the crystal-2f setup), depicting two magnificent panoramic maps in both real and reciprocal space.

Along the way, we observe that circular dichroism(CD), which also conforms to the haunting theorem pertaining to C points, can result in heart-formed L shorelines and infinite singularities arranged in disk-, ring-, and crescent-like shapes in 2D $\bar{k}_\uprho$ domain, implementing a bijective extension of the classical CO theory concerning singular axes, C points, and L lines; while in 3D $\bar{r}$ space, the genesis of Raman spikes during CR, double conical refraction(DCR) and optical knots(OKs) are discovered.

By integrating the custom-developed FO transfer functions of optical instruments, our trilogy (this LFCO model $+$ subsequent NFCO models) has also successfully reproduced and further explored numerous complicated experimental results in both LCO and NCO, unveiling the unified mathematical and physical essence underlying these diverse phenomena. This forward LFCO model is also inversely applied for inverse focal design in high N.A. laser writing. Lastly, we offer some initial demonstrations of the potential future applications of this powerful LCO model in (quantum) NCO, including spontaneous parametric down-conversion(SPDC) in 3D nonlinear photonic crystals(NPCs).


Such a novel paradigm (the M-M TC and FO OTFs with crystal-2f configuration) and the newly predicted $\bar{k}_\uprho$- and $\bar{r}$-space phenomena will inject theoretical, experimental and computational vitality into LCO \& NCO, as well as CO, FO, LO, NO, and QO, breathing new life into their future developments. It also lays a solid foundation for all upper-level architectures based on light-matter interactions.


In addition to developing NCO models based on this LCO framework, we are also actively exploring new solutions to the LCO model itself. This includes investigating alternative (non-rectangular, non-spherical) coordinate systems or retaining Berry \& Dennis's ``South-Pole Stereographic Projection'' coordinate system while utilizing non-uniform FT\cite{barnettParallelNonuniformFast2019} of type 2 (and 1) --- sampling non-uniform $\bar{k}_{\uprho}$ grid in 2D reciprocal space while uniformly sampling $\bar{\rho}$ grid in 2D real space to avoid solving transcendental equations.

As an advanced version of linear (wave) optics(LO) in isotropic free space (culminating in classical FO in air or vacuum) --- an area older than CO itself (in fact, LO and CO have each developed for over 200 years), LFCO (together with the NFCO derived from it) inherits all of its challenges and extensions. Examples include introducing the (Chriped) Z transform\cite{leuteneggerFastFocusField2006}, Bluestein algorithm\cite{huEfficientFullpathOptical2020a}, scaled angular spectrum\cite{heintzmannScalableAngularSpectrum2023}, and semi-analytical FT\cite{wangApplicationSemianalyticalFourier2019} into (L/N)FCO, etc., enabling adaptive field-of-view(FOV) or region-of-interest(ROI) scaling in real and reciprocal spaces as a function of propagation distance $z$, delaying the onset of aliasing errors, and reducing computational complexity in both time and space, all while satisfying the sampling theorem.

It is worth emphasizing that similar problems remain at the forefront of the intersection between applied mathematics and computer science (i.e. numerical analysis). That is, achieving fast and accurate semi-numerical implementation to PDEs of highly oscillatory fields and their integral solutions, even in the one-dimensional ordinary differential case, remains a relentless challenge and an enduring pursuit\cite{agocsAdaptiveSpectralMethod2024}.

As a longstanding issue spanning mathematics (linear algebra, PDEs, fractional FT\cite{pellat-finetFresnelDiffractionFractionalorder1994,dengDiffractionInterpretedFractional1996,pellat-finetSphericalAngularSpectrum2006,pellat-finetComplexOrderFractional2006,pellat-finetEffectDiffractionWigner2022}, topology), physics (Hamilton's five great legacies: manifolds, complex numbers, diabolic points, Hamiltonians, quaternions as a whole, Noether's theorem, and wave optics --- the mother of quantum mechanics), and computer science (sampling theorem, butterfly algorithm, Z transform, physical information and convolutional neural networks), along with a scientific ``toy gallery''(see Supplementary Note 9) where a simple laser pointer and a small crystal suffice for real-time experiments to verify surrounding objective reality, we hope the ``old yet new'' aspects of (L/N)(F)CO will once again attract attention across diverse domains.



\section{Methods}\label{sec:m}

\subsection{Boundary conditions for laboratory settings}\label{ssec:m-1}

For a commonly used homogeneous dielectric planar slab in laboratory settings, the space dependence of the major material quantity $\bar{\bar{\varepsilon}}^{\;\!\prime\omega}_{\mathrm{r}z} := \bar{\bar{\varepsilon}}^{\,\omega}_{\mathrm{r}z} + \frac{\mathbbm{i}}{\upvarepsilon_0 \omega} \bar{\bar{\sigma}}^{\;\!\omega}_{z}$ (see Supplementary Note 2) is characterized by two step functions\cite{chenWavevectorspaceMethodWave1993,nelsonDerivingTransmissionReflection1995,asoubarSimulationBirefringenceEffects2015}, assuming that the slab is surrounded by isotropic non-chiral transparent media. After establishing the 3D laboratory coordinate system (LCS), also referred to as the $\mathcal{Z}$ frame, whose +z-axis is aligned parallel to the inward normal of the front face of the slab with air as its surroundings, the dielectric tensor of the slab is expressed as\cite{chenWavevectorspaceMethodWave1993,nelsonDerivingTransmissionReflection1995}
\begin{align} \label{eq:m-1}
       \bar{\bar{\varepsilon}}^{\;\!\prime\omega}_{\mathrm{r}z} = 1 + \left( \bar{\bar{\varepsilon}}^{\;\!\prime\omega}_{\mathrm{r}} - 1 \right) \cdot \left[ \text{step}\left( z \right) - \text{step}\left( z - L \right) \right] = \begin{cases} \bar{\bar{\varepsilon}}^{\;\!\prime\omega}_{\mathrm{r}}, &0 < z < L \\ \text{undefined}, &z = 0\ \text{or}\ z = L \\ 1, &z < 0\ \text{or}\ z > L \end{cases}.
\end{align}
Then the wave equation (see Supplementary Note 2), together with \cref{eq:m-1}, contains all the necessary information for solving the distribution of electromagnetic field inside and outside the slab made of typical optical materials, provided that a two-dimensional(2D) distribution of the vector pump $\bar{\mathsfit{E}}^{\;\!\omega}_{z_0}$ ('s transverse components $\mathsfit{E}^{\;\!\omega}_{\mathrm{x}z_0},\mathsfit{E}^{\;\!\omega}_{\mathrm{y}z_0}$) in front of the slab ($z_0 < 0$) is given, which means neither divergence equations\cite{chenWavevectorspaceMethodWave1993,nelsonDerivingTransmissionReflection1995,landryCompleteMethodDetermine1995,abdulhalimExactMatrixMethod1999,mcleodVectorFourierOptics2014,asoubarSimulationBirefringenceEffects2015,changWavePropagationBianisotropic2014,chernChiralSurfaceWaves2017,chenCoordinateFreeApproach1982,gerardinConditionsVoigtWave2001,ossikovskiExtendedYehsMethod2017,changSimpleFormulasCalculating2001,zuAnalyticalNumericalModeling2022,zuOpticalSecondHarmonic2024} nor boundary conditions\cite{chenWavevectorspaceMethodWave1993,nelsonDerivingTransmissionReflection1995} are even required.

If we must assert which boundary conditions are the most fundamental, we would choose the tangential continuity of the electric field $\bar{\mathsfit{E}}$ and the generalized Snell's law\cite{changRayTracingAbsorbing2005,dupertuisGeneralizationComplexSnellDescartes1994} as the sole boundary conditions, due to the general failure of tangential continuity for the magnetic field $\bar{\mathsfit{H}}$\cite{chenWavevectorspaceMethodWave1993,nelsonDerivingTransmissionReflection1995,raabMultipoleTheoryElectromagnetism2004,grahamMultipoleSolutionMacroscopic2000,delangeElectromagneticBoundaryConditions2013} and the inconvenient use of the normal continuity for the magnetic induction field $\bar{\mathsfit{B}}$ as its substitute.

The trade-off is that, for each incident field, either two transmission fields passing through the anti-reflective(AR) coating/nanostructure or two reflected fields bounced back\cite{viottiCoherentPhaseTransfer2019,zhuTopologicalOpticalDifferentiator2021} by the high-reflective(HR) coating can be calculated relatively accurately, while finer effects such as the photon spin-Hall effect\cite{lingTopologyInducedPhaseTransitions2021,lingPhotonicSpinHallEffect2023} and the additional lateral shift\cite{zhangAlgorithmAccurateEfficient2024} cannot be revealed by this model.

Above boundary conditions adopted in our model naturally align with the standard configuration in modern nonlinear photonics laboratories where slab-shaped crystals with AR/HR coatings/nanostructures applied to both front and rear surfaces are frequently used, rendering our model applicable in the majority of cases.

Besides, regarding the central element of this work, i.e., the 3$\times$2 transition matrix of non-uniform LFCO, since it is defined solely within the material, no extra boundary conditions are required from this standpoint, apart from the generalized Snell's law, which mainly restrains the eigenvalues.

\subsection{Eigensystem corrections for phase continuity}\label{ssec:m-2}

The generalized Snell's law between transparent and dissipative/active media with non-Hermitian $\bar{\bar{\varepsilon}}^{\;\!\prime\omega}_{\mathrm{r}}$, will be violated, if one persists in employing uniform plane waves $\mathbbm{e}^{\mathbbm{i} \bar{k}^{\;\!\omega} \cdot \bar{r}} = \mathbbm{e}^{\mathbbm{i} k^{\;\!\omega} \hat{k} \cdot \bar{r}} = \mathbbm{e}^{\mathbbm{i} k_0^{\;\!\omega} n^{\omega} \hat{k} \cdot \bar{r}}$ with uniform complex wave vectors\cite{berryOpticalSingularitiesBirefringent2003,berryOpticalSingularitiesBianisotropic2005,yehElectromagneticPropagationBirefringent1979,chenCoordinateFreeApproach1982,gerardinConditionsVoigtWave2001,ossikovskiExtendedYehsMethod2017,changSimpleFormulasCalculating2001,zuAnalyticalNumericalModeling2022,kirillovEigenvalueSurfacesDiabolic2005,grundmannSingularOpticalAxes2016,brenierLasingConicalDiffraction2016}
\begin{align} \label{eq:m-2}
       \bar{k}^{\;\!\omega} = k^{\;\!\omega} \hat{k} \in \mathbb{C}_{\mathrm{r}} \times \mathbb{R}^2_{\varominus} \subsetneq \mathbb{C}^3_{\Yup}
\end{align}
across the slab material ($z<0 \to z>L$). Because when sticking to this form, transverse wave vectors $k^{\;\!\omega}_{\mathrm{x}},k^{\;\!\omega}_{\mathrm{y}} = k^{\;\!\omega} \hat{k}_{\mathrm{x}}, k^{\;\!\omega} \hat{k}_{\mathrm{y}} \in \mathbb{C}$ are generally complex and $\omega$-dispersive on the material side (with $\bar{\bar{\varepsilon}}^{\;\!\prime\omega}_{\mathrm{r}}$ and $0 < z < L$), whereas in air surroundings (where $\varepsilon_{\mathrm{r}} = 1$ and $z < 0$ or $> L$) and FT they remain real and non-$\omega$-dispersive, breaking the in-plane momentum conservation = phase continuity, one-to-one correspondence for each plane wave across interfaces and the requirement of FT.

By contrast, adhering to non-uniform complex wave vectors
\begin{subequations} \label{eq:m-3}
       \begin{align}
              \bar{k}^{\;\!\omega} &= \bar{k}_{\perp} + \bar{k}^{\;\!\omega}_{\mathrm{z}} \xrightarrow[\bar{k}^{\;\!\omega}_{\text{I}} = \bar{k}^{\;\!\omega}_{\mathrm{z}\text{I}}]{\bar{k}^{\;\!\omega}_{\text{R}} = \bar{k}^{\;\!\omega}_{\perp} + \bar{k}^{\;\!\omega}_{\mathrm{z}\text{R}}} \bar{k}^{\;\!\omega}_{\text{R}} + \mathbbm{i} \bar{k}^{\;\!\omega}_{\text{I}} \label{eq:m-3a} \\ &= \bar{k}^{\;\!\omega}_{\text{R}} + \mathbbm{i} \bar{k}^{\;\!\omega}_{\text{I}} \xrightarrow[\bar{k}^{\;\!\omega}_{\mathrm{z}} = \bar{k}^{\;\!\omega}_{\text{R}\mathrm{z}} + \mathbbm{i} \bar{k}^{\;\!\omega}_{\text{I}}]{\bar{k}_{\perp} = \bar{k}^{\;\!\omega}_{\text{R}\perp}} \bar{k}_{\perp} + \bar{k}^{\;\!\omega}_{\mathrm{z}} \in \mathbb{R}^2_{\perp} + \mathbb{C}_{\mathrm{Z}} \label{eq:m-3b}
       \end{align}
\end{subequations}
of the non-uniform plane waves $\mathbbm{e}^{\mathbbm{i} \bar{k}^{\;\!\omega} \cdot \bar{r}} = \mathbbm{e}^{\mathbbm{i} \left( \bar{k}_{\uprho} \cdot \bar{\rho} + k^{\;\!\omega}_{\mathrm{z}} z \right)}$ throughout $z<0 \to z>L$ naturally align with the generalized Snell's law\cite{changRayTracingAbsorbing2005,dupertuisGeneralizationComplexSnellDescartes1994,wangComplexRayTracing2008a} on the interface planes at $z = 0$ and $z = L$ with non-absorbing surroundings. Because through gluing to this form, the entire space ($z<0 \to z>L$) consistently takes $\omega$-dispersion-free spatial frequencies $\bar{k}_{\uprho} \in \mathbb{R}^2_{\perp}$ as transverse wave vectors, which further permit one-to-one correspondence across boundaries for each Fourier component = spatiotemporal spectrum while fulfilling the constraints of FT.

\begin{figure}[htbp!]
       \centering
       \includegraphics[width=1\textwidth]{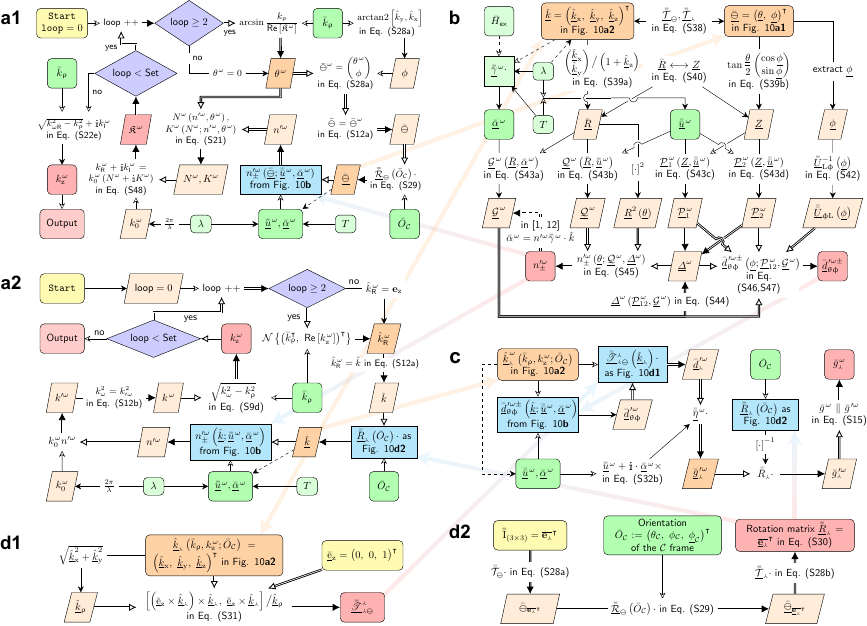}
       \caption{\textbf{Sub flowcharts for \cref{fig:m-2}.} \textbf{a1}-\textbf{a2} Two equivalent procedures for eigenvalue correction $n^{\prime\omega}(\hat{k}) \to k^{\;\!\omega}_{\mathrm{z}}(\bar{k}_{\uprho})$. \textbf{b} The functional form of Berry \& Dennis's eigensystem $n^{\prime\omega} ( \hat{\underline{k}} ), \bar{\underline{d}}^{\;\!\prime\omega}_{\uptheta\upphi} ( \hat{\underline{k}} )$. \textbf{c} Eigenvector transition $\bar{\underline{d}}^{\;\!\prime\omega}_{\uptheta\upphi} ( \hat{\underline{k}} ) \to \bar{g}^{\;\!\omega} (\bar{k}_{\uprho})$. \textbf{d1} Acquiring operator $\bar{\bar{\underline{\mathscr{T}}}}^{\Yup}_{\Yup\varominus}$ for $\bar{\underline{d}}^{\;\!\prime\omega}_{\uptheta\upphi} \to \bar{\underline{d}}^{\;\!\prime\omega}_{{\Yup}}$ transition. \textbf{d2} Building $\mathcal{Z} \to \mathcal{C}$ frame rotation matrix $\bar{\bar{\underline{R}}}_{\Yup}$ via spherical trigonometry. {\color{gray} For the meanings of styles of nodes \& arrows, see \cref{tab:m-1,tab:m-2}.} }\label{fig:m-1}
\end{figure}

Our approach to amending the complex eigenvalues $n^{\omega} (\hat{k}) \to k^{\;\!\omega}_{\mathrm{z}} (\bar{k}_{\uprho})$ in the aforementioned complex wave vectors $k_0^{\;\!\omega} n^{\omega} (\hat{k}) \hat{k} \to \bar{k}_{\perp} + \bar{k}^{\;\!\omega}_{\mathrm{z}} (\bar{k}_{\uprho})$, while retaining their analytical form, is quite straightforward (see Eq. (S12) in Supplementary Note 2):
\begin{subequations} \label{eq:m-4}
       \begin{align}
              \hat{k} &= \mathcal{N} \left\{ \text{Re} \left[ \begin{pmatrix} \bar{k}_{\uprho}, & k^{\;\!\omega}_{\mathrm{z}} \end{pmatrix}^{\intercal} \right] \right\}, \label{eq:m-4a} \\ k^{\;\!\omega}_{\mathrm{z}} &= \sqrt{k^2_{0\omega} n^{2}_{\;\!\omega} ( \hat{k} ) - k^2_\uprho}, \label{eq:m-4b}
       \end{align}
\end{subequations}
where both the input real ray direction $\hat{k}$ for uniform eigenvalue $n^{\omega}$ and the output non-uniform eigenvalue $k^{\;\!\omega}_{\mathrm{z}}$ are ultimately functions of $\bar{k}_{\uprho}$, and mutually coupled, as illustrated in \cref{fig:m-1}a2.

One can unravel this coupling by first discarding the imaginary part $\bar{k}^{\;\!\omega}_{\text{I}}$ of the complex non-uniform wave vector $\bar{k}^{\;\!\omega} = \bar{k}^{\;\!\omega}_{\text{R}} + \mathbbm{i} \bar{k}^{\;\!\omega}_{\text{I}}$ in \cref{eq:m-3b}, retaining only the real part $\bar{k}^{\;\!\omega}_{\text{R}}$ 's unit vector $\hat{k}$ from \cref{eq:m-4a} as the ray direction in $\boldsymbol{k}$ space. The input variable $\hat{k}$ is then inserted into \cref{eq:m-4b} to form a transcendental equation $k^{\;\!\omega}_{\mathrm{z}} = \sqrt{k^2_{0\omega} n^{2}_{\;\!\omega} ( \mathcal{N} \left\{ \text{Re} \left[ \begin{pmatrix} \bar{k}_{\uprho}, & k^{\;\!\omega}_{\mathrm{z}} \end{pmatrix}^{\intercal} \right] \right\} ) - k^2_\uprho}$ which asymptotically converges to the ground true eigenvalue $k^{\;\!\omega}_{\mathrm{z}} (\bar{k}_{\uprho})$ of the booker quartic on condition that
\begin{align} \label{eq:m-5}
       \lvert \bar{k}^{\;\!\omega}_{\text{I}} \rvert \ll \lvert \bar{k}^{\;\!\omega}_{\text{R}} \rvert,
\end{align}
where material absorption is sufficiently low with typically $k^{\;\!\omega}_{\text{I}} < k^{\;\!\omega}_{\text{R}} \cdot 10^{-3}$ when a transmission spectrum is still present, as exemplified by Brenier's laser crystal\cite{brenierVoigtWaveInvestigation2015,brenierLasingConicalDiffraction2016} in \cref{fig:MMTC-r}b,e.

Berry \& Dennis's input variable $\hat{\underline{k}}$ can be obtained from \cref{fig:m-1}a2 as an intermediate data. Inserting it into \cref{fig:m-1}b yields Berry \& Dennis's eigensystem $n^{\prime\omega} ( \hat{\underline{k}} ), \bar{\underline{d}}^{\;\!\prime\omega}_{\uptheta\upphi} ( \hat{\underline{k}} )$, where the eigenvector $ \bar{\underline{d}}^{\;\!\prime\omega}_{\uptheta\upphi} ( \hat{\underline{k}} )$ also converges to the ground truth $\bar{g}^{\;\!\omega} (\bar{k}_{\uprho})$ under the condition of \cref{eq:m-5} (see Supplementary Note 2.2), after undergoing the operation shown in \cref{fig:m-1}c.

\subsection{Overall flowchart of this LFCO model}\label{ssec:m-3}

\begin{figure}[htbp!] 
       \centering
       \includegraphics[width=0.9\textwidth]{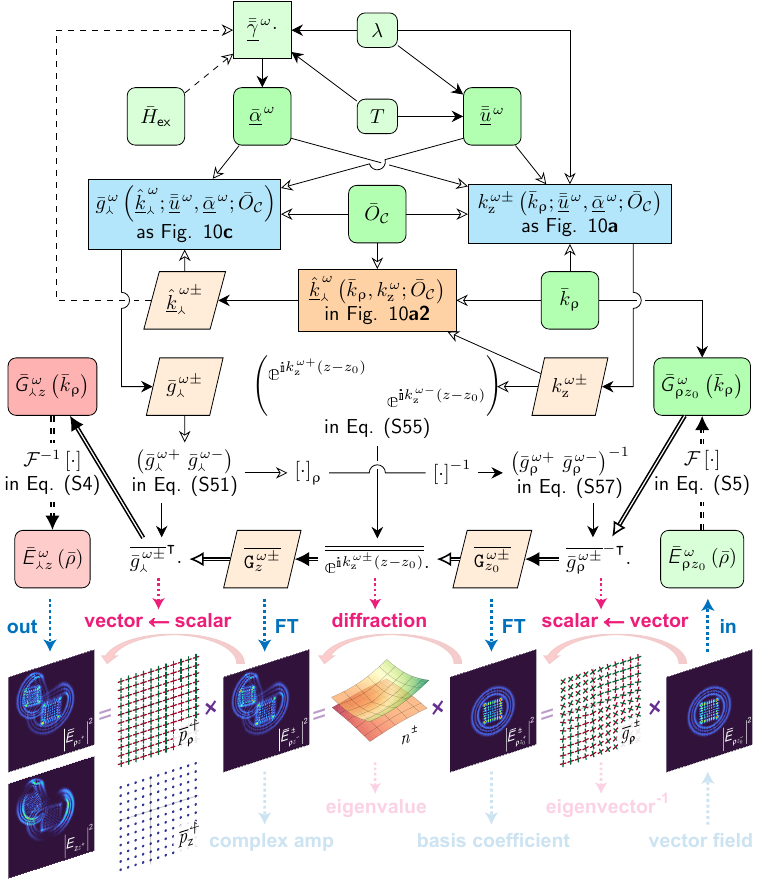}
       \caption{\textbf{Main flowchart to build the transition matrix in \cref{eq:r-5} and \cref{fig:3transition2matrix} from basic elements.} Atomic inputs include the orientation $\bar{O}_{\mathcal{C}}$ of the principal coordinate system(PCS) = the $\mathcal{C}$ frame, optical activity tensor $\bar{\bar{\gamma}}^{\;\!\omega}$\cite{berryOpticalSingularitiesBirefringent2003}, optical activity vector $\bar{\alpha}^{\;\!\omega}$\cite{landauCHAPTERXIELECTROMAGNETIC1984}, the symmetric part $\bar{\bar{u}}^{\;\!\omega}$ of $\bar{\bar{\varepsilon}}^{\;\!\prime-1}_{\mathrm{r}\omega}$, all initially in the $\mathcal{C}$ frame $\bar{\bar{\underline{\gamma}}}^{\;\!\omega}, \bar{\underline{\alpha}}^{\;\!\omega}, \bar{\bar{\underline{u}}}^{\;\!\omega}$, together with external magnetic field $\bar{H}_{\text{ex}}$, wavelength $\lambda$ of monochromatic light, temperature $T$ of the crystal, spatial frequency $\bar{k}_{\uprho}$, and the input vector pump $\bar{\mathsfit{E}}^{\;\!\omega}_{\uprho z_0} \left( \bar{\rho} \right)$ in 2D real space. {\color{gray} For the meanings of styles of nodes \& arrows, see \cref{tab:m-1,tab:m-2}.} }\label{fig:m-2}
\end{figure}

Starting from the atomic inputs, we built the comprehensive workflow diagram in \cref{fig:m-2} from the bottom up, referencing three submodules from \cref{fig:m-1}a,c. All elements in the flowcharts (\cref{fig:3transition2matrix,fig:m-1,fig:m-2}) of this article are accompanied by step-by-step derivations and detailed explanations provided in the Supplementary Notes.

\cref{tab:m-1,tab:m-2} briefly presents the categorical meanings of the graphical elements (nodes and arrows) as `CLASSes' in \cref{fig:3transition2matrix,fig:m-1,fig:m-2}. The physical/mathematical significance of the specific instantiated `OBJECTs' within each `CLASS' can be found in Supplementary Notes 2-5.

\begin{table}[htbp!]
       \centering
       \caption{The definitions of styles of nodes within \cref{fig:3transition2matrix,fig:m-1,fig:m-2}. }\label{tab:m-1}%
       \begin{tabular}{@{}l@{}}
              \includegraphics[width=1\textwidth]{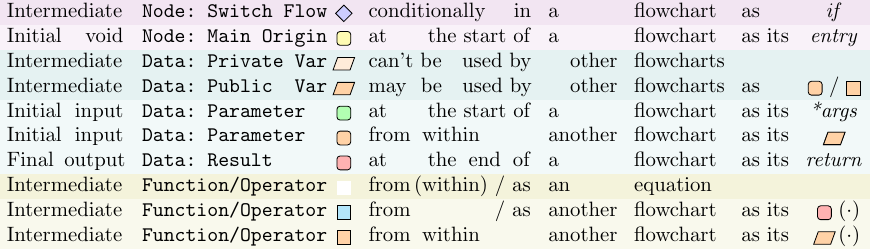}
       \end{tabular}
\end{table}

\begin{table}[htbp!]
       \centering
       \caption{The meanings of styles of arrows within \cref{fig:3transition2matrix,fig:m-1,fig:m-2}. }\label{tab:m-2}%
       \begin{tabular}{@{}l@{}}
              \includegraphics[width=1\textwidth]{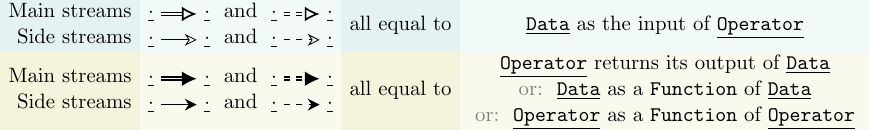}
       \end{tabular}
\end{table}

\ack{We thank Sir M.V. Berry for the insights shared through both our online and offline exchanges.}

\funding{This work was supported by the Natural Science Foundation of Jiangsu Province (BK20240005) and the Fundamental Research Funds for the Central Universities (021314380268).}

\roles{C.X. conceived, led, and built the entire trilogy starting from empty-nothing (down to atom, up to soul) independently, with his unique long-chain/tree/web (non)linear logic, interspersed with leap-like flashes of divine apocalypto, right before the AI era and his own liberation from VPN restrictions. Y.Z. warmly approved, consistently supported, and responsibly supervised the whole trilogy.}

\data{All data generated by this trilogy, along with the parameter file for reconstructing any (sub)figure, can be requested from the first author.}

\suppdata{{\one} \href{run:sn-Supplementary_Notes.pdf}{Supplementary Notes.pdf}, {\two} \href{https://www.youtube.com/watch?v=Inl1AkDAblo}{Supplementary Video 1.1: inner\_rings\_of\_Fig.\_S8\_\_0.1s.mp4}, {\three} \href{https://www.youtube.com/watch?v=CuFjKcdAIJQ}{Supplementary Video 1.2: inner\_rings\_of\_Fig.\_S8\_\_0.5s.mp4}, {\four} \href{https://www.youtube.com/watch?v=xkMWXW29HYc}{Supplementary Video 2.1: outer\_ring\_of\_Fig.\_S8\_\_0.1s.mp4}, {\five} \href{https://www.youtube.com/watch?v=PPZHhTvdfb0}{Supplementary Video 2.2: outer\_ring\_of\_Fig.\_S8\_\_0.5s.mp4}}


\providecommand{\newblock}{}


\begin{thebibliography}{100}
\expandafter\ifx\csname url\endcsname\relax
  \def\url#1{{\tt #1}}\fi
\expandafter\ifx\csname urlprefix\endcsname\relax\def\urlprefix{URL }\fi
\providecommand{\eprint}[2][]{\url{#2}}

\bibitem{ossikovskiConstitutiveRelationsOptically2021}
Ossikovski R, Arteaga O and Sturm C 2021 {\em Advanced Photonics Research\/}
  {\bf 2} 2100160 ISSN 2699-9293, 2699-9293

\bibitem{robertsSimplifiedCharacterizationUniaxial1992}
Roberts D 1992 {\em IEEE Journal of Quantum Electronics\/} {\bf 28} 2057--2074
  ISSN 00189197

\bibitem{dmitrievEffectiveNonlinearityCoefficients1993}
Dmitriev V and Nikogosyan D 1993 {\em Optics Communications\/} {\bf 95}
  173--182 ISSN 00304018

\bibitem{diesperovEffectiveNonlinearCoefficient1997}
Diesperov K~V and Dmitriev V~G 1997 {\em Quantum Electronics\/} {\bf 27}
  433--436 ISSN 1063-7818, 1468-4799

\bibitem{berryOpticalSingularitiesBirefringent2003}
Berry M~V and Dennis M~R 2003 {\em Proceedings of the Royal Society of London.
  Series A: Mathematical, Physical and Engineering Sciences\/} {\bf 459}
  1261--1292 ISSN 1364-5021, 1471-2946

\bibitem{wuVectorialNonlinearOptics2019}
Wu H~J, Yang H~R, {Rosales-Guzm{\'a}n} C, Gao W, Shi B~S and Zhu Z~H 2019 {\em
  Physical Review A\/} {\bf 100} 053840 ISSN 2469-9926, 2469-9934

\bibitem{zondyEffectsFocusingTypeI1998}
Zondy J~J 1998 {\em Optics Communications\/} {\bf 149} 181--206 ISSN 00304018

\bibitem{huoEffectiveMethodCalculating2015}
Huo G, Wang Y and Zhang M 2015 {\em Applied Physics B\/} {\bf 120} 239--246
  ISSN 0946-2171, 1432-0649

\bibitem{zondyTwincrystalWalkoffcompensatedTypeII1994}
Zondy J~J, Abed M and Khodja S 1994 {\em Journal of the Optical Society of
  America B\/} {\bf 11} 2368 ISSN 0740-3224, 1520-8540

\bibitem{vaicaitisCherenkovtypePhaseMatching2002}
Vai{\v c}aitis V 2002 {\em Optics Communications\/} {\bf 209} 485--490 ISSN
  00304018

\bibitem{turpinTypeTypeII2013}
Turpin A, Loiko Y~V, Kalkandjiev T~K, Trull J, Cojocaru C and Mompart J 2013
  {\em Optics Letters\/} {\bf 38} 2484 ISSN 0146-9592, 1539-4794

\bibitem{magnitskiyCharacterizationPolarizationangularSpectrum2015}
Magnitskiy S~A, Gostev P~P, Frolovtsev D~N and Firsov V~V 2015 {\em Moscow
  University Physics Bulletin\/} {\bf 70} 382--389 ISSN 0027-1349, 1934-8460

\bibitem{yaoAccurateCalculationOptimum1992}
Yao J, Shi W and Sheng W 1992 {\em Journal of the Optical Society of America
  B\/} {\bf 9} 891 ISSN 0740-3224, 1520-8540

\bibitem{wangComplexRayTracing2008a}
Wang Y, Liang L, Xin H and Wu L 2008 {\em Journal of the Optical Society of
  America A\/} {\bf 25} 653 ISSN 1084-7529, 1520-8532

\bibitem{dineiroComplexUnitaryVectors2007}
Di{\~n}eiro J~M, Berrogui M, Alfonso S, Alberdi C, Hern{\'a}ndez B and
  S{\'a}enz C 2007 {\em Journal of the Optical Society of America A\/} {\bf 24}
  1767 ISSN 1084-7529, 1520-8532

\bibitem{rafayelyanPlaneElectromagneticWaves2010}
Rafayelyan M~S and Gevorgyan A~H 2010 Plane electromagnetic waves in a
  homogeneous anisotropic uniaxial medium having a double anisotropy and an
  arbitrary orientation of its optical axis {\em International {{Conference}}
  on {{Laser Physics}} 2010\/} ed Papoyan A~V (Ashtarak, Armenia) p 79980K

\bibitem{yeTheoreticalAnalysisCorrection2024}
Ye S, Chang C, Liu X, Chen R, Wang C, Zhou Z, Wei D and Wang X 2024 {\em
  Physical Review Applied\/} {\bf 22} 044070 ISSN 2331-7019

\bibitem{wangGeometricalRayTracing2018}
Wang P 2018 {\em Journal of the Optical Society of America A\/} {\bf 35} 1114
  ISSN 1084-7529, 1520-8532

\bibitem{carnioAnalysisElectricField2018}
Carnio B~N and Elezzabi A~Y 2018 {\em Journal of Infrared, Millimeter, and
  Terahertz Waves\/} {\bf 39} 313--325 ISSN 1866-6906

\bibitem{berryOpticalSingularitiesBianisotropic2005}
Berry M 2005 {\em Proceedings of the Royal Society A: Mathematical, Physical
  and Engineering Sciences\/} {\bf 461} 2071--2098 ISSN 1364-5021, 1471-2946

\bibitem{shengNonlinearopticalCalculationsUsing1980}
Sheng S~C and Siegman A~E 1980 {\em Physical Review A\/} {\bf 21} 599--606 ISSN
  0556-2791

\bibitem{grundmannOpticallyAnisotropicMedia2017}
Grundmann M, Sturm C, Kranert C, Richter S, Schmidt-Grund R, Deparis C and
  Z{\'u}{\~n}iga-P{\'e}rez J 2017 {\em physica status solidi (RRL) -- Rapid
  Research Letters\/} {\bf 11} 1600295 ISSN 1862-6254, 1862-6270

\bibitem{xuFemtosecondLaserWriting2022}
Xu X, Wang T, Chen P, Zhou C, Ma J, Wei D, Wang H, Niu B, Fang X, Wu D, Zhu S,
  Gu M, Xiao M and Zhang Y 2022 {\em Nature\/} {\bf 609} 496--501 ISSN
  1476-4687

\bibitem{weiExperimentalDemonstrationThreedimensional2018}
Wei D, Wang C, Wang H, Hu X, Wei D, Fang X, Zhang Y, Wu D, Hu Y, Li J, Zhu S
  and Xiao M 2018 {\em Nature Photonics\/} {\bf 12} 596--600 ISSN 1749-4885,
  1749-4893

\bibitem{xuThreedimensionalNonlinearPhotonic2018}
Xu T, Switkowski K, Chen X, Liu S, Koynov K, Yu H, Zhang H, Wang J, Sheng Y and
  Krolikowski W 2018 {\em Nature Photonics\/} {\bf 12} 591--595 ISSN 1749-4885,
  1749-4893

\bibitem{keren-zurNewDimensionNonlinear2018}
{Keren-Zur} S and Ellenbogen T 2018 {\em Nature Photonics\/} {\bf 12} 575--577
  ISSN 1749-4885, 1749-4893

\bibitem{chenOpticallyInducedNonlinear2022}
Chen Y, Yang C, Liu S, Wang S, Wang N, Liu Y, Sheng Y, Zhao R, Xu T and
  Krolikowski W 2022 {\em Advanced Photonics Research\/} {\bf 3} 2100268 ISSN
  2699-9293, 2699-9293

\bibitem{zhangNonlinearPhotonicCrystals2021}
Zhang Y, Sheng Y, Zhu S, Xiao M and Krolikowski W 2021 {\em Optica\/} {\bf 8}
  372 ISSN 2334-2536

\bibitem{liuNonlinearVolumeHolography2020}
Liu S, Mazur L~M, Krolikowski W and Sheng Y 2020 {\em Laser \& Photonics
  Reviews\/} {\bf 14} 2000224 ISSN 1863-8880, 1863-8899

\bibitem{liuHighlyEfficient3D2023}
Liu S, Wang L, Mazur L~M, Switkowski K, Wang B, Chen F, Arie A, Krolikowski W
  and Sheng Y 2023 {\em Advanced Optical Materials\/} {\bf 11} 2300021 ISSN
  2195-1071, 2195-1071

\bibitem{neviereElectromagneticResonancesLinear1995}
Nevi{\`e}re M, Reinisch R and Popov E 1995 {\em Journal of the Optical Society
  of America A\/} {\bf 12} 513 ISSN 1084-7529, 1520-8532

\bibitem{linObservationDisorderedWave2013}
Lin Y~C, Tuan P~H, Yu Y~T, Liang H~C, Su K~W, Huang K~F and Chen Y~F 2013 {\em
  Physical Review B\/} {\bf 87} 045117 ISSN 1098-0121, 1550-235X

\bibitem{goulkovNewParametricScattering2003}
Goulkov M, Shinkarenko O, Ivleva L, Lykov P, Granzow T, Woike {\relax Th},
  Imlau M and W{\"o}hlecke M 2003 {\em Physical Review Letters\/} {\bf 91}
  243903 ISSN 0031-9007, 1079-7114

\bibitem{liConicalSecondHarmonic2015}
Li T, Zhao X, Zheng Y and Chen X 2015 {\em Optics Express\/} {\bf 23} 23827
  ISSN 1094-4087

\bibitem{rokeNonlinearOpticalScattering2004}
Roke S, Bonn M and Petukhov A~V 2004 {\em Physical Review B\/} {\bf 70} 115106

\bibitem{katzFocusingCompressionUltrashort2011}
Katz O, Small E, Bromberg Y and Silberberg Y 2011 {\em Nature Photonics\/} {\bf
  5} 372--377 ISSN 1749-4885, 1749-4893

\bibitem{liuScatteringassistedSecondHarmonic2016}
Liu H, Li J, Fang X, Zhao X, Zheng Y and Chen X 2016 {\em Optics Express\/}
  {\bf 24} 24137 ISSN 1094-4087

\bibitem{baudrier-raybautRandomQuasiphasematchingBulk2004}
{Baudrier-Raybaut} M, Ha{\"i}dar R, Kupecek {\relax Ph}, Lemasson {\relax Ph}
  and Rosencher E 2004 {\em Nature\/} {\bf 432} 374--376 ISSN 0028-0836,
  1476-4687

\bibitem{xuConicalSecondHarmonic2004}
Xu P, Ji S~H, Zhu S~N, Yu X~Q, Sun J, Wang H~T, He J~L, Zhu Y~Y and Ming N~B
  2004 {\em Physical Review Letters\/} {\bf 93} 133904 ISSN 0031-9007,
  1079-7114

\bibitem{anConicalSecondHarmonic2013}
An N, Zheng Y, Ren H, Deng X and Chen X 2013 {\em Applied Physics Letters\/}
  {\bf 102} 201112 ISSN 0003-6951, 1077-3118

\bibitem{ebersLightDiffractionSlab2020}
Ebers L, Hammer M and F{\"o}rstner J 2020 {\em Optics Express\/} {\bf 28} 36361
  ISSN 1094-4087

\bibitem{popovMaxwellEquationsFourier2001}
Popov E and Nevi{\`e}re M 2001 {\em Journal of the Optical Society of America
  A\/} {\bf 18} 2886 ISSN 1084-7529, 1520-8532

\bibitem{saltielMultiorderNonlinearDiffraction2009}
Saltiel S~M, Neshev D~N, Krolikowski W, Arie A, Bang O and Kivshar Y~S 2009
  {\em Optics Letters\/} {\bf 34} 848 ISSN 0146-9592, 1539-4794

\bibitem{zhongKdomainMethodFast2020}
Zhong H, Zhang S, {Baladron-Zorita} O, Shi R, Hellmann C and Wyrowski F 2020
  {\em Optics Express\/} {\bf 28} 11074 ISSN 1094-4087

\bibitem{saltielCerenkovTypeSecondHarmonicGeneration2009}
Saltiel S~M, Sheng Y, {Voloch-Bloch} N, Neshev D~N, Krolikowski W, Arie A,
  Koynov K and Kivshar Y~S 2009 {\em IEEE Journal of Quantum Electronics\/}
  {\bf 45} 1465--1472 ISSN 0018-9197, 1558-1713

\bibitem{liReformulationFourierModal1998}
Li L 1998 {\em Journal of Modern Optics\/} {\bf 45} 1313--1334 ISSN 0950-0340,
  1362-3044

\bibitem{shiPhysicalopticsPropagationCurved2019}
Shi R, Hellmann C and Wyrowski F 2019 {\em Journal of the Optical Society of
  America A\/} {\bf 36} 1252 ISSN 1084-7529, 1520-8532

\bibitem{gerkeAperiodicVolumeOptics2010}
Gerke T~D and Piestun R 2010 {\em Nature Photonics\/} {\bf 4} 188--193 ISSN
  1749-4885, 1749-4893

\bibitem{xuLargeFieldofviewNonlinear2024}
Xu X, Chen P, Ma T, Ma J, Zhou C, Su Y, Lv M, Fan W, Zhai B, Sun Y, Wang T, Hu
  X, Zhu S~N, Xiao M and Zhang Y 2024 {\em Nano Letters\/} {\bf 24} 1303--1308
  ISSN 1530-6984, 1530-6992

\bibitem{chenQuasiphasematchingdivisionMultiplexingHolography2021b}
Chen P, Wang C, Wei D, Hu Y, Xu X, Li J, Wu D, Ma J, Ji S, Zhang L, Xu L, Wang
  T, Xu C, Chu J, Zhu S, Xiao M and Zhang Y 2021 {\em Light: Science \&
  Applications\/} {\bf 10} 146 ISSN 2047-7538

\bibitem{chenLaserNanoprinting3D2023}
Chen P, Xu X, Wang T, Zhou C, Wei D, Ma J, Guo J, Cui X, Cheng X, Xie C, Zhang
  S, Zhu S, Xiao M and Zhang Y 2023 {\em Nature Communications\/} {\bf 14} 5523
  ISSN 2041-1723

\bibitem{gutierrez-cuevasRayCausticStructure2024}
{Guti{\'e}rrez-Cuevas} R, Dennis M~R and Alonso M~A 2024 {\em New Journal of
  Physics\/} {\bf 26} 013011 ISSN 1367-2630

\bibitem{fogretLightrayApproachFractional2023}
Fogret {\'E} and {Pellat-Finet} P 2023 A light-ray approach to fractional
  fourier optics (\textit{Preprint} \eprint{2304.03500})

\bibitem{wyrowskiApproximateSolutionMaxwell2015}
Wyrowski F, Zhong H, Zhang S and Hellmann C 2015 Approximate solution of
  {{Maxwell}}'s equations by geometrical optics {\em {{SPIE Optical Systems
  Design}}\/} ed Smith D~G, Wyrowski F and Erdmann A (Jena, Germany) p 963009

\bibitem{avendano-alejoCausticsCausedRefraction2008}
{Avenda{\~n}o-Alejo} M, {D{\'i}az-Uribe} R and Moreno I 2008 {\em Journal of
  the Optical Society of America A\/} {\bf 25} 1586 ISSN 1084-7529, 1520-8532

\bibitem{shenFastFouriertransformBasedNumerical2006}
Shen F and Wang A 2006 {\em Applied Optics\/} {\bf 45} 1102 ISSN 0003-6935,
  1539-4522

\bibitem{ochoaAlternativeApproachEvaluate2017}
Ochoa N~A 2017 {\em Optics Express\/} {\bf 25} 12008 ISSN 1094-4087

\bibitem{mansuripurDistributionLightFocus1986}
Mansuripur M 1986 {\em Journal of the Optical Society of America A\/} {\bf 3}
  2086 ISSN 1084-7529, 1520-8532

\bibitem{khoninaAnalogRayleighSommerfeld2013}
Khonina S and Kharitonov S 2013 {\em Journal of Modern Optics\/} {\bf 60}
  814--822 ISSN 0950-0340, 1362-3044

\bibitem{iqbalNoncircularlyShapedConical2022}
Iqbal M~W, Marsal N and Montemezzani G 2022 {\em Scientific Reports\/} {\bf 12}
  7317 ISSN 2045-2322

\bibitem{shimobabaEfficientDiffractionCalculations2018}
Shimobaba T, Takahashi T, Yamamoto Y, Nishitsuji T, Shiraki A, Hoshikawa N,
  Kakue T and Ito T 2018 {\em OSA Continuum\/} {\bf 1} 642 ISSN 2578-7519

\bibitem{zhangPropagationElectromagneticFields2016}
Zhang S, Asoubar D, Hellmann C and Wyrowski F 2016 {\em Applied Optics\/} {\bf
  55} 529 ISSN 0003-6935, 1539-4522

\bibitem{pellat-finetSphericalAngularSpectrum2006}
{Pellat-Finet} P, Durand P~E and Fogret {\'E} 2006 {\em Optics Letters\/} {\bf
  31} 3429 ISSN 0146-9592, 1539-4794

\bibitem{odateAngularSpectrumCalculations2011}
Odate S, Koike C, Toba H, Koike T, Sugaya A, Sugisaki K, Otaki K and Uchikawa K
  2011 {\em Optics Express\/} {\bf 19} 14268 ISSN 1094-4087

\bibitem{mansuripurCertainComputationalAspects1989}
Mansuripur M 1989 {\em Journal of the Optical Society of America A\/} {\bf 6}
  786 ISSN 1084-7529, 1520-8532

\bibitem{leuteneggerFastFocusField2006}
Leutenegger M, Rao R, Leitgeb R~A and Lasser T 2006 {\em Optics Express\/} {\bf
  14} 11277 ISSN 1094-4087

\bibitem{huEfficientFullpathOptical2020a}
Hu Y, Wang Z, Wang X, Ji S, Zhang C, Li J, Zhu W, Wu D and Chu J 2020 {\em
  Light: Science \& Applications\/} {\bf 9} 119 ISSN 2047-7538

\bibitem{heintzmannScalableAngularSpectrum2023}
Heintzmann R, Loetgering L and Wechsler F 2023 {\em Optica\/} {\bf 10}
  1407--1416 ISSN 2334-2536

\bibitem{wangApplicationSemianalyticalFourier2019}
Wang Z, Zhang S, {Baladron-Zorita} O, Hellmann C and Wyrowski F 2019 {\em
  Optics Express\/} {\bf 27} 15335 ISSN 1094-4087

\bibitem{liuFastGenerationArbitrary2023}
Liu X, Hu Y, Tu S, Kuang C, Liu X and Hao X 2023 {\em Optics and Lasers in
  Engineering\/} {\bf 162} 107405 ISSN 0143-8166

\bibitem{matsushimaFastCalculationMethod2003}
Matsushima K, Schimmel H and Wyrowski F 2003 {\em JOSA A\/} {\bf 20} 1755--1762
  ISSN 1520-8532

\bibitem{weiModelingOffaxisDiffraction2023}
Wei H, Liu X, Hao X, Lam E~Y and Peng Y 2023 {\em Optica\/} {\bf 10} 959 ISSN
  2334-2536

\bibitem{huDiffractionModelingArbitrary2025}
Hu Y, Liu X, Liu X and Hao X 2025 {\em Optica\/} {\bf 12} 39--45 ISSN 2334-2536

\bibitem{chenCoordinateFreeApproach1982}
Chen H~C 1982 {\em Journal of Applied Physics\/} {\bf 53} 4606--4609 ISSN
  0021-8979, 1089-7550

\bibitem{gerardinConditionsVoigtWave2001}
Gerardin J and Lakhtakia A 2001 {\em Optik\/} {\bf 112} 493--495 ISSN 0030-4026

\bibitem{ossikovskiExtendedYehsMethod2017}
Ossikovski R and Arteaga O 2017 {\em Optics Letters\/} {\bf 42} 3690 ISSN
  0146-9592, 1539-4794

\bibitem{changSimpleFormulasCalculating2001}
Chang C~M and Shieh H~P~D 2001 {\em Japanese Journal of Applied Physics\/} {\bf
  40} 6391 ISSN 0021-4922, 1347-4065

\bibitem{zuAnalyticalNumericalModeling2022}
Zu R, Wang B, He J, Wang J~J, Weber L, Chen L~Q and Gopalan V 2022 {\em npj
  Computational Materials\/} {\bf 8} 246 ISSN 2057-3960

\bibitem{kirillovEigenvalueSurfacesDiabolic2005}
Kirillov O, Mailybaev A and Seyranian A 2005 On eigenvalue surfaces near a
  diabolic point {\em Proceedings. 2005 {{International Conference Physics}}
  and {{Control}}, 2005.\/} (Saint Petersburg, Russia: IEEE) pp 319--325 ISBN
  978-0-7803-9235-9

\bibitem{grundmannSingularOpticalAxes2016}
Grundmann M and Sturm C 2016 {\em Physical Review A\/} {\bf 93} 053839 ISSN
  2469-9926, 2469-9934 (\textit{Preprint} \eprint{1601.03760})

\bibitem{brenierLasingConicalDiffraction2016}
Brenier A 2016 {\em Applied Physics B\/} {\bf 122} 237 ISSN 0946-2171,
  1432-0649

\bibitem{mackayExorcizingGhostWaves2019}
Mackay T~G and Lakhtakia A 2019 {\em Optik\/} {\bf 192} 162926 ISSN 00304026

\bibitem{muralidharAlgebraComplexVectors2015}
Muralidhar K 2015 {\em Mathematics\/} {\bf 3} 781--842 ISSN 2227-7390

\bibitem{dupertuisGeneralizationComplexSnellDescartes1994}
Dupertuis M~A, Proctor M and Acklin B 1994 {\em JOSA A\/} {\bf 11} 1159--1166
  ISSN 1520-8532

\bibitem{alfonsoComplexUnitaryVectors2004}
Alfonso S, Alberdi C, Di{\~n}eiro J~M, Berrogui M, Hern{\'a}ndez B and
  S{\'a}enz C 2004 {\em Journal of the Optical Society of America A\/} {\bf 21}
  1776 ISSN 1084-7529, 1520-8532

\bibitem{changRayTracingAbsorbing2005}
Chang P~C, Walker J and Hopcraft K 2005 {\em Journal of Quantitative
  Spectroscopy and Radiative Transfer\/} {\bf 96} 327--341 ISSN 00224073

\bibitem{wangComplexRayTracing2008}
Wang Y, Shi P, Xin H and Wu L 2008 {\em Journal of Optics A: Pure and Applied
  Optics\/} {\bf 10} 075009 ISSN 1464-4258, 1741-3567

\bibitem{waseerNonuniformPlaneWaves2019}
Waseer W~I, Naqvi Q~A and Mughal M~J 2019 {\em Optics Communications\/} {\bf
  453} 124334 ISSN 00304018

\bibitem{zhangRigorousModelingLaser2015}
Zhang S, Asoubar D and Wyrowski F 2015 Rigorous modeling of laser light
  propagation through uniaxial and biaxial crystals {\em {{SPIE LASE}}\/} ed
  Glebov A~L and Leisher P~O (San Francisco, California, United States) p
  93460N

\bibitem{asoubarSimulationBirefringenceEffects2015}
Asoubar D, Zhang S and Wyrowski F 2015 {\em Optics Express\/} {\bf 23} 13848
  ISSN 1094-4087

\bibitem{mcleodVectorFourierOptics2014}
McLeod R~R and Wagner K~H 2014 {\em Advances in Optics and Photonics\/} {\bf 6}
  368 ISSN 1943-8206

\bibitem{sturmElectromagneticWavesCrystals2024}
Sturm C 2024 {\em Advanced Photonics Research\/} {\bf 5} 2300235 ISSN 2699-9293

\bibitem{zhangFullyVectorialSimulation2016}
Zhang S and Wyrowski F 2016 Fully vectorial simulation of light propagation
  through uniaxial and biaxial crystals {\em {{SPIE Photonics Europe}}\/} ed
  Wyrowski F, Sheridan J~T and Meuret Y (Brussels, Belgium) p 988909

\bibitem{borzdovWavesLinearQuadratic1996}
Borzdov G~N 1996 {\em Pramana\/} {\bf 46} 245--257 ISSN 0304-4289, 0973-7111

\bibitem{berremanOpticsStratifiedAnisotropic1972}
Berreman D~W 1972 {\em Journal of the Optical Society of America\/} {\bf 62}
  502 ISSN 0030-3941

\bibitem{stallingaBerreman4x4Matrix1999}
Stallinga S 1999 {\em Journal of Applied Physics\/} {\bf 85} 3023--3031 ISSN
  0021-8979, 1089-7550

\bibitem{molerNineteenDubiousWays2003}
Moler C and Van~Loan C 2003 {\em SIAM Review\/} {\bf 45} 3--49 ISSN 0036-1445,
  1095-7200

\bibitem{zarifiPlaneWaveReflection2014}
Zarifi D, Soleimani M and Abdolali A 2014 {\em Iranian Journal of Electrical
  and Electronic Engineering\/} {\bf 10} 250--255 ISSN 2383-3890

\bibitem{hernandezScalableComputationJordan2017}
Hern{\'a}ndez F, Pick A and Johnson S~G 2017 Scalable computation of jordan
  chains (\textit{Preprint} \eprint{1704.05837})

\bibitem{seyranianCouplingEigenvaluesComplex2005}
Seyranian A~P, Kirillov O~N and Mailybaev A~A 2005 {\em Journal of Physics A:
  Mathematical and General\/} {\bf 38} 1723--1740 ISSN 0305-4470, 1361-6447

\bibitem{kirillovUnfoldingEigenvalueSurfaces2005}
Kirillov O~N, Mailybaev A~A and Seyranian A~P 2005 {\em Journal of Physics A:
  Mathematical and General\/} {\bf 38} 5531--5546 ISSN 0305-4470, 1361-6447

\bibitem{mailybaevStrongWeakCoupling2005}
Mailybaev A, Kirillov O and Seyranian A 2005 Strong and weak coupling of
  eigenvalues of complex matrices {\em Proceedings. 2005 {{International
  Conference Physics}} and {{Control}}, 2005.\/} (Saint Petersburg, Russia:
  IEEE) pp 312--318 ISBN 978-0-7803-9235-9

\bibitem{kirillovGeometricalOpticsStability2025}
Kirillov O 2025 {\em Mathematics\/} {\bf 13} 382 ISSN 2227-7390

\bibitem{kirillovDissipationinducedInstabilitiesMagnetized2018}
Kirillov O~N 2018 {\em Journal of Mathematical Sciences\/} {\bf 235} 455--472
  ISSN 1573-8795

\bibitem{kirillovFindingStrongestStable2021}
Kirillov O~N and Overton M~L 2021 {\em The Quarterly Journal of Mechanics and
  Applied Mathematics\/} {\bf 74} 223--250 ISSN 0033-5614, 1464-3855

\bibitem{kirillovLocatingSetsExceptional2018}
Kirillov O~N 2018 {\em Entropy\/} {\bf 20} 502 ISSN 1099-4300

\bibitem{wiersigDistanceExceptionalPoints2022}
Wiersig J 2022 {\em Physical Review Research\/} {\bf 4} 033179 ISSN 2643-1564

\bibitem{wiersigRevisitingHierarchicalConstruction2022}
Wiersig J 2022 {\em Physical Review A\/} {\bf 106} 063526 ISSN 2469-9926,
  2469-9934

\bibitem{wiersigReviewExceptionalPointbased2020}
Wiersig J 2020 {\em Photonics Research\/} {\bf 8} 1457 ISSN 2327-9125

\bibitem{pessoaAvoidingMatrixExponentials2024}
Pessoa P, Schweiger M and Presse S 2024 {\em The Journal of Chemical Physics\/}
  {\bf 160} 094109 ISSN 0021-9606, 1089-7690 (\textit{Preprint}
  \eprint{2312.05647})

\bibitem{barnettParallelNonuniformFast2019}
Barnett A~H, Magland J~F and af~Klinteberg L 2019 A parallel non-uniform fast
  fourier transform library based on an "exponential of semicircle" kernel
  (\textit{Preprint} \eprint{1808.06736})

\bibitem{rodriguezvarelaApplicationNonuniformFFT2020}
Rodr{\'i}guez~Varela F, Irag{\"u}en B~G and {Sierra-Casta{\~n}er} M 2020 {\em
  IEEE Transactions on Antennas and Propagation\/} {\bf 68} 7571--7579 ISSN
  1558-2221

\bibitem{ballantineConicalDiffractionDispersion2014}
Ballantine K~E, Donegan J~F and Eastham P~R 2014 {\em Physical Review A\/} {\bf
  90} 013803 ISSN 1050-2947, 1094-1622

\bibitem{wuBandGapEngineering2020}
Wu F, Guo Z~W, Wu J~J, Jiang H~T and Du G~Q 2020 {\em Acta Physica Sinica\/}
  {\bf 69} 154205 ISSN 1000-3290, 1000-3290

\bibitem{qianCouplingInteractionsAnisotropic2023}
Qian L~M, Sun M~R and Zheng G~G 2023 {\em Acta Physica Sinica\/} {\bf 72}
  077101 ISSN 1000-3290, 1000-3290

\bibitem{passlerGeneralized442017}
Passler N~C and Paarmann A 2017 {\em JOSA B\/} {\bf 34} 2128--2139 ISSN
  1520-8540

\bibitem{passlerLayerresolvedAbsorptionLight2020}
Passler N~C, Jeannin M and Paarmann A 2020 {\em Physical Review B\/} {\bf 101}
  165425

\bibitem{harutyunyanAntidiffractionLight2015}
Harutyunyan H 2015 {\em Nature Photonics\/} {\bf 9} 213--214 ISSN 1749-4885,
  1749-4893

\bibitem{highVisiblefrequencyHyperbolicMetasurface2015}
High A~A, Devlin R~C, Dibos A, Polking M, Wild D~S, Perczel J, De~Leon N~P,
  Lukin M~D and Park H 2015 {\em Nature\/} {\bf 522} 192--196 ISSN 0028-0836,
  1476-4687

\bibitem{wuStrongExtrinsicChirality2021}
Wu B, Wang M, Wu F and Wu X 2021 {\em Applied Optics\/} {\bf 60} 4599 ISSN
  1559-128X, 2155-3165

\bibitem{wuNarrowbandDirectionalChiral2025}
Wu B, Huang X, Liu H and Wu X 2025 {\em Optics Express\/} {\bf 33} 16965--16975

\bibitem{salamaFreeSpaceSuper2020}
Salama N~A, Desouky M, Obayya S~S~A and Swillam M~A 2020 {\em Scientific
  Reports\/} {\bf 10} 11529 ISSN 2045-2322

\bibitem{berryOpticalPolarizationEvolution2011}
Berry M~V 2011 {\em Journal of Optics\/} {\bf 13} 115701 ISSN 2040-8978,
  2040-8986

\bibitem{hernandezExceptionalPointsNonHermitian2011}
Hern{\'a}ndez E, J{\'a}uregui A and Mondrag{\'o}n A 2011 {\em Physical Review
  E\/} {\bf 84} 046209 ISSN 1539-3755, 1550-2376

\bibitem{songBreakupRecoveryTopological2019}
Song W, Sun W, Chen C, Song Q, Xiao S, Zhu S and Li T 2019 {\em PHYSICAL REVIEW
  LETTERS\/}

\bibitem{shuChiralTransmissionOpen2024}
Shu X, Zhong Q, Hong K, You O, Wang J, Hu G, Al{\`u} A, Zhang S,
  Christodoulides D~N and Chen L 2024 {\em Light: Science \& Applications\/}
  {\bf 13} 65 ISSN 2047-7538

\bibitem{burkeNonhermitianScatteringTightbinding2020}
Burke P~C, Wiersig J and Haque M 2020 {\em Physical Review A\/} {\bf 102}
  012212 ISSN 2469-9926, 2469-9934

\bibitem{suDirectMeasurementNonhermitian2021}
Su R, Estrecho E, Biega{\'n}ska D, Huang Y, Wurdack M, Pieczarka M, Truscott
  A~G, Liew T~C~H, Ostrovskaya E~A and Xiong Q 2021 {\em Science Advances\/}
  {\bf 7} eabj8905 ISSN 2375-2548

\bibitem{xiaNonlinearTuningPT2021a}
Xia S, Kaltsas D, Song D, Komis I, Xu J, Szameit A, Buljan H, Makris K~G and
  Chen Z 2021 {\em Science\/} {\bf 372} 72--76 ISSN 0036-8075, 1095-9203

\bibitem{tuRenyiEntropiesNegative2022}
Tu Y~T, Tzeng Y~C and Chang P~Y 2022 {\em SciPost Physics\/} {\bf 12} 194 ISSN
  2542-4653

\bibitem{hanExceptionalEntanglementPhenomena2023}
Han P~R, Wu F, Huang X~J, Wu H~Z, Zou C~L, Yi W, Zhang M, Li H, Xu K, Zheng D,
  Fan H, Wen J, Yang Z~B and Zheng S~B 2023 {\em Physical Review Letters\/}
  {\bf 131} 260201 ISSN 0031-9007, 1079-7114 (\textit{Preprint}
  \eprint{2210.04494})

\bibitem{tuGeneralPropertiesFidelity2023}
Tu Y~T, Jang I, Chang P~Y and Tzeng Y~C 2023 {\em Quantum\/} {\bf 7} 960

\bibitem{brenierPhaseDistributionsAccompanying2022}
Brenier A, Majchrowski A and Michalski E 2022 {\em Optical Materials\/} {\bf
  128} 112353 ISSN 09253467

\bibitem{berryConicalDiffractionComplexified2006}
Berry M~V and Jeffrey M~R 2006 {\em Journal of Optics A: Pure and Applied
  Optics\/} {\bf 8} 1043--1051 ISSN 1464-4258, 1741-3567

\bibitem{ikramovTakagisDecompositionSymmetric2012}
Ikramov {\relax Kh}~D 2012 {\em Computational Mathematics and Mathematical
  Physics\/} {\bf 52} 1--3 ISSN 0965-5425, 1555-6662

\bibitem{houdeMatrixDecompositionsQuantum2024}
Houde M, McCutcheon W and Quesada N 2024 {\em Canadian Journal of Physics\/}
  {\bf 102} 497--507 ISSN 0008-4204, 1208-6045 (\textit{Preprint}
  \eprint{2403.04596})

\bibitem{berryProximityDegeneraciesChiral2006}
Berry M~V 2006 {\em Journal of Physics A: Mathematical and General\/} {\bf 39}
  10013--10018 ISSN 0305-4470, 1361-6447

\bibitem{brenierPolarizationPropertiesLasing2014a}
Brenier A 2014 {\em Laser Physics Letters\/} {\bf 11} 115819 ISSN 1612-2011,
  1612-202X

\bibitem{brenierVoigtWaveInvestigation2015}
Brenier A 2015 {\em Journal of Optics\/} {\bf 17} 075603 ISSN 2040-8978,
  2040-8986

\bibitem{chatterjeeRevisitingFresnelCoefficients2003}
Chatterjee M~R and Nema S 2003 Revisiting the fresnel coefficients for uniform
  plane wave propagation across a nonchiral, reciprocal and chiral,
  nonreciprocal interface {\em Optical {{Science}} and {{Technology}},
  {{SPIE}}'s 48th {{Annual Meeting}}\/} ed Ambs P and Beyette Jr F~R (San
  Diego, California, USA) p~22

\bibitem{wangAsymmetricWavefrontShaping2023}
Wang B, Li Y, Shen X and Krolikowski W 2023 {\em Optics Express\/} {\bf 31}
  25143--25152 ISSN 1094-4087

\bibitem{yehGeneralizedModelWire1982}
Yeh P 1982 Generalized {{Model For Wire Grid Polarizers}} {\em 25th {{Annual
  Technical Symposium}}\/} ed Trapani G~B (San Diego) pp 13--21

\bibitem{pereyraTransferMatrixMethod2022}
Pereyra P 2022 {\em physica status solidi (b)\/} {\bf 259} 2100405 ISSN
  0370-1972, 1521-3951

\bibitem{berryChiralConicalDiffraction2006}
Berry M~V and Jeffrey M~R 2006 {\em Journal of Optics A: Pure and Applied
  Optics\/} {\bf 8} 363--372 ISSN 1464-4258, 1741-3567

\bibitem{linFastVectorialCalculation2012}
Lin J, {Rodr{\'i}guez-Herrera} O~G, Kenny F, Lara D and Dainty J~C 2012 {\em
  Optics Express\/} {\bf 20} 1060 ISSN 1094-4087

\bibitem{liAnalyticalVectorialStructure2013}
Li J, Chen Y and Cao Q 2013 {\em Optics \& Laser Technology\/} {\bf 45}
  734--747 ISSN 00303992

\bibitem{martinez-herreroEvanescentFieldVectorial2008}
{Mart{\'i}nez-Herrero} R, Mej{\'i}as P~M and Carnicer A 2008 {\em Optics
  Express\/} {\bf 16} 2845 ISSN 1094-4087

\bibitem{chaumetFullyVectorialHighly2006}
Chaumet P~C 2006 {\em Journal of the Optical Society of America A\/} {\bf 23}
  3197 ISSN 1084-7529, 1520-8532

\bibitem{chenVectorialOpticalFields2018}
Chen J, Wan C and Zhan Q 2018 {\em Science Bulletin\/} {\bf 63} 54--74 ISSN
  20959273

\bibitem{napeRevealingInvarianceVectorial2022}
Nape I, Singh K, Klug A, Buono W, {Rosales-Guzman} C, McWilliam A,
  {Franke-Arnold} S, Kritzinger A, Forbes P, Dudley A and Forbes A 2022 {\em
  Nature Photonics\/} {\bf 16} 538--546 ISSN 1749-4885, 1749-4893

\bibitem{wuConformalFrequencyConversion2022}
Wu H~J, Yu B~S, Zhu Z~H, Gao W, Ding D~S, Zhou Z~Y, Hu X~P,
  {Rosales-Guzm{\'a}n} C, Shen Y and Shi B~S 2022 {\em Optica\/} {\bf 9} 187
  ISSN 2334-2536

\bibitem{gaoTopologicalPhotonicPhase2015}
Gao W, Lawrence M, Yang B, Liu F, Fang F, B{\'e}ri B, Li J and Zhang S 2015
  {\em Physical Review Letters\/} {\bf 114} 037402 ISSN 0031-9007, 1079-7114

\bibitem{merkulovPossibleConicalSingularities2011}
Merkulov V~S 2011 {\em JETP Letters\/} {\bf 94} 353--355 ISSN 0021-3640,
  1090-6487

\bibitem{sturmSingularOpticalAxes2016}
Sturm C and Grundmann M 2016 {\em Physical Review A\/} {\bf 93} 053839 ISSN
  2469-9926, 2469-9934

\bibitem{grundmannAngularPositionSingular2021}
Grundmann M and Sturm C 2021 {\em Physical Review A\/} {\bf 103} 053510 ISSN
  2469-9926, 2469-9934

\bibitem{sturmPropagationElectromagneticWaves}
Sturm C and Grundmann M {\em physica status solidi (RRL) -- Rapid Research
  Letters\/} {\bf n/a} 2400402 ISSN 1862-6270

\bibitem{brenierPolarizationPropertiesLasing2014}
Brenier A 2014 {\em Laser Physics Letters\/} {\bf 11} 115819 ISSN 1612-2011,
  1612-202X

\bibitem{pancharatnamPropagationLightAbsorbing1955}
Pancharatnam S 1955 {\em Proceedings of the Indian Academy of Sciences -
  Section A\/} {\bf 42} 235--248 ISSN 0370-0089

\bibitem{schellLaserStudiesInternal1978b}
Schell A~J and Bloembergen N 1978 {\em Journal of the Optical Society of
  America\/} {\bf 68} 1098 ISSN 0030-3941

\bibitem{brenierChiralityDichroismCompetition2017}
Brenier A, Majchrowski A and Michalski E 2017 {\em Optical Materials\/} {\bf
  72} 813--820 ISSN 09253467

\bibitem{berryConicalDiffractionAsymptotics2004}
Berry M~V 2004 {\em Journal of Optics A: Pure and Applied Optics\/} {\bf 6}
  289--300 ISSN 1464-4258, 1741-3567

\bibitem{schellLaserStudiesInternal1978a}
Schell A~J and Bloembergen N 1978 {\em Journal of the Optical Society of
  America\/} {\bf 68} 1093 ISSN 0030-3941

\bibitem{schellLaserStudiesInternal1978}
Schell A~J and Bloembergen N 1978 {\em Physical Review A\/} {\bf 18} 2592--2602
  ISSN 0556-2791

\bibitem{grantFrequencydoubledConicallyrefractedGaussian2014}
Grant S~D, Zolotovskaya S~A, Kalkandjiev T~K, Gillespie W~A and Abdolvand A
  2014 {\em Optics Express\/} {\bf 22} 21347 ISSN 1094-4087

\bibitem{zolotovskayaSecondharmonicConicalRefraction2011}
Zolotovskaya S~A, Abdolvand A, Kalkandjiev T~K and Rafailov E~U 2011 {\em
  Applied Physics B\/} {\bf 103} 9--12 ISSN 0946-2171, 1432-0649

\bibitem{kroupaSecondharmonicConicalRefraction2010}
Kroupa J 2010 {\em Journal of Optics\/} {\bf 12} 045706 ISSN 2040-8978,
  2040-8986

\bibitem{alekseevaShadowConicalRefraction1999}
Alekseeva L, Kidyarov B~I, Meshalkina S and Stroganov V~A 1999 Shadow conical
  refraction when the optical harmonics are generated {\em {{ICONO}} '98:
  {{Laser Spectroscopy}} and {{Optical Diagnostics--Novel Trends}} and
  {{Applications}} in {{Laser Chemistry}}, {{Biophysics}}, and
  {{Biomedicine}}\/} ed Chesnokov S~S, Kandidov V~P and Koroteev N~I (Moscow,
  Russia) p 465

\bibitem{shihConicalRefractionSecondHarmonic1969}
Shih H and Bloembergen N 1969 {\em Physical Review\/} {\bf 184} 895--904 ISSN
  0031-899X

\bibitem{velichkinaDemonstrationPhenomenaConical1980}
Velichkina T~S, Vasil'eva O~I, Israilenko A~I and Yakovlev I~A 1980 {\em Soviet
  Physics Uspekhi\/} {\bf 23} 176--177 ISSN 0038-5670

\bibitem{stroganovConicalRefractionSecond1980}
Stroganov V~I, Illarionov A~I and Kidyarov B~I 1980 {\em Journal of Applied
  Spectroscopy\/} {\bf 32} 341--344 ISSN 0021-9037, 1573-8647

\bibitem{illarionovExperimentalObservationConical1979}
Illarionov A and Stroganov V 1979 {\em Optics Communications\/} {\bf 31}
  239--241 ISSN 00304018

\bibitem{maSumfrequencyGenerationFemtosecond2018}
Ma J, Yuan P, Wang J, Xie G, Zhu H and Qian L 2018 {\em Optics Letters\/} {\bf
  43} 3670 ISSN 0146-9592, 1539-4794

\bibitem{peetFrequencyDoublingLaser2011}
Peet V and Shchemelyov S 2011 {\em Journal of Optics\/} {\bf 13} 055205 ISSN
  2040-8978, 2040-8986

\bibitem{olyslagerElectromagneticsExoticMedia2002}
Olyslager F and Lindell I 2002 {\em IEEE Antennas and Propagation Magazine\/}
  {\bf 44} 48--58 ISSN 1045-9243

\bibitem{berryConicalDiffractionObservations2006}
Berry M, Jeffrey M and Lunney J 2006 {\em Proceedings of the Royal Society A:
  Mathematical, Physical and Engineering Sciences\/} {\bf 462} 1629--1642 ISSN
  1364-5021, 1471-2946

\bibitem{saadGeneralStudyInternal2016}
Saad F and Belafhal A 2016  {\bf 2} 13

\bibitem{brenierLightPropagationProperties2019}
Brenier A, Majchrowski A and Michalski E 2019 {\em Optical Materials\/} {\bf
  91} 286--291 ISSN 09253467

\bibitem{brenierRevealingModesControlling2017}
Brenier A 2017 {\em Laser Physics\/} {\bf 27} 105001 ISSN 1054-660X, 1555-6611

\bibitem{favaroNonbirefringentLimitAll2011}
Favaro A and Bergamin L 2011 {\em Annalen der Physik\/} {\bf 523} 383--401 ISSN
  0003-3804, 1521-3889

\bibitem{eimerlQuantumElectrodynamicsOptical1988}
Eimerl D 1988 {\em Journal of the Optical Society of America B\/} {\bf 5} 1453
  ISSN 0740-3224, 1520-8540

\bibitem{berryChapterConicalDiffraction2007}
Berry M and Jeffrey M 2007 Chapter 2 {{Conical}} diffraction: {{Hamilton}}'s
  diabolical point at the heart of crystal optics {\em Progress in
  {{Optics}}\/} vol~50 (Elsevier) pp 13--50 ISBN 978-0-444-53023-3

\bibitem{mackayElectromagneticAnisotropyBianisotropy2010}
Mackay T~G and Lakhtakia A 2010 {\em Electromagnetic Anisotropy and
  Bianisotropy: A Field Guide\/} (New Jersey, NJ: World Scientific) ISBN
  978-981-4289-61-0

\bibitem{lakhtakiaCovariancesInvariancesMaxwell1995}
Lakhtakia A 1995 {\em Covariances and Invariances of the Maxwell Postulates\/}
  (WORLD SCIENTIFIC) pp 390--410 ISBN 978-981-02-2095-2 978-981-283-132-3

\bibitem{favaroRecentAdvancesClassical2012}
Favaro A 2012 Recent advances in classical electromagnetic theory

\bibitem{lakhtakiaWhenDoesChoice2007}
Lakhtakia A, Geddes~Iii J~B and Mackay T~G 2007 {\em Optics Express\/} {\bf 15}
  17709 ISSN 1094-4087

\bibitem{mackayModernAnalyticalElectromagnetic2020}
Mackay D~T~G 2020 {\em Modern {{Analytical Electromagnetic Homogenization}}
  with {{Mathematica}} ({{Second Edition}})\/} (IOP Publishing Ltd) ISBN
  978-0-7503-3423-5

\bibitem{baeklerKummerTensorDensity2014}
Baekler P, Favaro A, Itin Y and Hehl F~W 2014 {\em Annals of Physics\/} {\bf
  349} 297--324 ISSN 00034916

\bibitem{favaroLightPropagationLocal2016}
Favaro A and Hehl F~W 2016 {\em Physical Review A\/} {\bf 93} 013844

\bibitem{favaroElectromagneticWavePropagation2016}
Favaro A 2016 Electromagnetic wave propagation in metamaterials: {{A}} visual
  guide to fresnel-kummer surfaces and their singular points {\em 2016 {{URSI
  International Symposium}} on {{Electromagnetic Theory}} ({{EMTS}})\/} pp
  622--624

\bibitem{laxLinearNonlinearElectrodynamics1971}
Lax M and Nelson D~F 1971 {\em Physical Review B\/} {\bf 4} 3694--3731

\bibitem{chenWavevectorspaceMethodWave1993}
Chen B and Nelson D~F 1993 {\em Physical Review B\/} {\bf 48} 15365--15371 ISSN
  0163-1829, 1095-3795

\bibitem{matosConicalRefractionGeneralized2011}
Matos S~A, Paiva C~R and Barbosa A~M 2011 Conical refraction in generalized
  biaxial media: {{A}} geometric algebra approach {\em 2011 {{IEEE EUROCON}} -
  {{International Conference}} on {{Computer}} as a {{Tool}}\/} (Lisbon: IEEE)
  pp 1--3 ISBN 978-1-4244-7487-5 978-1-4244-7486-8 978-1-4244-7485-1

\bibitem{zeunerOpticalAnaloguesMassless2012}
Zeuner J~M, Efremidis N~K, Keil R, Dreisow F, Christodoulides D~N,
  T{\"u}nnermann A, Nolte S and Szameit A 2012 {\em Physical Review Letters\/}
  {\bf 109} 023602 ISSN 0031-9007, 1079-7114

\bibitem{berryConicalDiffractionNcrystal2010}
Berry M~V 2010 {\em Journal of Optics\/} {\bf 12} 075704 ISSN 2040-8986

\bibitem{raabMultipoleTheoryElectromagnetism2004}
Raab R~E and De~Lange O~L 2004 {\em Multipole Theory in Electromagnetism\/}
  (Oxford University Press) ISBN 978-0-19-856727-1

\bibitem{mikhailichenkoConicalRefractionExperiments2007}
Mikhailichenko {\relax Yu}~P 2007 {\em Russian Physics Journal\/} {\bf 50}
  788--795 ISSN 1064-8887, 1573-9228

\bibitem{peetFarfieldStructureGaussian2013}
Peet V 2013 {\em Optics Communications\/} {\bf 311} 150--155 ISSN 00304018

\bibitem{peetExperimentalStudyInternal2014}
Peet V 2014 {\em Journal of Optics\/} {\bf 16} 075702 ISSN 2040-8978, 2040-8986

\bibitem{mylnikovPartiallyCoherentConical2022}
Mylnikov V~{\relax Yu}, Dudelev V~V, Rafailov E~U and Sokolovskii G~S 2022 {\em
  Scientific Reports\/} {\bf 12} 16863 ISSN 2045-2322

\bibitem{mylnikovCloseRelationshipBessel2020}
Mylnikov V~{\relax Yu}, Rafailov E~U and Sokolovskii G~S 2020 {\em Optics
  Express\/} {\bf 28} 33900 ISSN 1094-4087

\bibitem{sokolovskiiConicalRefractionNew2013}
Sokolovskii G~S, Carnegie D~J, Kalkandjiev T~K and Rafailov E~U 2013 {\em
  Optics Express\/} {\bf 21} 11125 ISSN 1094-4087

\bibitem{abdolvandConicalRefractionNd2010}
Abdolvand A, Wilcox K~G, Kalkandjiev T~K and Rafailov E~U 2010 {\em Optics
  Express\/} {\bf 18} 2753 ISSN 1094-4087

\bibitem{phelanConicalDiffractionBessel2009}
Phelan C~F, O'Dwyer D~P, Rakovich Y~P, Donegan J~F and Lunney J~G 2009 {\em
  Optics Express\/} {\bf 17} 12891 ISSN 1094-4087

\bibitem{peinadoInterferometricCharacterizationStructured2015}
Peinado A, Turpin A, Iemmi C, M{\'a}rquez A, Kalkandjiev T~K, Mompart J and
  Campos J 2015 {\em Optics Express\/} {\bf 23} 18080 ISSN 1094-4087

\bibitem{odwyerCreationAnnihilationOptical2011}
O'Dwyer D~P, Phelan C~F, Rakovich Y~P, Eastham P~R, Lunney J~G and Donegan J~F
  2011 {\em Optics Express\/} {\bf 19} 2580 ISSN 1094-4087

\bibitem{odwyerGenerationContinuouslyTunable2010}
O'Dwyer D~P, Phelan C~F, Rakovich Y~P, Eastham P~R, Lunney J~G and Donegan J~F
  2010 {\em Optics Express\/} {\bf 18} 16480 ISSN 1094-4087

\bibitem{odwyerOpticalTrappingUsing2012}
O'Dwyer D~P, Ballantine K~E, Phelan C~F, Lunney J~G and Donegan J~F 2012 {\em
  Optics Express\/} {\bf 20} 21119 ISSN 1094-4087

\bibitem{qianQuasiAiryBeamsTunable2016}
Qian Y and Zhang S 2016 {\em Optics Express\/} {\bf 24} 9489 ISSN 1094-4087

\bibitem{peinadoConicalRefractionTool2013}
Peinado A, Turpin A, Lizana A, Fern{\'a}ndez E, Mompart J and Campos J 2013
  {\em Optics Letters\/} {\bf 38} 4100 ISSN 0146-9592, 1539-4794

\bibitem{turpinPolarizationTailoredNovel2015}
Turpin A, Loiko {\relax Yu}~V, Peinado A, Lizana A, Kalkandjiev T~K, Campos J
  and Mompart J 2015 {\em Optics Express\/} {\bf 23} 5704 ISSN 1094-4087

\bibitem{ribes-pleguezueloMethodSimulateAnalyse2017}
{Ribes-Pleguezuelo} P, Zhang S, Beckert E, Eberhardt R, Wyrowski F and
  T{\"u}nnermann A 2017 {\em Optics Express\/} {\bf 25} 5927 ISSN 1094-4087

\bibitem{peetImprovingDirectivityLaser2010}
Peet V 2010 {\em Optics Express\/} {\bf 18} 19566 ISSN 1094-4087

\bibitem{peetBiaxialCrystalVersatile2010}
Peet V 2010 {\em Journal of Optics\/} {\bf 12} 095706 ISSN 2040-8978, 2040-8986

\bibitem{turpinConicalRefractionFundamentals2016}
Turpin A, Loiko Y~V, Kalkandjiev T~K and Mompart J 2016 {\em Laser \& Photonics
  Reviews\/} {\bf 10} 750--771 ISSN 18638880

\bibitem{ciattoniCircularlyPolarizedBeams2003}
Ciattoni A, Cincotti G and Palma C 2003 {\em Journal of the Optical Society of
  America A\/} {\bf 20} 163 ISSN 1084-7529, 1520-8532

\bibitem{ciattoniAngularMomentumDynamics2003}
Ciattoni A, Cincotti G and Palma C 2003 {\em Physical Review E\/} {\bf 67}
  036618 ISSN 1063-651X, 1095-3787

\bibitem{berryOrbitalSpinAngular2005}
Berry M~V, Jeffrey M~R and Mansuripur M 2005 {\em Journal of Optics A: Pure and
  Applied Optics\/} {\bf 7} 685--690 ISSN 1464-4258, 1741-3567

\bibitem{odwyerConicalDiffractionLinearly2010}
O'Dwyer D~P, Phelan C~F, Ballantine K~E, Rakovich Y~P, Lunney J~G and Donegan
  J~F 2010 {\em Optics Express\/} {\bf 18} 27319 ISSN 1094-4087

\bibitem{luSpinorbitInteractionsGaussian2012}
Lu X and Chen L 2012 {\em Optics Express\/} {\bf 20} 11753 ISSN 1094-4087

\bibitem{qianGenerationAcceleratingBeams2019}
Qian Y, Zhang S and Ren Z 2019 {\em Annalen der Physik\/} {\bf 531} 1800473
  ISSN 0003-3804, 1521-3889

\bibitem{belyiPropagationHighorderCircularly2011}
Belyi V~N 2011 {\em Optical Engineering\/} {\bf 50} 059001 ISSN 0091-3286

\bibitem{khoninaComparativeInvestigationNonparaxial2015}
Khonina S and Kharitonov S 2015 {\em Journal of Modern Optics\/} {\bf 62}
  125--134 ISSN 0950-0340, 1362-3044

\bibitem{belyiSpintoorbitalAngularMomentum2013}
Belyi V~N, Khilo N~A, Kurilkina S~N and Kazak N~S 2013 {\em Journal of
  Optics\/} {\bf 15} 044018 ISSN 2040-8978, 2040-8986

\bibitem{brenierAspectsScalingOrbital2021}
Brenier A, Majchrowski A and Michalski E 2021 {\em Optik\/} {\bf 227} 166020
  ISSN 00304026

\bibitem{brenierInvestigationSumOrbital2020}
Brenier A 2020 {\em Journal of Optics\/} {\bf 22} 045603 ISSN 2040-8978,
  2040-8986

\bibitem{brenierEvolutionVorticesCreated2020}
Brenier A 2020 {\em Optical Materials\/} {\bf 110} 110504 ISSN 09253467

\bibitem{ahnConicalRefractionElastic2017}
Ahn Y~K, Lee H~J and Kim Y~Y 2017 {\em Scientific Reports\/} {\bf 7} 10072 ISSN
  2045-2322

\bibitem{mohammadiDesignOpticalDiode2019}
Mohammadi G, Jahangiri F and Amini T 2019 {\em Engineering Research Express\/}
  {\bf 1} 025047 ISSN 2631-8695

\bibitem{iqbalSituHologramsTwowave2024}
Iqbal M~W, Shiposh Y, Kohutych A, Marsal N, Grabar A~A and Montemezzani G 2024
  {\em JOSA B\/} {\bf 41} 1091--1098 ISSN 1520-8540

\bibitem{jonesNewCalculusTreatment1941}
Jones R~C 1941 {\em Journal of the Optical Society of America\/} {\bf 31} 488
  ISSN 0030-3941

\bibitem{landryCompleteMethodDetermine1995}
Landry G~D and Maldonado T~A 1995 {\em Journal of the Optical Society of
  America A\/} {\bf 12} 2048 ISSN 1084-7529, 1520-8532

\bibitem{asoubarEfficientSemianalyticalPropagation2014}
Asoubar D, Zhang S, Wyrowski F and Kuhn M 2014 {\em Journal of the Optical
  Society of America A\/} {\bf 31} 591 ISSN 1084-7529, 1520-8532

\bibitem{yangLightShapingFreeform2020a}
Yang L, Badar I, Hellmann C and Wyrowski F 2020 {\em Optics Express\/} {\bf 28}
  16202 ISSN 1094-4087

\bibitem{pflaumFarFieldCalculation2022a}
Pflaum C, Rall P and Lindlein N 2022 {\em Journal of the Optical Society of
  America A\/} {\bf 39} 198 ISSN 1084-7529, 1520-8532

\bibitem{yehElectromagneticPropagationBirefringent1979}
Yeh P 1979 {\em Journal of the Optical Society of America\/} {\bf 69} 742 ISSN
  0030-3941

\bibitem{ariyawansaObliquePropagationLight2018}
Ariyawansa A and Brown T~G 2018 {\em Optics Express\/} {\bf 26} 18832 ISSN
  1094-4087

\bibitem{zhangAnalysisPulseFront2014}
Zhang S, Asoubar D, Kammel R, Nolte S and Wyrowski F 2014 {\em Journal of the
  Optical Society of America A\/} {\bf 31} 2437 ISSN 1084-7529, 1520-8532

\bibitem{zhangAlgorithmPropagationElectromagnetic2017}
Zhang S, Hellmann C and Wyrowski F 2017 {\em Applied Optics\/} {\bf 56} 4566
  ISSN 0003-6935, 1539-4522

\bibitem{zhangNonparaxialIdealizedPolarizer2018}
Zhang S, Partanen H, Hellmann C and Wyrowski F 2018 {\em Optics Express\/} {\bf
  26} 9840 ISSN 1094-4087

\bibitem{abdulhalimExactMatrixMethod1999}
Abdulhalim I 1999 {\em Journal of Optics A: Pure and Applied Optics\/} {\bf 1}
  655--661 ISSN 1464-4258, 1741-3567

\bibitem{grechinFourierSpaceMethod2014}
Grechin S~G, Nikolaev P~P and Okhrimchuk A~G 2014 {\em Quantum Electronics\/}
  {\bf 44} 34--41 ISSN 1063-7818, 1468-4799

\bibitem{zhangEfficientRigorousPropagation2012}
Zhang S, Asoubar D, Wyrowski F and Kuhn M 2012 Efficient and rigorous
  propagation of harmonic fields through plane interfaces {\em {{SPIE Photonics
  Europe}}\/} ed Wyrowski F, Sheridan J~T, Tervo J and Meuret Y (Brussels,
  Belgium) p 84290J

\bibitem{zhangAlgorithmAccurateEfficient2024}
Zhang Y, Zhang S, Zheng Y and Wang S 2024 {\em Optics Express\/} {\bf 32} 29279
  ISSN 1094-4087

\bibitem{anelliOriginDependenceMaterial2015}
Anelli M, Jonsson D, Fliegl H and Ruud K 2015 {\em Molecular Physics\/} {\bf
  113} 1899--1913 ISSN 0026-8976, 1362-3028

\bibitem{wangSequentialThreeDimensionalNonlinear2023}
Wang C, Chen P, Wei D, Zhang L, Zhang Z, Xu L, Hu Y, Li J, Zhang Y, Xiao M, Chu
  J and Wu D 2023 {\em ACS Photonics\/}

\bibitem{jinCompactEngineeringPathEntangled2013}
Jin H, Xu P, Luo X~W, Leng H~Y, Gong Y~X, Yu W~J, Zhong M~L, Zhao G and Zhu S~N
  2013 {\em Physical Review Letters\/} {\bf 111} 023603 ISSN 0031-9007,
  1079-7114

\bibitem{tangHarmonicSpinOrbit2020}
Tang Y, Li K, Zhang X, Deng J, Li G and Brasselet E 2020 {\em Nature
  Photonics\/} ISSN 1749-4885, 1749-4893

\bibitem{kalkandjievConicalRefractionExperimental2008}
Kalkandjiev T~K and Bursukova M~A 2008 Conical refraction: An experimental
  introduction {\em Photonics {{Europe}}\/} ed Sheridan J~T and Wyrowski F
  (Strasbourg, France) p 69940B

\bibitem{abdolvandConicalDiffractionMulticrystal2011}
Abdolvand A 2011 {\em Applied Physics B\/} {\bf 103} 281--283 ISSN 0946-2171,
  1432-0649

\bibitem{mateosSimultaneousGenerationSecond2012}
Mateos L, Molina P, Galisteo J, L{\'o}pez C, Baus{\'a} L~E and Ram{\'i}rez M~O
  2012 {\em Optics Express\/} {\bf 20} 29940 ISSN 1094-4087

\bibitem{phelanConicalDiffractionGaussian2012}
Phelan C~F, Ballantine K~E, Eastham P~R, Donegan J~F and Lunney J~G 2012 {\em
  Optics Express\/} {\bf 20} 13201 ISSN 1094-4087

\bibitem{fangConicalThirdharmonicGeneration2017}
Fang X, Wei D, Wang Y, Wang H, Zhang Y, Hu X, Zhu S and Xiao M 2017 {\em
  Applied Physics Letters\/} {\bf 110} 111105 ISSN 0003-6951, 1077-3118

\bibitem{jangMulticycleTerahertzPulse2020}
Jang D and Kim K~Y 2020 {\em Optics Express\/}  16

\bibitem{chenQuasiphasematchingdivisionMultiplexingHolography2021}
Chen P, Wang C, Wei D, Hu Y, Xu X, Li J, Wu D, Ma J, Ji S, Zhang L, Xu L, Wang
  T, Xu C, Chu J, Zhu S, Xiao M and Zhang Y 2021 {\em Light: Science \&
  Applications\/} {\bf 10} 146 ISSN 2047-7538

\bibitem{pinheirodasilvaSpinOrbitalAngular2022}
Pinheiro Da~Silva B, Buono W~T, Pereira L~J, Tasca D~S, Dechoum K and Khoury
  A~Z 2022 {\em Nanophotonics\/} {\bf 11} 771--778 ISSN 2192-8614

\bibitem{liNonlinearMetasurfaceSimultaneous2017}
Li G, Wu L, Li K~F, Chen S, Schlickriede C, Xu Z, Huang S, Li W, Liu Y, Pun
  E~Y~B, Zentgraf T, Cheah K~W, Luo Y and Zhang S 2017 {\em Nano Letters\/}
  {\bf 17} 7974--7979 ISSN 1530-6984, 1530-6992

\bibitem{yuSimultaneousForwardBackward2008}
Yu N~E, Kang C, Yoo H~K, Jung C, Lee Y~L, Kee C~S, Ko D~K, Lee J, Kitamura K
  and Takekawa S 2008 {\em Applied Physics Letters\/} {\bf 93} 041104 ISSN
  0003-6951, 1077-3118

\bibitem{zhangSimultaneousNegativeRefraction2015}
Zhang J and Zhang X 2015 {\em Journal of Applied Physics\/} {\bf 118} 123103
  ISSN 0021-8979, 1089-7550

\bibitem{chenPhaseMatchingControlledOrbital2020a}
Chen Y, Ni R, Wu Y, Du L, Hu X, Wei D, Zhang Y and Zhu S 2020 {\em Physical
  Review Letters\/} {\bf 125} 143901 ISSN 0031-9007, 1079-7114

\bibitem{hehlSpacetimeMetricLocal2006}
Hehl F and Obukhov Y 2006 Spacetime metric from local and linear
  electrodynamics: {{A}} new axiomatic scheme {\em Special {{Relativity}}:
  {{Will}} It {{Survive}} the {{Next}} 101 {{Years}}?\/} ed Ehlers J and
  L{\"a}mmerzahl C (Berlin, Heidelberg: Springer) pp 163--187 ISBN
  978-3-540-34523-7

\bibitem{mackayElectromagneticAnisotropyBianisotropy2019}
Mackay T~G and Lakhtakia A 2019 {\em Electromagnetic {{Anisotropy}} and
  {{Bianisotropy}}: {{A Field Guide}}\/} 2nd ed (WORLD SCIENTIFIC) ISBN
  978-981-12-0313-8 978-981-12-0314-5

\bibitem{berryJohnFrederickNye2020}
Berry M 2020 {\em Biographical Memoirs of Fellows of the Royal Society\/} {\bf
  69} 425--441 ISSN 0080-4606, 1748-8494

\bibitem{nyePhysicalPropertiesCrystals2012}
Nye J~F 2012 {\em Physical Properties of Crystals: Their Representation by
  Tensors and Matrices\/} reprinted ed Oxford Science Publications (Oxford:
  Clarendon Press) ISBN 978-0-19-851165-6

\bibitem{yaoWavefrontPhasemodulationControl2013}
Yao C, Rodriguez F~J, {Bravo-Abad} J and Martorell J 2013 {\em Physical Review
  A\/} {\bf 87} 063804 ISSN 1050-2947, 1094-1622

\bibitem{chenPhaseMatchingControlledOrbital2020}
Chen Y, Ni R, Wu Y, Du L, Hu X, Wei D, Zhang Y and Zhu S 2020 {\em PHYSICAL
  REVIEW LETTERS\/}  6

\bibitem{zoldiParallelImplementationsSplitStep1997}
Zoldi S~M, Ruban V, Zenchuk A and Burtsev S 1997 {\em arXiv:physics/9711012\/}
  (\textit{Preprint} \eprint{physics/9711012})

\bibitem{trajtenberg-millsSimulatingCorrelationsStructured2020}
Trajtenberg-Mills S, Karnieli A, Voloch-Bloch N, Megidish E, Eisenberg H~S and
  Arie A 2020 {\em Laser \& Photonics Reviews\/} {\bf 14} 1900321 ISSN
  1863-8880, 1863-8899

\bibitem{barsiImagingNonlinearMedia2009}
Barsi C, Wan W and Fleischer J~W 2009 {\em Nature Photonics\/} {\bf 3} 211--215
  ISSN 1749-4885, 1749-4893

\bibitem{ellenbogenNonlinearGenerationManipulation2009}
Ellenbogen T, {Voloch-Bloch} N, {Ganany-Padowicz} A and Arie A 2009 {\em Nature
  Photonics\/} {\bf 3} 395--398 ISSN 1749-4885, 1749-4893

\bibitem{rozenbergInverseDesignSpontaneous2022}
Rozenberg E, Karnieli A, Yesharim O, {Foley-Comer} J, {Trajtenberg-Mills} S,
  Freedman D, Bronstein A~M and Arie A 2022 {\em Optica\/} {\bf 9} 602 ISSN
  2334-2536

\bibitem{yesharimObservationAllopticalStern2022}
Yesharim O, Karnieli A, Jackel S, Di~Domenico G, {Trajtenberg-Mills} S and Arie
  A 2022 {\em Nature Photonics\/} {\bf 16} 582--587 ISSN 1749-4885, 1749-4893

\bibitem{treebyNonlinearUltrasoundSimulation2020}
Treeby B~E, Wise E~S, Kuklis F, Jaros J and Cox B~T 2020 {\em The Journal of
  the Acoustical Society of America\/} {\bf 148} 2288 ISSN 1520-8524

\bibitem{treebyModelingNonlinearUltrasound2012}
Treeby B~E, Jaros J, Rendell A~P and Cox B~T 2012 {\em The Journal of the
  Acoustical Society of America\/} {\bf 131} 4324--4336 ISSN 1520-8524

\bibitem{firouziSpacePseudospectralMethod2017}
Firouzi K and {Khuri-Yakub} B~T 2017 {\em IEEE Transactions on Ultrasonics,
  Ferroelectrics, and Frequency Control\/} {\bf 64} 749--760 ISSN 1525-8955

\bibitem{zhongFastPropagationElectromagnetic2018}
Zhong H, Zhang S, Shi R, Hellmann C and Wyrowski F 2018 {\em Journal of the
  Optical Society of America A\/} {\bf 35} 661 ISSN 1084-7529, 1520-8532

\bibitem{zhaoNontrivialPhaseMatching2022}
Zhao X, Long H, Xu H, Kougo J, Xia R, Li J, Huang M and Aya S 2022 {\em
  Proceedings of the National Academy of Sciences\/} {\bf 119} e2205636119 ISSN
  0027-8424, 1091-6490

\bibitem{midwinterEffectsPhaseMatching1965}
Midwinter J~E and Warner J 1965 {\em British Journal of Applied Physics\/} {\bf
  16} 1135--1142 ISSN 0508-3443

\bibitem{chenCoordinatefreeApproachWave1981}
Chen H~C 1981 {\em Radio Science\/} {\bf 16} 1213--1215 ISSN 00486604

\bibitem{samlanChiralDynamicsExceptional2018}
Samlan C~T and Viswanathan N~K 2018 {\em Optics Letters\/} {\bf 43} 3538 ISSN
  0146-9592, 1539-4794

\bibitem{takenakaUnifiedFormalismPolarization1973}
Takenaka H 1973 {\em Japanese Journal of Applied Physics\/} {\bf 12} 1729--1731
  ISSN 0021-4922, 1347-4065

\bibitem{liSpintoorbitalAngularMomentum2020}
Li S, Zhang X, Gong W, Bu Z and Shen B 2020 {\em New Journal of Physics\/} {\bf
  22} 013054 ISSN 1367-2630

\bibitem{agocsAdaptiveSpectralMethod2024}
Agocs F~J and Barnett A~H 2024 {\em SIAM Journal on Numerical Analysis\/} {\bf
  62} 295--321 ISSN 0036-1429

\bibitem{wangTheoryAlgorithmHomeomorphic2020}
Wang Z, {Baladron-Zorita} O, Hellmann C and Wyrowski F 2020 {\em Optics
  Express\/} {\bf 28} 10552 ISSN 1094-4087

\bibitem{wrightDeepPhysicalNeural2022}
Wright L~G, Onodera T, Stein M~M, Wang T, Schachter D~T, Hu Z and McMahon P~L
  2022 {\em Nature\/} {\bf 601} 549--555 ISSN 0028-0836, 1476-4687

\bibitem{zusinBesselBeamTransformation2010}
Zusin D~H, Maksimenka R, Filippov V~V, Chulkov R~V, Perdrix M, Gobert O and
  Grabtchikov A~S 2010 {\em Journal of the Optical Society of America A\/} {\bf
  27} 1828 ISSN 1084-7529, 1520-8532

\bibitem{boydNonlinearOptics2019}
Boyd R~W 2019 {\em Nonlinear Optics\/} 4th ed (San Diego: Academic Press is an
  imprint of Elsevier) ISBN 978-0-12-811002-7

\bibitem{matosAnisotropyDoneRight2010}
Matos S~A, Paiva C~R and Barbosa A~M 2010 {\em The European Physical Journal
  Applied Physics\/} {\bf 49} 33006 ISSN 1286-0042, 1286-0050

\bibitem{pellat-finetFresnelDiffractionFractionalorder1994}
{Pellat-Finet} P 1994 {\em Optics Letters\/} {\bf 19} 1388--1390 ISSN 1539-4794

\bibitem{dengDiffractionInterpretedFractional1996}
Deng X, Li Y, Qiu Y and Fan D 1996 {\em Optics Communications\/} {\bf 131}
  241--245 ISSN 00304018

\bibitem{pellat-finetComplexOrderFractional2006}
{Pellat-Finet} P and Fogret {\'E} 2006 {\em Optics Communications\/} {\bf 258}
  103--113 ISSN 00304018

\bibitem{pellat-finetEffectDiffractionWigner2022}
{Pellat-Finet} P and Fogret {\'E} 2022 Effect of diffraction on wigner
  distributions of optical fields and how to use it in optical resonator
  theory. {{I}} -- stable resonators and gaussian beams (\textit{Preprint}
  \eprint{2005.13430})

\bibitem{nelsonDerivingTransmissionReflection1995}
Nelson D~F 1995 {\em Physical Review E\/} {\bf 51} 6142--6153 ISSN 1063-651X,
  1095-3787

\bibitem{changWavePropagationBianisotropic2014}
Chang P~H, Kuo C~Y and Chern R~L 2014 {\em Optics Express\/} {\bf 22} 25710
  ISSN 1094-4087

\bibitem{chernChiralSurfaceWaves2017}
Chern R~L and Yu Y~Z 2017 {\em Optics Express\/} {\bf 25} 11801 ISSN 1094-4087

\bibitem{zuOpticalSecondHarmonic2024}
Zu R, Wang B, He J, Weber L, Saha A, Chen L~Q and Gopalan V 2024 {\em npj
  Computational Materials\/} {\bf 10} 64 ISSN 2057-3960

\bibitem{grahamMultipoleSolutionMacroscopic2000}
Graham E~B and Raab R~E 2000 {\em Proceedings of the Royal Society A\/} {\bf
  456} 1193--1215

\bibitem{delangeElectromagneticBoundaryConditions2013}
{de Lange} O~L and Raab R~E 2013 {\em Journal of Mathematical Physics\/} {\bf
  54} 093513 ISSN 0022-2488

\bibitem{viottiCoherentPhaseTransfer2019}
Viotti A~L, Laurell F, Zukauskas A, Canalias C and Pasiskevicius V 2019 {\em
  Optics Letters\/} {\bf 44} 3066 ISSN 0146-9592, 1539-4794

\bibitem{zhuTopologicalOpticalDifferentiator2021}
Zhu T, Guo C, Huang J, Wang H, Orenstein M, Ruan Z and Fan S 2021 {\em Nature
  Communications\/} {\bf 12} 680 ISSN 2041-1723

\bibitem{lingTopologyInducedPhaseTransitions2021}
Ling X, Guan F, Cai X, Ma S, Xu H~X, He Q, Xiao S and Zhou L 2021 {\em Laser \&
  Photonics Reviews\/} {\bf 15} 2000492 ISSN 1863-8880, 1863-8899

\bibitem{lingPhotonicSpinHallEffect2023}
Ling X, Zhang Z, Dai Z, Wang Z, Luo H and Zhou L 2023 {\em Laser \& Photonics
  Reviews\/} {\bf 17} 2200783 ISSN 1863-8880, 1863-8899

\bibitem{landauCHAPTERXIELECTROMAGNETIC1984}
Landau L~D and Lifshitz E~M 1984 {{CHAPTER XI}} - {{ELECTROMAGNETIC WAVES IN
  ANISOTROPIC MEDIA}} {\em Electrodynamics of {{Continuous Media}} ({{Second
  Edition}})\/} ({\em Course of {{Theoretical Physics}}\/} vol~8) ed Landau L~D
  and Lifshitz E~M (Amsterdam: Pergamon) pp 331--357 ISBN 978-0-08-030275-1

\end{thebibliography}
\end{document}